# Optimization of Amorphous Germanium Electrical Contacts and Surface Coatings on High Purity Germanium Radiation Detectors

Mark Amman

Lawrence Berkeley National Laboratory, Berkeley, CA 94720 USA (retired)

## Abstract

Semiconductor detector fabrication technologies developed decades ago are widely employed today to commercially produce gamma-ray detectors from large volume, single crystals of high purity Ge (HPGe). Most all of these detectors are used exclusively for spectroscopy measurements and are of simple designs with only two impurity based electrical contacts produced with B implantation and Li diffusion. Though these technologies work well for the simple spectroscopy detectors, the Li contact in particular is thick and lacks room temperature stability in a manner that makes it inappropriate for many of the more complex detectors needed for gamma-ray imaging and particle tracking applications. These applications typically require detectors in which the radiation interaction position as well as the deposited energy is measured. Such detectors are created by dividing one or both of the standard electrical contacts into many segments. Thin films of amorphous semiconductors such as amorphous Ge (a-Ge) are the basis for an alternative electrical contact that is easy to fabricate, thin, and can be finely segmented. The a-Ge also functions well as a passivation coating on the HPGe surfaces not covered by the electrical contacts. The success of the a-Ge electrical contact and surface passivation technology is evidenced by the large number of detectors produced using the technology within the last decade and the wide application space for these detectors. The a-Ge film at the heart of this technology is typically deposited using RF sputtering. The properties of the a-Ge affect the performance of the resultant detectors, and these properties substantially depend on and are controllable through the sputter deposition process parameters. The subject of this paper is this interconnection of fabrication process parameters, a-Ge properties, and detector performance. The properties of a-Ge thin film electrical resistance, a-Ge contact electron injection, and room temperature storage stability were evaluated as a function of the sputter process parameters of sputter gas pressure and sputter gas $H_2$ composition. Two different sputter deposition systems were used to produce a-Ge resistors and HPGe detectors with a-Ge electrical contacts. These samples were electrically characterized as a function of temperature, and the relationships between the measurement results and the sputter process parameters that were common between the two sputter deposition systems were identified. A summary of this study and discussion of the relevance of the findings to the optimization of detector performance are given in this paper. Also included are an introduction to the a-Ge electrical contact and surface coating technology, a historical review of its development, a presentation of a theoretical model of the contact, and an overview of the detector physics of importance when optimizing the HPGe detectors with a-Ge contacts and surface coatings.

M. Amman, 2018, "Optimization of Amorphous Germanium Electrical Contacts and Surface Coatings on High Purity Germanium Radiation Detectors"

# 1. Introduction

Germanium is the detector material of choice for gamma-ray spectroscopy when good detection efficiency and excellent energy resolution are required. For this application, a single crystal of Ge is fabricated into a diode and operated under reverse voltage bias in order to produce a large region in the Ge that is depleted of free charge carriers. This depletion region forms the active volume in the detector that is used for the absorption and energy measurement of the incident radiation. The success of Ge is a result of a combination of favorable charge carrier generation statistics, excellent charge carrier transport, high atomic number and density, and ability to grow large single crystals of the material that have the required electrical properties [Pehl 1977]. Germanium's importance as a gamma-ray detector material emerged in the early 1960s when the Li drift process was first applied to Ge by Freck and Wakefield [Freck 1962] and then further developed to produce large detector volumes [Tavendale 1963] [Malm 1967] [Henck 1968] [Bertolini 1968 and references therein]. Lithium drifting allows for the creation of extremely electrically pure material through the precise introduction of the electron donor Li into an initially p-type semiconductor in an amount that closely matches the original electrically active acceptors in the semiconductor. This compensation method was used on Si by Pell [Pell 1960] prior to its implementation with Ge, and even today remains a useful process for Si. Due in part to the significant mobility of Li in Ge at room temperature, great difficulties were encountered in producing and storing the detectors [Bertolini 1968]. These problems were later overcome with the development of high purity Ge (HPGe) in the 1970s [Hall 1971] [Hansen 1971] [Llacer 1972] [Haller 1981]. This material is grown with a net electrically active impurity concentration at the $10^{10}$ impurities/cm$^3$ level. Such a purity level is sufficiently low for large depletion depths to be achieved without the need for Li compensation. This opened the door to sophisticated detector geometries and arrays, and a wider spectrum of applications [Pehl 1977] [Lee 2003] [Vetter 2007] [Eberth 2008].

Several challenges exist in furthering the HPGe based detector technology and expanding its application space. One of the main drawbacks of Ge is that it must be cooled to below about 150 K for large volume detectors [Nakano 1971] [Armantrout 1972] [Pehl 1973] [Nakano 1977] [McCabe 2015]. This is necessary in order to reduce the thermal generation of charge carriers that would act to mask the charge signal generated by the incident radiation as well as to improve charge carrier transport and to reduce other deleterious effects such as surface channels [Kingston 1956] [Llacer 1964] [Dinger 1975] [Malm 1976] [Hansen 1980] [Hull 1995]. In the past, nearly all HPGe based detector systems were cooled through the use of liquid $N_2$, which limits the application space for the technology due to the bulk, hazard, and supply issues associated with the cryogenic liquid. Fortunately, in more recent years, this challenge has been substantially met through the development and use of mechanical cooling [Smith 2002] [Vedrenne 2003] [Becker 2003] [Upp 2005] [Goldsten 2007] [Chiu 2017] [Ortec 2018] [Canberra 2018]. Increasing the crystal size of HPGe is also a challenge and is highly desirable in order to improve the efficiency of single detector systems and reduce the complexity of detector arrays. The crystal size and resultant detector volumes have advanced continuously since the development of HPGe [Sangsingkeow 2003] [Luke 2005], and efforts along this line of work continue today in both industry and academia. Another ongoing challenge for HPGe is the general improvement of performance to meet the needs of increasingly demanding applications. The value of a radiation detector is dictated by its performance as characterized by properties such as energy resolution, energy threshold, entrance window thickness, spatial resolution, efficiency, and response uniformity. These characteristics are determined in large part by the properties of the detector material but also by the detector design and technologies used to produce the detector. Additionally, factors such as ease of detector fabrication, fabrication yield, and production cost figure into the viability of a technology. The focus of this paper is on the fabrication technologies used to produce modern HPGe based detectors. Specifically, the use and optimization of thin film layers of amorphous semiconductors (typically Ge and Si) for the production of the detectors are presented.

The remainder of this first section of the paper is an introduction to the amorphous Ge (a-Ge) electrical contact and surface passivation technology. The technology is described, and its advantages and disadvantages as compared to conventional technologies are given. To facilitate this comparison, the basic physical construction and operation of a conventional HPGe based radiation detector are first presented.

The structure and function of one of the simplest HPGe based radiation detectors are illustrated in Figure 1.1a. The schematic cross-sectional diagram shown is of a simple planar detector. The basic function of the detector is to convert the energy of any absorbed radiation into an electrical signal. This detector consists of a single crystal block of HPGe material with two electrical contacts that have been fabricated onto opposing crystal faces. The HPGe is used to absorb the incident radiation while the electrical contacts allow for the application of a high voltage bias and the measurement of the electrical signal. The detection sequence in such a detector is as follows. Radiation such as a gamma ray enters the HPGe and can interact through a variety of physical processes [Debertin 1988] [Knoll 1989] [Gilmore 2008] resulting in the partial or complete absorption of the gamma-ray energy. Eventually, this deposited energy is partially converted into free charge carriers near the interaction site(s) in the form of electron-hole pairs. The number of electron-hole pairs generated is on average directly proportional to the energy deposited by the gamma ray. At this point in the detection sequence, no measurable signal is present since the electrons and holes are in equal numbers and in close proximity to each other, thereby resulting in a net charge neutral region as seen by the detector contacts. To form a signal, the detector is operated with a high voltage applied across the HPGe. This detector voltage ($V_d$) introduces an electric field within the HPGe and causes the electrons and holes to drift and separate. The spatial separation of the electrons from the holes induces charge on the detector's electrical contacts. This induced charge is then typically converted to a voltage by a charge sensitive preamplifier connected to one of the contacts. If the electrons are completely collected to the contact held at positive voltage and the holes to the one with the relatively negative voltage, the induced charge will be equivalent to the electron-hole charge generated by the gamma-ray interaction event and therefore serves as a measure of the deposited energy. This then is the basis for gamma-ray spectroscopy in which the energy distribution of a gamma-ray source is measured. The magnitude of the detector voltage used is typically chosen, at a minimum, to fully deplete the HPGe of free charge carriers since any undepleted HPGe will have no internal electric field, and, as a result, radiation interaction events occurring in such regions will not produce appropriate signals. Furthermore, it is desirable to over deplete the detector so that the generated charge carriers are collected as fast as possible in order to minimize energy resolution degradation due to charge carrier trapping. This desire to over deplete to the point where the charge carriers achieve saturation velocity [Ottaviani 1975] [Jacoboni 1981] necessitates a typical applied voltage of one to a few thousand volts and fields at the electrical contacts of about 1 kV·cm$^{-1}$.

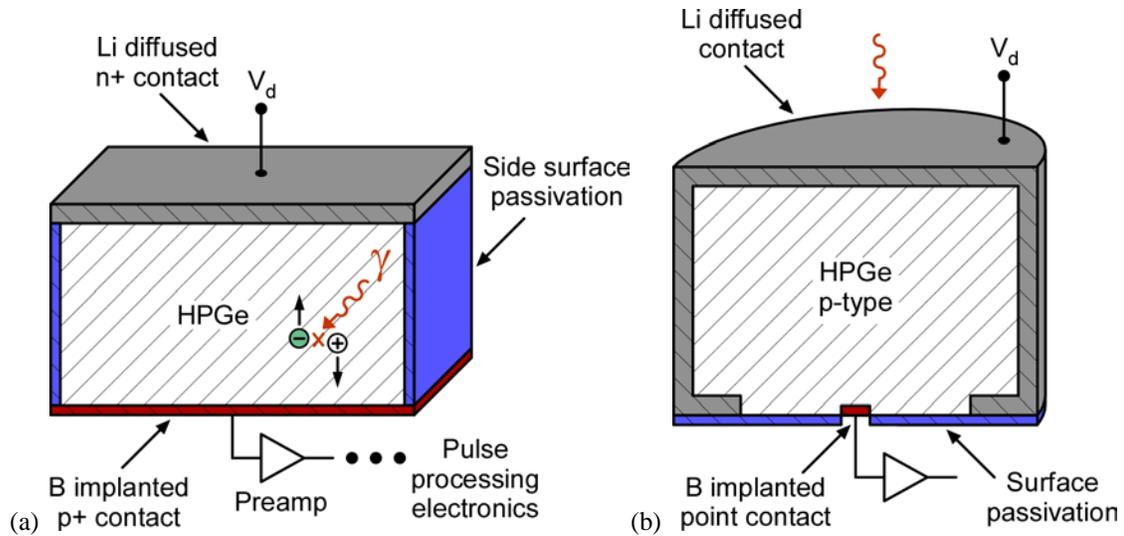

**Figure 1.1** Schematic cross-sectional drawings of HPGe based radiation detectors constructed using conventional impurity doped electrical contacts. **(a)** Simple planar detector. **(b)** P-type point contact detector.

From the above description of the basic HPGe detector operation, the desired behavior of the detector's electrical contacts can be identified. Consider the situation in which the electrical contacts on the HPGe are ohmic, that is, they freely exchange charge carriers with the HPGe. With such contacts, a large electrical current would flow through the HPGe due to the applied voltage [Pehl 1977], and the fluctuations in this current would easily mask the signal pulses generated by radiation interactions in the detector. Consequently, the electrical contacts cannot be ohmic and must instead block charge carrier injection into the HPGe while at the same time provide little to no barrier to charge collection through the HPGe and the contact. More specifically, the positive contact (the one attached to the positive high voltage in the Figure 1.1a example) must block hole injection and not inhibit electron collection, and, likewise, the relatively negative contact (the one held near ground potential by the charge sensitive preamplifier in the Figure 1.1a example) must block electron injection and not inhibit hole collection. Additional desirable features of the contacts are that they should withstand high electrical fields (~ few kV·cm$^{-1}$) without suffering an electrical breakdown, be mechanically robust, introduce only a thin dead layer, and be stable with time, temperature cycling, and environmental exposure.

Fabricating electrical contacts on HPGe with the appropriate characteristics is key to the successful production of high performance HPGe based detectors. This is, however, not sufficient. As is evident from the simple planar detector example of Figure 1.1a, in addition to the HPGe surfaces covered by electrical contacts, there are surfaces between these contacts, and these inter-contact surfaces must also be processed to achieve certain characteristics. These characteristics include low surface leakage current, ability to withstand high electrical fields without breaking down, charge neutrality or charge state that does not disturb the electric field uniformity within the HPGe, mechanical robustness, introduction of only a thin dead layer, and stability with time, temperature cycling, and environmental exposure. To obtain these characteristics, a detector fabrication process will normally include steps that chemically treat the inter-contact surfaces and coat the surfaces with thin films of various materials.

Though this introduction to the structure and function of a simple HPGe based spectroscopy detector has used the example of a simple planar configuration, the discussion also applies equally well to other detector geometries. This includes the widely used cylindrical shape that comprises the coaxial detector and the point contact detector (schematically shown in Figure 1.1b). Such detectors have a geometrical advantage that, when compared to the planar detector, enables larger fully depleted volumes to be achieved. The cylindrical geometry also benefits from a reduction in the size of the problematic inter-contact surface area. In common with the planar detector, these detectors also rely on the fabrication of a hole blocking contact, an electron blocking contact, and passivated inter-contact surfaces.

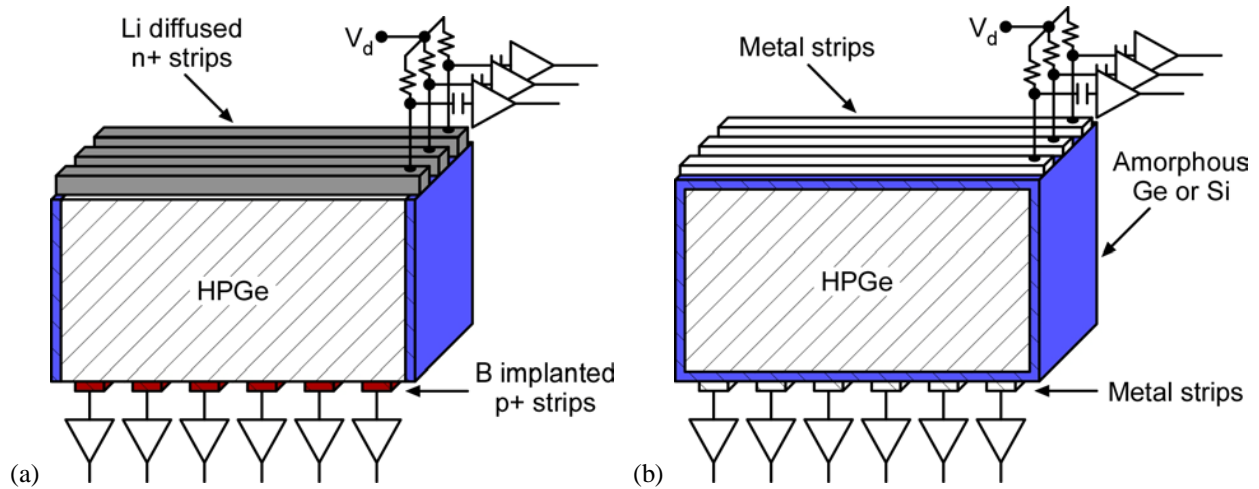

**Figure 1.2** Schematic cross-sectional drawings of example position sensitive HPGe based radiation detectors. Both detectors shown are of an orthogonal strip configuration. **(a)** Detector produced with conventional doped electrical contacts. **(b)** Detector produced with amorphous semiconductor electrical contacts.

    Well established and reliable processes exist for manufacturing the electrical contacts on simple detectors of the type shown in Figure 1.1. The industry standard technologies that have been used for many decades on HPGe are impurity doped contacts. Doped contacts are created by introducing elemental impurities near the surface of the HPGe in order to generate substantial numbers of charge carriers of either electrons (referred to as an n+ contact) or holes (p+ contact) depending on whether the impurity is an electron donor or acceptor, respectively. The blocking behavior of such a contact can be understood in part through the mass action law [Aschcroft 1976] [Sze 1981] [Blakemore 1987]. For example, by creating an n+ layer that has a high concentration of electrons, the layer must as a result have a low concentration of holes and, as such, would limit hole injection when used as a positive contact. In this case, the hole injection is a result of the generation and diffusion of holes within the neutral portion of the n+ layer [Shockley 1949] [Aschcroft 1976] [Sze 1981]. The analogous situation is true for the use of a p+ layer as an electron blocking contact.

    The two standard doped contacts, as shown in Figure 1.1, are the Li diffused contact and the B ion implanted contact for the positive and negative sides of the detector, respectively. These contacts have nearly all of the desirable characteristics discussed previously and perform well when used with simple detector geometries in which only the measurement of deposited energy is needed. However, for applications such as gamma-ray imaging and particle tracking, it is desired to measure the position of the radiation interaction events in the detector as well as the energy. This position measurement is often achieved by dividing or segmenting the electrical contacts into many strips or pixels and then reading out the signals from each of the contact segments. An example of such a position sensitive detector is the orthogonal strip detector shown schematically in Figure 1.2a when implemented with the Li and B doped contacts. The position of an interaction event in this detector is determined in the two dimensions parallel to the electrical contact face by the spatial intersection of the specific Li strip collecting the electrons from the event with the B strip collecting the holes. The location can be further localized in the dimension perpendicular to the contacts based on the difference in the charge arrival times at the two strips [Momayezi 1999] [Amman 2000a] [Amman 2000b]. Unfortunately, the production and reliable use of the Figure 1.2a detector are substantial challenges. Though the conventional B implanted contact can be readily segmented [Luke 1984] [Protic 1985] [Gutknecht 1990], the Li diffused contact presents a problem as a result of its thickness and significant diffusion of Li at room temperature in HPGe. Practically, the segmentation of the Li contact is limited to a granularity of no better than about 1 mm. Modern applications demand increasingly finer position resolution, meaning smaller contact structures, thereby limiting the applicability of Li diffused contacts. Although thin n+ ion implanted contacts have been developed as a replacement for the Li diffused one [Hubbard 1977] [Riepe 1979] [Pehl 1985], they are difficult to produce and are less robust than the B implanted contacts, making segmentation impractical. Furthermore, the segmentation of the implanted contact typically consists of etching through the implant to the undoped HPGe thereby leaving a sensitive inter-contact surface. Detectors have been successfully produced in which this inter-contact surface is not chemically treated or coated, but it remains unclear as to whether such processes would be fully successful in complex instruments in which detectors experience significant environmental exposure and lengthy storage times at room temperature.

    In addition to the impurity doped electrical contacts, other contact technologies have been used with HPGe to produce detectors. One such technology is the metal on semiconductor (MS) contact (also known as surface barrier or Schottky contact) that has been extensively used for electron blocking. To produce this contact, a metal such as Pt, Pd, Au, or Ni is deposited directly onto the properly prepared HPGe surface [Malm 1975a]. The electron blocking behavior is the result of an energy difference or barrier that

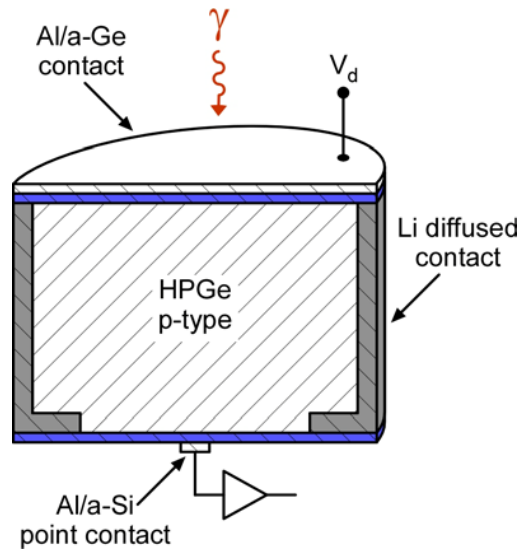

**Figure 1.3** Schematic cross-sectional drawing of a p-type point contact detector with a thin entrance window produced with an a-Ge electrical contact and a point contact made with a-Si.

exists and must be overcome to move electrons from the metal to the lowest electron conduction band state in the HPGe. Though the MS technology is successfully used with HPGe for simple detector configurations, its applicability to more complex geometries is hindered due to a number of disadvantages. First, the charge blocking behavior in this contact occurs very near the surface of the detector at the metal to HPGe interface. As a result, the electrical characteristics of the contact can easily be degraded through mechanical damage to the metal layer. Consequently, this fragile nature leads to less robust detector fabrication and necessitates greater handling care during fabrication and use. Second, much like the doped contacts, directly segmenting the MS contact leaves sensitive inter-contact surfaces in need of additional passivation processing steps. Finally, though in theory the proper selection of the metal and predeposition HPGe surface treatment should produce a contact that effectively blocks hole injection, in practice, Fermi level pinning appears to dominate the electrical behavior. This leads to contacts that are generally good at blocking electron injection rather than hole injection. The MS contact would therefore not be a suitable replacement for the hole blocking Li diffused contact. However, a possible exception to this could be sputtered Yt [Hull 2011].

Over the last two decades, another contact technology has emerged as an advantageous replacement for the conventional doped contacts on HPGe. This technology is based on amorphous semiconductors and is capable of providing finely segmented contacts on HPGe detectors with both electron blocking and hole blocking properties. Consequently, the amorphous semiconductor contact can replace both the B implanted contact and the problematic Li diffused contact [Hansen 1977]. As an example, a schematic cross-sectional drawing of an orthogonal strip detector based on amorphous semiconductor electrical contacts is shown in Figure 1.2b. The technology is simple and is implemented by first coating the properly prepared HPGe crystal with a high resistivity thin film of an amorphous semiconductor such as amorphous Ge (a-Ge) or amorphous Si (a-Si). This is then followed by depositing a patterned layer of metal, typically Al, on top of the amorphous film. The amorphous film dictates the charge blocking behavior of the contact, whereas the low resistivity metal defines the physical area of the contact. As will be discussed in detail later in this paper, the amorphous semiconductor to HPGe interface behaves much like the MS contact presented previously except that in the case of the amorphous semiconductor contact, good hole blocking is routinely achieved. In addition to fabrication simplicity, the amorphous semiconductor contact technology has a number of advantages. First, since the metal that defines the charge sensing contact area can be patterned using standard photolithographic processes, very fine contact segmentation can be achieved [Luke 1994a]. Second, the contact can be used as an entrance window capable of passing low energy x rays and gamma rays. This is because the layers of metal and amorphous semiconductor that compose the contact are thin, typically tens to hundreds of nanometers thick, and there appears to be comparatively little near contact dead layer [Luke 1994b]. For some detector geometries, such as the p-type coaxial or p-type point contact (as shown in Figure 1.1b), the radiation typically enters the detector through the positive contact. For a detector with conventional contacts, this would be through the Li diffused contact, which is thick and therefore absorbs low energy photons thereby limiting the detector efficiency at these low energies. Replacing at least part of this Li contact with one made of a-Ge, for example, creates a detector with the combined benefits of a thin radiation entrance window along with a mechanically robust surface for handling and mounting the detector (see Figure 1.3). The third benefit of amorphous semiconductor contacts is that the fabrication process naturally creates a detector in which all surfaces are passivated thereby significantly reducing performance changes caused by environmental exposure. This is because a-Ge acts as an excellent protective surface coating on HPGe [Hansen 1980] and, as is shown in the Figure 1.2b example, detectors based on the amorphous semiconductor contact technology are often completely coated with such a layer. A fourth advantage of the amorphous semiconductor contact is that a contact can be produced such that it both effectively blocks hole injection when positively biased and electron injection when negatively biased [Hansen 1977] [Luke 1992]. This bipolar blocking can simplify detector fabrication since a single contact process can in principle be used for the entire detector. It also enables the simple manufacture of more complex detector configurations such as those incorporating field shaping contacts [Amman 2000a] [Amman 2000b]. An example of this is given in Figure 1.4a. The detector schematically shown in this figure makes use of a positively biased grid contact surrounding a set of cathode strips. The grid improves the charge collection to the strips and, as a result, allows narrower, lower capacitance strips to be used [Cooper 2015] [Cooper 2018]. A final advantage of the amorphous semiconductor contact is that its combination of high film resistivity and charge injection blocking properties allow it to form the basis for unique detectors such as those relying on proximity charge sensing [Luke 2009] [Amman 2013]. An example of such a detector is schematically shown in Figure 1.4b. With this detector, the signals from the radiation interaction events are measured using metal electrodes placed near, but not electrically connected to, the HPGe crystal. The a-Ge layer coating the side of the detector that faces the readout electrodes is critical to the success of this detector. This layer must be transparent to the field of the radiation generated charge but still conductive enough that the collected charge and detector leakage current efficiently drain to the on-detector periphery strip contacts. The potential advantages of proximity charge sensing are simplified detector fabrication, expanded readout electrode geometry options, and improved position resolution through signal interpolation.

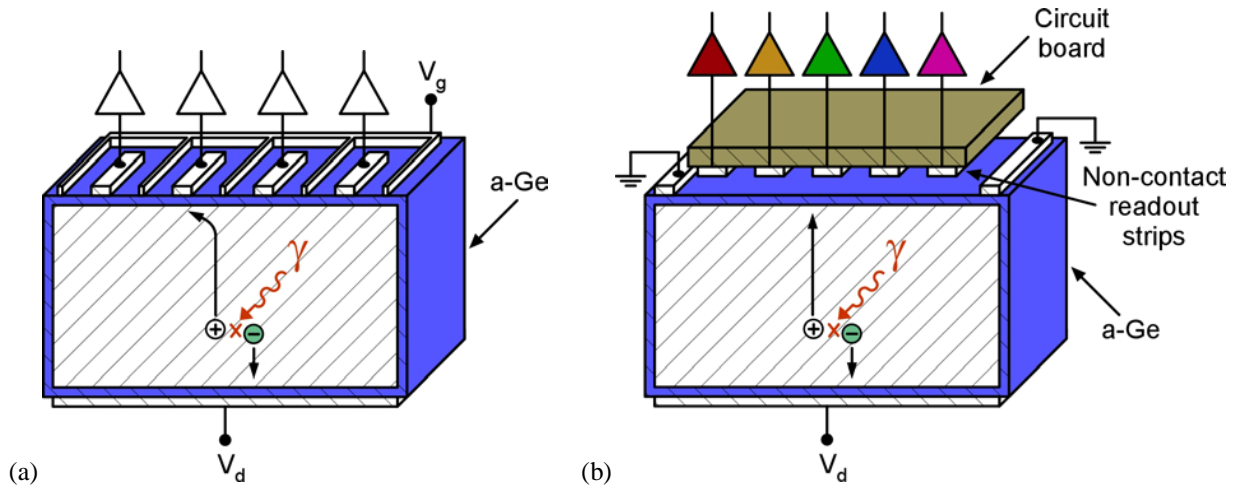

**Figure 1.4** Schematic cross-sectional drawings of example HPGe based radiation detectors that make use of the unique characteristics of the amorphous semiconductor technology. **(a)** Strip detector with a field shaping grid. The grid improves the hole collection to the strips and makes it possible to use narrower, lower capacitance strips [Amman 2000a] [Cooper 2015] [Cooper 2018]. **(b)** Proximity electrode signal readout detector [Luke 2009] [Amman 2013]. This example illustrates a simple single sided strip readout configuration. Readout from both detector sides is also possible as well as more complex electrode configurations.

 Though the amorphous semiconductor contact technology clearly offers significant advantages, it also has a couple of drawbacks. First, the contacts are not as effective at blocking charge injection as the doped contacts. This becomes an issue only when it is desired to operate the detectors at elevated temperatures. The meaning of elevated temperature is detector and application dependent but can be assessed based on the leakage current densities measured at various temperatures as reported later in this paper. Another downside of the amorphous contact is that it is not as robust as the Li diffused contact. The charge injection behavior of the Li contact is controlled by the highly doped Li diffused layer that is typically a significant fraction of a millimeter in thickness. To negatively affect the electrical properties of this contact through mechanical damage, the damage must be extensive enough to break through this thick layer. In contrast, the electrical behavior of the amorphous semiconductor contact is dictated by the amorphous semiconductor to HPGe interface. This interface is buried beneath the detector surface by the metal and amorphous semiconductor layers which combined will typically amount to a thickness of a micron or less. Such a thin layer is an advantage when the contact must be used as an entrance window for the radiation, but it is also much easier to damage during detector fabrication and end-use handling. Nonetheless, the contact has been shown to be robust enough to withstand processes such as photolithographic patterning and ultrasonic wire bonding without incurring degradation in its electrical characteristics.

 The amorphous semiconductor electrical contact technology has been the basis for many of the fine contact segmentation and thin window advancements associated with HPGe detectors in recent years. This success has led to a desire for further advancement through a better understanding of the contact physics and improved engineering. One goal then is to fully understand the relationship between the process parameters used when fabricating the contacts and surface coatings, and the performance characteristics of the resultant detector. Such detector fabrication optimization work has been pursued at Lawrence Berkeley National Laboratory (LBNL) and is the subject of this paper. The focus of this paper is on contacts and surface coatings that are based on RF sputtered a-Ge, which is the most common amorphous semiconductor used with HPGe. Specifically, measurements of a-Ge film resistance, a-Ge contact leakage current, and stability with room temperature storage all made as a function of the a-Ge deposition process parameters are summarized in this paper. Based on the results from these measurements, detectors with better optimized a-Ge contacts can be produced.

The outline for the remainder of this paper is the following. In the next section, Section 2, a historical background of the development and use of the amorphous semiconductor technology for the production of semiconductor based radiation detectors is given. Included in this section are descriptions of and results from example detectors fabricated at LBNL that demonstrate many of the capabilities of the technology. In Section 3, a theory that describes the physics of the amorphous semiconductor contact is reviewed. Also given in this section is a description of the analysis method based on this contact model that is used to analyze the measured detector data. In order to improve the performance of a detector, it is necessary to understand how the electrical contact and passivation layer properties affect the detector performance. In Section 4, this topic is covered through the presentation of the basic detector physics that governs performance. In Section 5, extensive details are provided on the fabrication and testing of the a-Ge thin film samples and a-Ge contact HPGe detectors. Then, in Sections 6 and 7, the results from the a-Ge thin film resistance and HPGe detector leakage current measurements are summarized. Finally, in Section 8, a summary of the findings from the a-Ge study and an outline of possible future work are given.

## 2. Amorphous Semiconductor Electrical Contact Historical Background and Modern Examples

Though the use of amorphous semiconductors as electrical contacts on semiconductor based radiation detectors has expanded dramatically only in recent years, the technology is not new, and its invention and development date back to the 1960s and 1970s. In this section, a historical review of the amorphous semiconductor contact is given. Following this, the value of the technology is illustrated through a number of example detectors produced at LBNL.

Amorphous Ge contacts on crystalline Ge and Si were first experimentally investigated by Grigorovici et al. [Grigorovici 1964]. Their work consisted of measuring the electrical characteristics of a-Ge films vacuum evaporated onto glass and the junctions formed between these same films deposited onto single crystals of Ge and Si. Significant findings from this work included that the charge transport through their a-Ge films at low temperatures was a thermally activated impurity/defect type conduction, the a-Ge to crystalline Ge or Si junction was rectifying (blocks charge injection), and the impurity/defect density of the a-Ge was sufficiently large so that when used as a blocking contact on crystalline Ge or Si, most of the depletion occurred within the crystalline semiconductor rather than the a-Ge layer. The first use of a-Ge as an electrical contact on a semiconductor based radiation detector was demonstrated by England and Hammer [England 1971]. They produced detectors consisting of single crystal Si with at least one contact composed of vacuum evaporated a-Ge with an overlayer of Al or other high conductivity metal. These detectors were successfully operated over-depleted, and the a-Ge contact was shown to be low noise, capable of blocking both hole and electron injection, and less sensitive to surface radiation damage than the standard Au surface barrier contact. The successful use of the a-Ge contact on Ge based radiation detectors was first shown by Hansen and Haller at LBNL [Hansen 1977]. As part of their effort to develop a thin hole blocking contact on HPGe capable of replacing the standard thick Li diffused contact, they discovered that vacuum evaporated a-Ge with an overlayer of Al could be used as both a hole blocking contact and an electron blocking contact on HPGe. Also noted in this initial work was an undesirable wide variation from detector to detector in the high field leakage current. A few years later, Hansen et al. [Hansen 1980] investigated the use of sputtered hydrogenated a-Ge as a protective surface coating on HPGe. In this work, the a-Ge was either DC or RF diode sputtered in an Ar-$H_2$ gas mixture onto the non-contact surfaces of planar detectors, and then the impacts on the detectors' side surface channels, leakage current, electronic noise, spectral performance, and surface stability were measured. The dependencies of the results on the HPGe impurity concentration, HPGe presputter surface treatment, and sputter gas $H_2$ concentration were explored. Several significant findings came out of this study. It was shown that the type and strength of the side surface channel on the detectors depended on the HPGe impurity type and concentration. Furthermore, by adjusting the $H_2$ concentration of the gas used to sputter the a-Ge coating on the detectors, an optimum surface condition could be obtained in which the surface channel effects were minimized and complete charge collection would result for radiation interaction events occurring very near the coated surface. The study also concluded that the a-Ge coating did not increase detector noise or negatively affect the detector spectral response and that the coated detectors were stable with exposure to a variety of ambient conditions.

Concurrent with this early experimental work on the use of a-Ge as an electrical contact and passivation layer was the experimental and theoretical work leading to an understanding of the charge transport in amorphous semiconductors [Mott 1979 and references therein] [Overhof 1989 and references therein] and the development of a theoretical model for the amorphous semiconductor to crystalline semiconductor junction [Dohler 1974] [Brodsky 1975a] [Brodsky 1975b]. The developments from this work that are relevant to the study summarized in this paper are presented later in Section 3.

As described above, the early studies of amorphous semiconductor contacts on radiation detectors all made use of vacuum evaporated films to produce the contacts. However, the successful demonstration of sputtered hydrogenated a-Ge as an HPGe detector passivation layer highlighted the potential benefits of such films over evaporated a-Ge and provided motivation to explore the use of the sputtered films as electrical contacts. Such work with sputtered hydrogenated films of a-Ge and a-Si was pursued at LBNL in the 1990s and later. In 1992, Luke et al. [Luke 1992] successfully demonstrated HPGe detectors that made use of RF sputtered hydrogenated a-Ge for both an electrical contact and the passivation coating on the non-contact surfaces. Planar detectors from both n-type and p-type HPGe were produced with one or both contacts consisting of sputtered a-Ge covered by a metal layer of either Au or Al. During the a-Ge deposition of the contacts, the sides of the detectors were also covered with a-Ge. This film was left in place and served as a surface passivation coating. These detectors functioned well with low leakage currents at 79 K and low electronic noise. The spectroscopic performance of the detectors was found to be as good as that obtained with detectors having conventional contacts, and the a-Ge contact was shown to have only a thin dead layer. The detectors could also be operated with the a-Ge contact biased with either voltage polarity even at high electric fields. However, the leakage current as a function of temperature data given in the paper indicates that the contact was likely better at blocking hole injection than it was at blocking electron injection. When the a-Ge contact was used to block holes, the detector exhibited leakage currents comparable to that obtained from detectors with conventional electrical contacts. Also shown in this paper was the first demonstration of the ease with which the a-Ge contact could be segmented. A two segment contact was formed on one side of a detector by vacuum evaporating the contact metal through a shadow mask. The potential of the a-Ge contact for enabling simple, finely segmented detector fabrication was demonstrated by Luke et al. [Luke 1994a] in 1994. In this work, a single sided strip detector was produced for x-ray computed tomography and consisted of 140 strips with a 0.5 mm center to center spacing and a 0.2 mm gap. Also in 1994, Luke et al. [Luke 1994b] further investigated the low energy entrance window behavior of the sputtered a-Ge contact. In this study, detectors with a-Ge contacts were produced and characterized for their low energy x-ray response. These detectors exhibited a bias voltage dependent spectral background that decreased with increasing voltage. At high detector voltages, the spectral background of the detectors was shown to be superior to that of detectors with either Pd surface barrier or B implanted contacts and to compare favorably to that obtained with Si(Li) surface barrier detectors.

Starting around 2000, much of the a-Ge contact based HPGe detector development work was directed towards position sensitive detectors for gamma-ray imaging and particle tracking. Orthogonal strip detectors produced completely using a-Ge contacts were demonstrated first as small prototypes [Luke 2000] [Amman 2000a] [Amman 2000b] and then as large area devices appropriate for imaging applications [Phlips 2002] [Coburn 2002] [Coburn 2005]. With these detectors it was shown that gamma-ray interaction positions within the detector could be determined in all three spatial dimensions [Momayezi 1999] [Amman 2000a] [Amman 2000b] and that the a-Ge contact made it possible to easily implement field shaping contacts in order to improve charge collection [Amman 2000a] [Amman 2000b]. During this period, it also became apparent that there was a need to better understand the relationship between the process parameters used when fabricating the amorphous semiconductor contacts and the performance characteristics of the resultant detectors. Following such a strategy of basic device development allowed better optimized contacts to be produced. For example, it was commonly observed in the a-Ge contact based strip detectors that for some fraction of the radiation interaction events, the generated charge would be collected directly to an inter-strip surface rather than to a strip [Luke 2000]. For such events, the detected energy is less than that actually deposited in the detector despite summing the signals measured by all nearby strips. However, it was demonstrated that this undesirable behavior could be substantially eliminated by adjusting the a-Ge sputter deposition process parameters [Amman 2000b] [Looker 2015b]. Another undesirable characteristic that was observed with some of the a-Ge contact detectors was that the leakage current would change substantially when the detectors were stored for extended periods of time at room temperature. This issue becomes a more apparent and significant problem when constructing complex detector arrays since this often necessitates storing the detectors for a year or more before final integration into the instrument. Fortunately, it was also shown that this issue can be eliminated and stable leakage current with room temperature storage be achieved with an appropriate adjustment of the a-Ge sputter deposition process parameters [Amman 2007] [Looker 2015a]. Another useful finding from the basic contact studies was that the charge injection blocking characteristics of a contact are dependent on the semiconductor and sputter process parameters used to produce the contact. For example, it was shown that adding $H_2$ to the Ar sputter gas can improve the hole blocking of the resultant a-Ge contact at the expense of the contact's ability to block electron injection. As a result, the hydrogenated a-Ge contact was shown to be much better at blocking hole injection than it was at blocking electron injection. In contrast to this was the contact formed from hydrogenated a-Si, which was observed to be better at blocking electron injection [Amman 2007] [Looker 2015a].

The substantial success of the HPGe detectors based on amorphous semiconductor contacts is evidenced by their use for a wide range of applications. Such detectors have been successfully employed or are being developed for space science [Coburn 2005] [Bandstra 2009] [Bandstra 2011] [Shih 2012] [Kierans 2016] [Chiu 2017], material science [Headspith 2007], nuclear and particle physics [Larson 2013], medical imaging [Dilmanian 1993], nuclear nonproliferation and homeland security [Ziock 2002] [Ziock 2003] [Cunningham 2005] [Mihailescu 2007], and environmental remediation [Phlips 2002], and are at the heart of commercially produced imaging instruments [PHDS 2018]. Example detectors fabricated at LBNL along with select measurements made with the detectors are shown in Figures 2.1 through 2.6. The HPGe crystals used to produce most all of these detectors were procured from the company Ortec [Ortec 2018]. Orthogonal strip detectors are the subject of Figures 2.1 through 2.3. These large volume detectors were produced from slices of HPGe nominally 10 cm in diameter and 1.5 cm thick that were cut at LBNL to an approximately square area of size 8 cm by 8 cm. All surfaces of the detectors were coated with either a-Ge or a-Si, and then Al metallization was evaporated onto each of the two contact faces to form the strips and a surrounding guard ring. The patterning of the strips and guard ring was accomplished through the use of shadow masks placed on the detector during the Al evaporation or through the use of photolithography. The photograph of Figure 2.1a is of a detector produced for the Gamma-Ray Imager/Polarimeter for Solar flares (GRIPS) instrument developed for the study of solar flare gamma-ray and hard x-ray emissions [Shih 2012]. Ten detectors were produced for the first iteration of this instrument. Each of these detectors was fabricated with a-Ge coating the non-contact sides of the detector and the positive contact face. On the negative contact face, a-Si was used so as to take advantage of the superior electron blocking of a-Si over that of a-Ge. The segmentation of each contact face consists of 149 strips with a center to center strip spacing of 0.5 mm and a gap of 0.06 mm between each strip. These strips are surrounded by a guard ring approximately 3 mm in width. The purpose of the guard ring is to take up surface leakage current along the sides of the detector and to reduce the negative effects associated with any side surface channel. A summary of the detector characterization measurements made on the set of GRIPS detectors prior to their integration into the flight instrument is given in Figure 2.2. The top graph displays for each detector the full depletion voltage along with the operating detector voltage used for the spectroscopy measurements. Relatively pure HPGe was used for all of the detectors so that low full depletion voltages would be achieved, thereby enabling operating voltages near 1 kV. Such low operating voltages simplify the instrument's high voltage system and improve the detector yield. The middle graph of Figure 2.2 is a plot of the average strip leakage current. These measurements were made with only a partial IR shield covering the detector. Consequently, the typically measured leakage current of 1 pA/strip is a result of both electrical contact injection and IR radiation charge generation and therefore does not provide a good measure of the low leakage current performance capability of the contact technology. Nonetheless, this level of leakage current is well below that needed to ensure that the current does not significantly contribute to the total detector electronic noise. For the typical operating conditions of the detectors, the detector electronic noise is dominated by series noise that results primarily from the readout amplifier noise voltage (and is dependent on the strip capacitance) rather than the parallel noise that results in part from detector leakage currents and resistances in parallel with the detector [Radeka 1972] [Goulding 1972] [Goulding 1982] [Spieler 2005]. A noise characteristic obtained with a GRIPS detector that supports this conclusion is presented later in this paper in Section 4. The lower graph of Figure 2.2 contains the average gamma-ray peak widths measured with each detector at the energies of 60 keV and 662 keV. The average peak widths were measured separately for the hole collecting strips and the electron collecting strips so that any dependence on the charge carrier type could be identified. At the energy of 60 keV, the peak widths are dominated by the detector electronic noise. At the higher energy of 662 keV, charge generation statistics and charge trapping effects contribute significantly to the peak widths. These data provide interesting information on the nature of the charge trapping in the HPGe. Based on the small electrode effect [Malm 1975b] [Luke 1996], it can be concluded that the electron collecting strips will be more sensitive to electron trapping, whereas the hole collecting strips will be more sensitive to hole trapping. From the 662 keV peak widths of Figure 2.2, it is clear that the widths are generally better for the hole collecting strips, implying that the trapping of holes is less significant than that of the electrons in the HPGe detector material. However, there is significant detector to detector variation in this behavior, including detectors that exhibit nearly the same resolution for all strips. It should also be noted that the electric field profile with depth in the detector may also play a role in the energy resolution difference observed between the hole collecting and electron collecting strips.

Another set of space science detectors is shown in the photograph of Figure 2.1b, and a summary of the characterization measurements made on the detectors prior to their integration into the flight instrument is given in Figure 2.3. These detectors were produced for the Compton Spectrometer and Imager (COSI) instrument developed for astrophysical observations from a high altitude balloon [Kierans 2016]. Versions of this instrument including those under its former name of the Nuclear Compton Telescope (NCT) were successfully flown in 2005 [Coburn 2005], 2009 [Bandstra 2009] [Bandstra 2011], and 2016 [Kierans 2016] [Chiu 2017]. A total of 14 detectors were produced for the latest version of the instrument, and 12 of these were assembled into the array shown in Figure 2.1b that is at the heart of the instrument. The COSI detectors are much like the GRIPS detectors just described except that the non-contact sides and both contact faces of the COSI detectors were all coated with a-Ge, and the strip segmentation is coarser. The segmentation of each contact face consists of 37 strips with a center to center strip spacing of 2 mm and a gap of 0.25 mm between each strip. These strips are surrounded by a guard ring 2 mm in width along with a gap approximately 1 mm in size between the edge of the guard ring and the edge of the detector. As can be seen from the detector summary of Figure 2.3, both p-type and n-type HPGe were used for the detectors, and a low typical strip leakage current of about 3 pA was achieved. The energy resolutions obtained with these detectors are similar to those of the GRIPS detectors, and, also like the GRIPS detectors, the hole collecting strips of the COSI detectors generally had better resolutions at the higher energy than the electron collecting strips.

A detector example demonstrating the ability to produce electrical contacts that have very fine segmentation is shown in Figure 2.4. This is a single sided strip detector designed for energy dispersive extended x-ray absorption fine structure measurements [Headspith 2007]. One contact face of this detector consists of 1024 strips produced with a-Ge and Al metallization, while the opposing contact face is a full area B implant coated with Al. The strips have a 50 μm center to center spacing and are 5 mm long. The detector is 1 mm thick and is operated at a temperature of about 100 K. The detector strips are read out in a current integrating mode using custom mixed signal ASICs.

The detector shown in Figure 2.5 implements a field shaping grid similar to that shown schematically in Figure 1.4a. The goal with this detector was to realize high event throughput at high counting rates while maintaining good energy resolution [Cooper 2015] [Cooper 2018]. Such characteristics are needed for applications in areas such as nuclear physics and spent nuclear fuel assay where the source radiation flux is high yet the target spectral features to be analyzed are weak. The HPGe crystal geometry of this detector is the same as that previously described for the GRIPS and COSI detectors. Unlike the orthogonal strip detectors, this high rate detector has only one of the electrical contact faces divided into segments. This contact is divided into ten strips, and each strip is separately read out. Each of these strips is surrounded by a narrow grid that is voltage biased in order to ensure complete charge collection to the strips. On the opposing face of the detector is a full area Li diffused contact coated with Al. The strip readout of this detector is not done for the purpose of position detection but rather to divide the total event rate among all strips thereby increasing the overall event processing rate of the detector. To further enhance the event throughput, short pulse processing shaping times are used. This strategy, however, will degrade energy resolution due to increasing detector series noise with decreasing shaping time. Since the series noise scales with detector capacitance, good energy resolution is maintained by keeping this capacitance relatively low through the use of large gaps between the strips and grid. The large gaps then necessitate the use of the voltage biased grid.

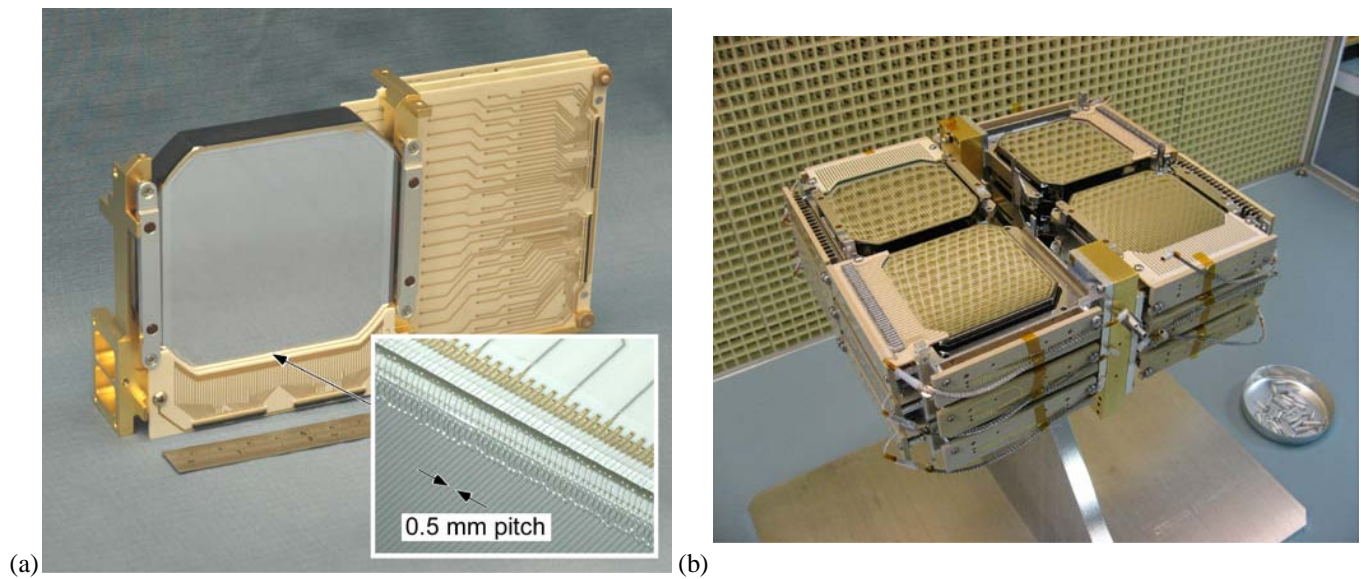

**Figure 2.1** Photographs of HPGe orthogonal strip detectors of the type schematically shown in Figure 1.2b. These detectors were produced for space science instruments. Each detector has an area of approximately 8 cm by 8 cm and is approximately 1.5 cm thick. **(a)** Example detector produced for the Gamma-Ray Imager/Polarimeter for Solar flares (GRIPS) instrument. Each side of the detector has 149 strips with a center to center strip spacing of 0.5 mm and a gap of 0.06 mm between each strip. The detector is held in an Al mount, and attached to this mount are circuit boards used to route the detector signals to flex circuits (not shown). Connections between the detector strips and the circuit board traces were made with ultrasonic wedge wire bonding. The larger of the two circuit boards is the board for the high voltage side of the detector. This circuit board is actually a stack of three circuit boards, and this stack houses the high voltage coupling capacitor and bias resistor for each readout channel. **(b)** Array of 12 detectors produced for the Compton Spectrometer and Imager (COSI) instrument. Each side of each detector has 37 strips with a center to center strip spacing of 2 mm and a gap of 0.25 mm between each strip.

The final example is the p-type point contact (PPC) detector shown in Figure 2.6a [Luke 1989] [Barbeau 2007] [Amman 2009]. This detector has a configuration much like that shown schematically in Figure 1.3. A Li diffused layer coated with Al forms the positive contact on the cylindrical shaped p-type HPGe crystal except for the top/front portion of the crystal that instead has a contact consisting of a-Ge coated with Al. The a-Ge contact area with its associated thin window dead layer enables the measurement of low energy gamma and x-rays. This particular detector was created as part of an effort to implement position detection in the PPC detector configuration and as such has the a-Ge contact electrically separated from the Li contact. The point contact on this detector consists of an a-Si layer with an approximately 2 mm diameter Al circle evaporated on top of it. The measured capacitance of this point contact and associated wiring was determined to be about 0.9 pF. An example energy spectrum measured from the point contact of this detector is shown in Figure 2.6b.

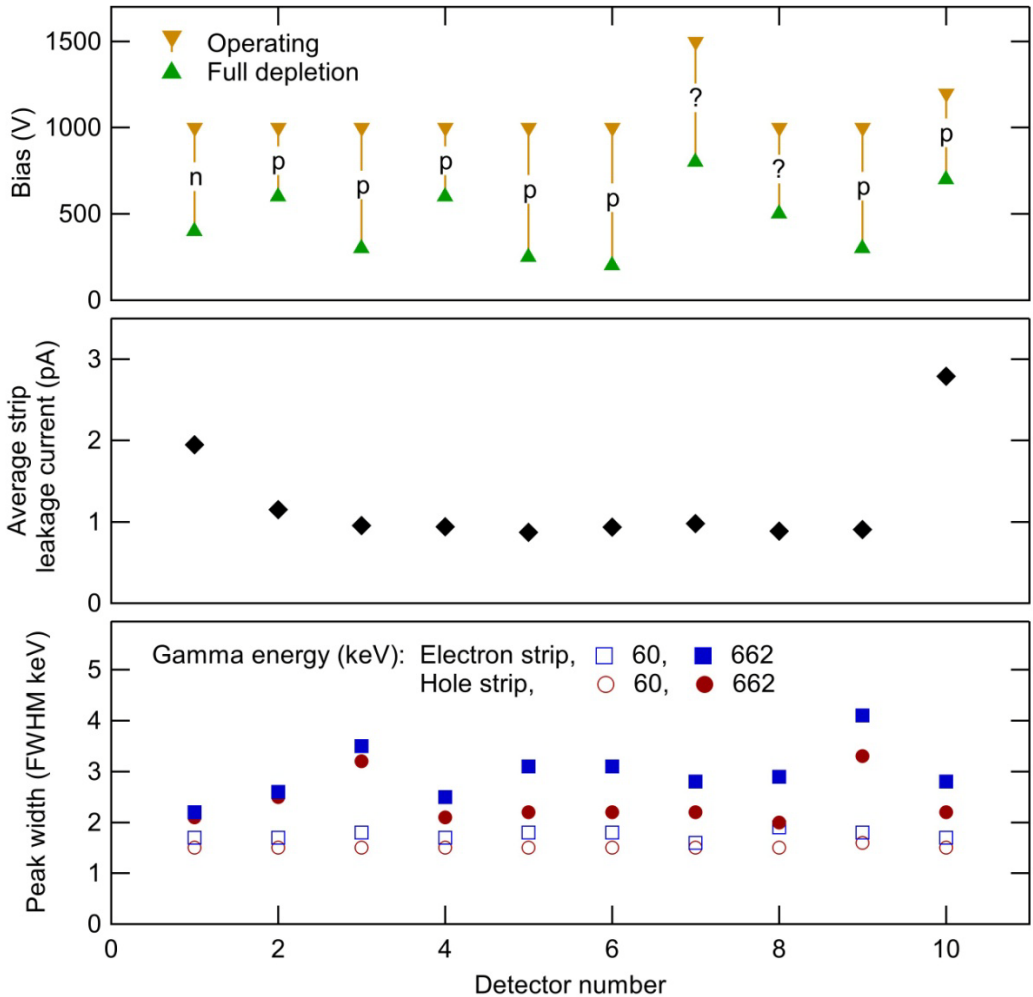

**Figure 2.2** Summary of the initial characterization measurements made on the set of ten orthogonal strip HPGe detectors produced for the GRIPS instrument. **(Top)** Full depletion voltage and operating voltage for each detector. The full depletion voltages were determined based on detector capacitance measured as a function of the applied voltage. Also indicated in this graph is the HPGe impurity type as determined from low energy gamma-ray count rate measurements made with the detector when operated in a partially depleted state. A question mark indicates that the result of the measurement was inconclusive. **(Middle)** Average strip leakage current for each detector. These measurements were made in a test cryostat that had only a partial IR shield covering the detector and achieved a typical detector cold stage temperature of about 81 K. **(Bottom)** Average energy resolutions of each detector measured at the gamma-ray energies of 60 keV (Am-241 source) and 662 keV (Cs-137 source). The averages were determined separately for electron and hole collecting strips. These spectroscopy measurements were made using compact, low power preamps operated at room temperature followed by conventional pulse processing electronics. A pulse shaping time of 8 µs was used.

M. Amman, 2018, "Optimization of Amorphous Germanium Electrical Contacts and Surface Coatings on High Purity Germanium Radiation Detectors"

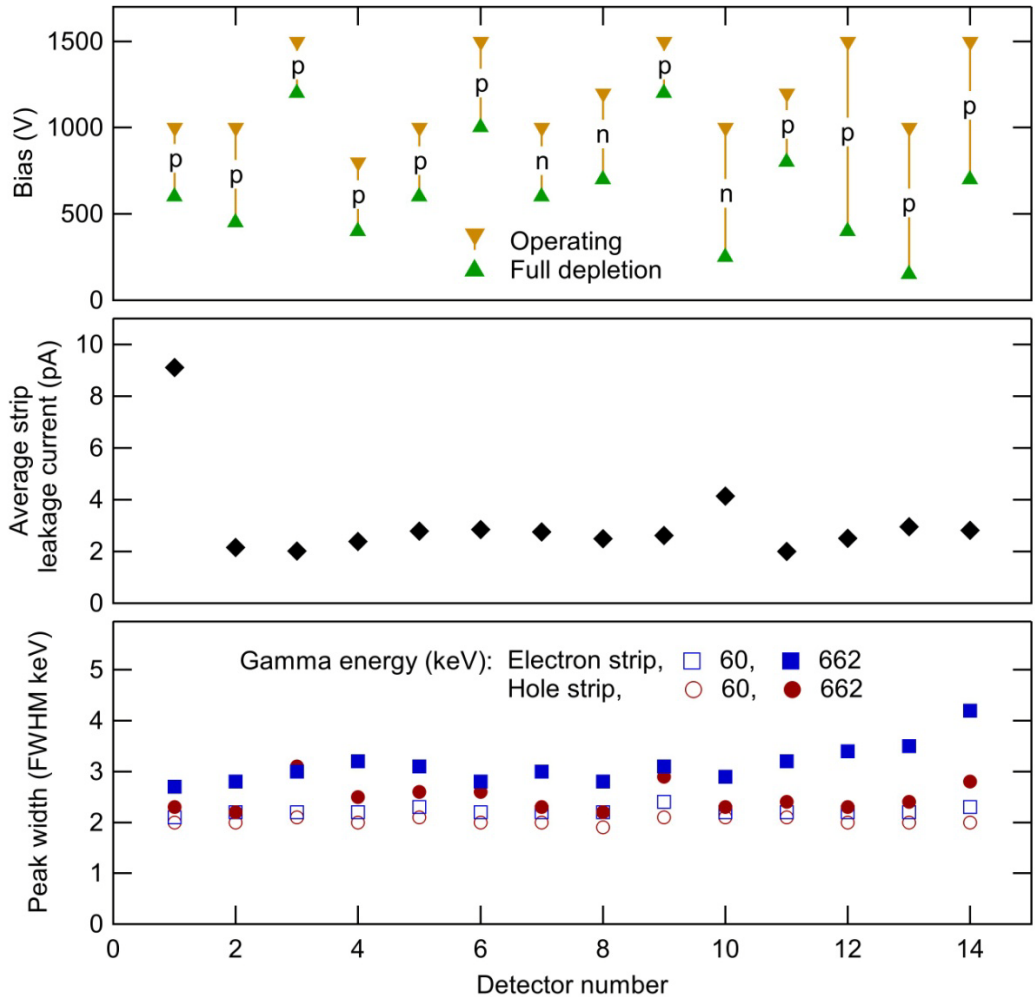

**Figure 2.3** Summary of the initial characterization measurements made on the set of 14 orthogonal strip HPGe detectors produced for the COSI instrument. (**Top**) Full depletion voltage and operating voltage for each detector. The full depletion voltages were determined based on detector capacitance measured as a function of the applied voltage. Also indicated in this graph is the HPGe impurity type as determined from low energy gamma-ray count rate measurements made with the detector when operated in a partially depleted state. (**Middle**) Average strip leakage current for each detector. These measurements were made in a test cryostat that achieved a typical detector cold stage temperature of about 80 K. (**Bottom**) Average energy resolutions of each detector measured at the gamma-ray energies of 60 keV (Am-241 source) and 662 keV (Cs-137 source). The averages were determined separately for electron and hole collecting strips. These spectroscopy measurements were made using compact, low power preamps operated at room temperature followed by conventional pulse processing electronics. A pulse shaping time of 2 µs was used.

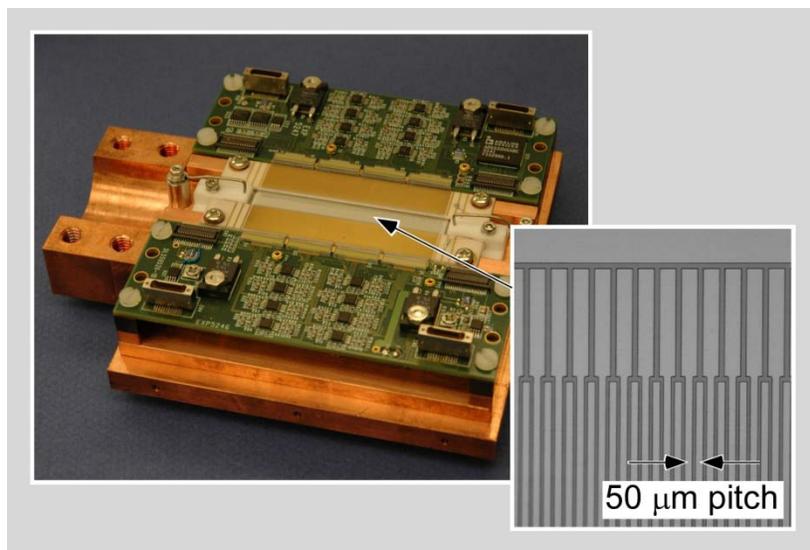

**Figure 2.4** Photograph of a fine pitched single sided strip HPGe detector developed for energy dispersive extended x-ray absorption fine structure measurements. The detector is shown clamped to a Cu cold plate that also supports two interface boards and two readout electronics boards. The interface boards are adjacent to the detector and, in conjunction with wire bonds, serve to connect the detector strips to the green readout electronics boards. The electrical contact on the strip side of the detector consists of a-Ge covered by a finely patterned layer of Al (see inset). The strips have a 50 μm center to center spacing and are 5 mm long. The electrical contact on the opposing side of the detector consists of a non-segmented, full area B implant coated with Al. The detector is 1 mm thick and is operated at a temperature of about 100 K.

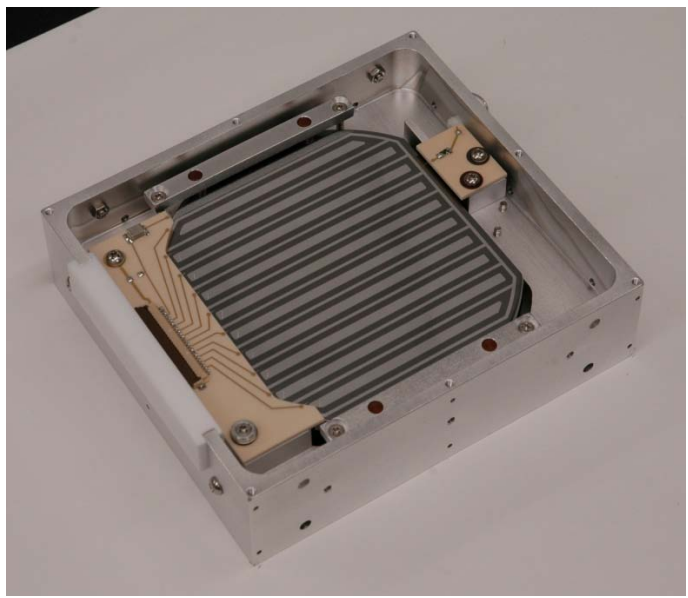

**Figure 2.5** Photograph of an HPGe strip detector with a field shaping grid of the type schematically shown in Figure 1.4a. This detector was developed for applications requiring gamma-ray spectroscopy with high event throughput at high counting rates while maintaining good energy resolution. The detector has an area of approximately 8 cm by 8 cm and is approximately 1.5 cm thick. The electrical contacts on the strip face of the detector are a-Ge with Al metallization. The opposing face of the detector has a full area contact consisting of a Li diffused layer coated with Al. The typical operating bias conditions of this particular detector are 1000 V applied to the Li contact, 200 V applied to the grid, and each strip held at near ground potential through connection to a charge sensitive preamplifier.

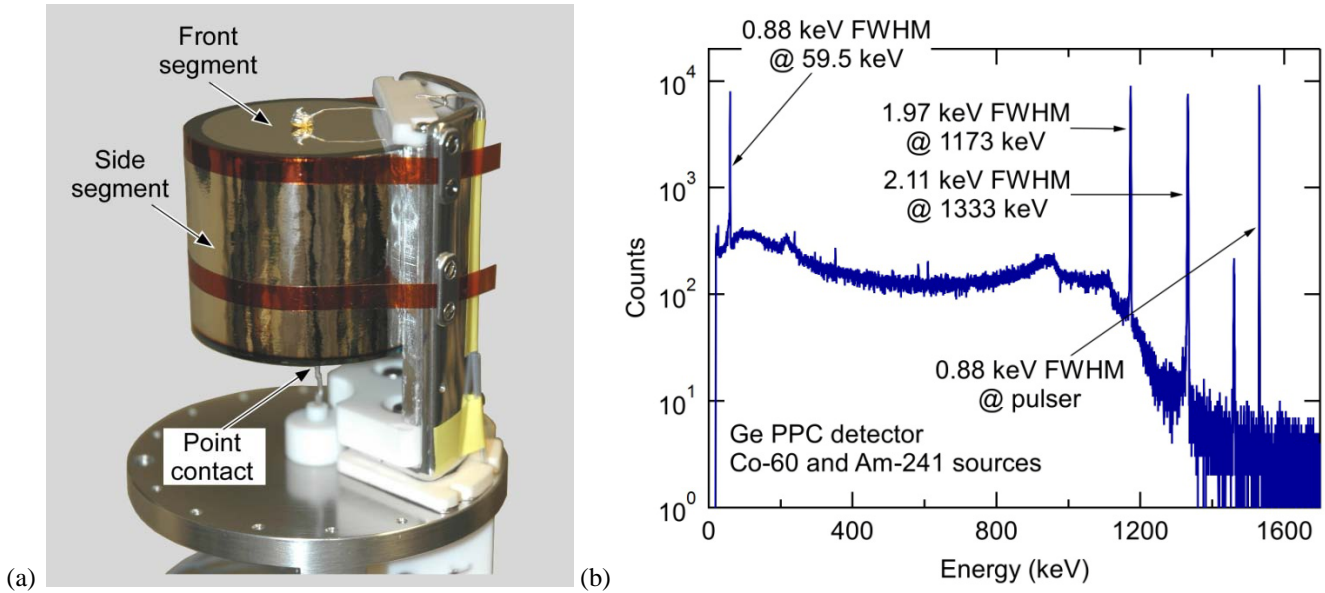

**Figure 2.6 (a)** Photograph of an HPGe p-type point contact detector. The detector has a thin entrance window produced with an a-Ge electrical contact and a point contact made with a-Si. The detector configuration is much like that shown schematically in Figure 1.3 except that the detector shown in this photograph has the a-Ge entrance window electrically separated from the Li diffused side contact [Amman 2009]. The detector diameter is 61 mm, its length is 50 mm, and the point contact diameter is about 2 mm. The detector is shown strapped to the cold stage of a test cryostat. **(b)** Energy spectrum measured with a p-type point contact detector much like that shown schematically in Figure 1.3 in the presence of Am-241 and Co-60 sources. The detector was operated at a bias of 3000 V, and standard analog pulse processing with a pulse shaping time of 12 µs was used to obtain the spectrum. Note that the front end signal readout electronics used were not optimized with temperature or for the small detector capacitance.

## 3. Amorphous Semiconductor Electrical Contact Theory

Much of the data analysis in this paper is straightforward in that the detector leakage current is simply evaluated as a function of a sputter process parameter, temperature, or time spent at room temperature in order to identify any trend in the data. It can, however, be of value to analyze the data based on a physical model of the amorphous semiconductor electrical contact. This can provide additional insight into the physical mechanisms dictating the detector performance. In principle, an understanding of this underlying physics of the electrical contact can assist in devising and implementing strategies to improve and optimize the contact. For this reason, in this section, the theory of the charge transport within the a-Ge film and through the a-Ge to crystalline HPGe interface are presented.

To begin, consider the case of an electrical contact consisting of a-Ge sputtered onto n-type crystalline HPGe with Al deposited onto the a-Ge. Schematic electron energy diagrams for such a contact are shown in Figures 3.1 and 3.2. The diagrams are plots of electron energy as a function of spatial coordinate perpendicular to the plane of the electrical contact. Allowed energy states are shaded, and the shading is green when occupied and gray when unoccupied. The HPGe portion of the energy diagram depicts the well-known band gap of forbidden energy levels separating the nearly fully occupied valence band of allowed levels from the nearly fully unoccupied conduction band of allowed levels. Both of these energy bands consist of energy states that extend throughout the crystal. Note that this diagram is overly simplified in part because the impurity states and partial occupancy of the valence and conduction bands are not depicted. In contrast to the electron energy level structure of single crystal HPGe is that of a-Ge. A simplified picture of the a-Ge energy structure consists of spatially extended electron energy states contained in separate valence and conduction bands with the energy gap between these two bands populated by spatially localized energy states. The band gap, which is similar to that of the crystalline HPGe, results from spatial short range atomic order in the a-Ge, whereas the localized states result from the disorder of the atomic bonding (bond length and angle) and defects including dangling bonds, voids, and impurities. For this example, the Fermi energy level $E_f$ in the a-Ge is assumed to be pinned near mid energy gap [Mott 1979] in contrast to the crystalline HPGe in which $E_f$ is dictated by the impurity concentration and the effective density of states in the valence and conduction bands [Sze 1981]. Electrical conduction through the a-Ge layer can take place through several different physical processes. These include thermal activation of carriers to the extended states and then conduction through these states, thermal activation to the localized energy states near the band edge and then hopping conduction between these states, and hopping conduction through the localized defect energy states near the $E_f$ [Mott 1979 and references therein]. It is expected that a-Ge films deposited through vacuum evaporation or

M. Amman, 2018, "Optimization of Amorphous Germanium Electrical Contacts and Surface Coatings on High Purity Germanium Radiation Detectors"

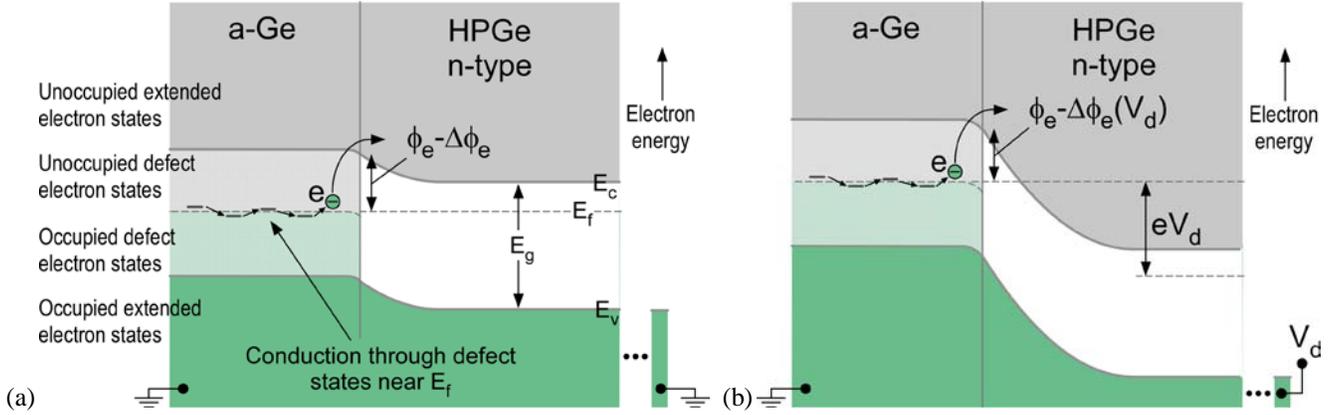

**Figure 3.1** Schematic electron energy diagrams of an a-Ge electrical contact on n-type crystalline HPGe. The diagrams show only the a-Ge to crystalline HPGe junction, and do not include the metal layer covering the a-Ge and the electrical contact on the opposing side of the HPGe that are both necessary to produce a complete radiation detector. **(a)** Energy diagram with zero voltage applied across the device. The energies noted in the diagram, $E_c$, $E_v$, $E_g$, $E_f$, and $\phi_e$, are of the conduction band edge, valence band edge, energy gap, Fermi level, and electron barrier. **(b)** Energy diagram with a relatively negative voltage applied to the a-Ge contact. This results in an increase of the depletion region near the contact. A reduction $\Delta\phi_e$ of the electron energy barrier occurs due to the penetration of the field into the a-Ge.

sputtering have a significant density of defect states near $E_f$. For this reason, the charge transport through such films at the low temperatures used for HPGe detector operation is likely dominated by conduction through the spatially localized defect energy states near $E_f$. This expectation is supported by measurements made on evaporated and sputtered a-Ge [Mott 1979 and references therein]. For the case of the sputtered films, it has also been demonstrated that the conductivity of the a-Ge at low temperatures can be reduced by several orders of magnitude through the addition of $H_2$ to the Ar sputter gas [Lewis 1976] [Connell 1976] [Moustakas 1977]. Furthermore, the conductivity can also be greatly reduced through annealing. The conductivity reduction with hydrogen addition can be attributed to the ability of hydrogen to compensate individual dangling Ge bonds that would otherwise contribute to energy gap defect states. Annealing, likewise, can reduce energy gap defect states by, for example, reducing the Ge bond angle distortion and improving the coordination number.

Much like a metal, the charge transport through sputtered a-Ge at typical HPGe detector temperatures predominantly takes place through electronic energy levels near the Fermi energy. This similarity in conduction then motivates the use of metal on semiconductor (MS) contact theory [Sze 1981] to describe the amorphous semiconductor contacts. A model of the amorphous on crystalline semiconductor (ACS) junction of this type was developed by Dohler and Brodsky [Dohler 1974] [Brodsky 1975a] [Brodsky 1975b] and then later successfully applied to the a-Ge contact on HPGe [Hull 2005] [Looker 2015a]. The basic elements of this model are depicted in the Figure 3.1 energy diagrams. Not shown in the diagrams is the metal layer on top of the a-Ge. Since the charge transport in the a-Ge is dominated by that through states near the Fermi level, there should be a barrier free exchange of charge between the metal and the a-Ge. As a result, this interface should not play a significant role in the electrical properties of the contact. The Figure 3.1a diagram shows the state of thermal equilibrium in which the Fermi levels in the a-Ge and HPGe coincide and assumes an ideal contact without interface energy states. From this diagram, it is evident that a potential energy barrier of $\phi_e$ exists that inhibits electron injection from the a-Ge to the conduction band of the HPGe. This is the physical mechanism leading to the desired charge injection blocking behavior of the electrical contact. The effectiveness of this blocking depends in part on the magnitude of the barrier, which based on MS theory should be dictated by the work function of a-Ge, the electron affinity of HPGe, and the localized energy states existing at the interface between the two materials [Sze 1981]. The magnitude and energy distribution of the interface states likely depend on the preparation of the HPGe surface and the a-Ge deposition process. A theoretical determination of the barrier is therefore not straightforward, and, for the work of this paper, the barrier is treated as a parameter to be extracted from electrical measurements. For an electron at the Fermi level to be injected from the a-Ge into the HPGe, it must gain an amount of thermal energy equal to the barrier height. This thermal activation mechanism then leads to an exponential dependence of the injected leakage current on the barrier height. In the ACS model, the positive current flow through the device normalized by the contact area is given by

$$J = J_\infty(T)\exp(-\phi_e/kT)[1 - \exp(-qV_d/kT)]f(V_d) \,, \tag{3.1}$$

where

M. Amman, 2018, "Optimization of Amorphous Germanium Electrical Contacts and Surface Coatings on High Purity Germanium Radiation Detectors"

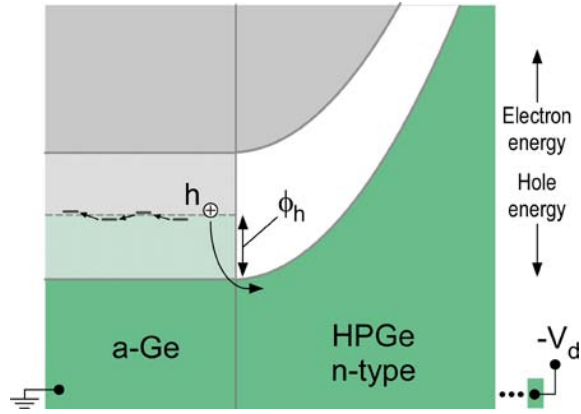

**Figure 3.2** Schematic electron energy diagram of an a-Ge electrical contact on n-type crystalline HPGe. The diagram shows only the a-Ge to crystalline HPGe junction, and does not include the metal layer covering the a-Ge and the electrical contact on the opposing side of the HPGe that are both necessary to produce a complete radiation detector. This diagram is for the case of a relatively positive voltage applied to the a-Ge contact that is of sufficient magnitude to fully deplete the HPGe. For this voltage bias polarity, depletion begins at the contact not shown in the diagram and then extends through the HPGe to the contact shown as the voltage is increased.

$$f(V_d) = \exp\left(\left\{\left[(2qV_{bi} + 2qV_d + N/N_f)N/N_f\right]^{1/2} - N/N_f\right\}/kT\right), \tag{3.2}$$

and $T$ is the temperature, $\phi_e$ is the energy barrier to electron injection, $k$ is the Boltzmann constant, $V_d$ is the reverse bias voltage applied across the device, $N$ is the net ionized impurity concentration of the HPGe, $N_f$ is the density of localized energy states near the Fermi level in the a-Ge, $q$ is the magnitude of the electron charge, and $V_{bi}$ is the built-in voltage at the contact.

The current density equation obtained from the standard MS theory differs from that of Equation (3.1) in the terms $J_\infty(T)$ and $f(V_d)$. In the ideal MS theory, $J_\infty(T) = A^*T^2$, where $A^*$ is the effective Richardson constant [Sze 1981]. Since this term is dictated in part by the transition probabilities between energy states on each side of the contact, it is to be expected that the term will be different between the MS and ACS contacts. In the ACS model of Dohler and Brodsky, $J_\infty(T)$ was left as a parameter to be determined from measurements. In the data analysis of this paper, however, an explicit form for the temperature dependence is needed, and therefore the $T^2$ dependence predicted by the MS model is retained. Consequently, the following is assumed:

$$J_\infty(T) = J_o T^2, \tag{3.3}$$

where the prefactor $J_o$ is a constant to be determined from the measurements. The other term that is different between the two models, $f(V_d)$, accounts for the lowering of the energy barrier as a result of electric field penetration into the a-Ge. This is schematically shown in the energy diagrams of Figure 3.1. In the case of a metal, this penetration is negligible, and, as a result, $f(V_d) = 1$ for an MS contact.

For the analysis in this paper in which the detector is assumed to be operated with a large reverse bias voltage, Equations (3.1) and (3.2) are simplified by assuming that $V_d \gg kT/q$, $V_{bi}$, and $N/qN_f$. Combining Equations (3.1) through (3.3) and applying the large detector voltage approximation, the following is obtained for the current density:

$$J = J_o T^2 \exp(-\phi_e/kT) \exp\left[(2qV_d N/N_f)^{1/2}/kT\right]. \tag{3.4}$$

The above equation was derived for the case where the HPGe is partially depleted. The electric field magnitude at the contact in this case is

$$E = (2qV_d N/\varepsilon)^{1/2}, \tag{3.5}$$

where $\varepsilon$ is the dielectric constant of HPGe, and the large detector voltage approximation has again been applied. Combining Equations (3.4) and (3.5), the following is obtained:

$$J = J_o T^2 \exp(-\phi_e/kT) \exp\left[(\varepsilon/N_f)^{1/2} E/kT\right], \tag{3.6}$$

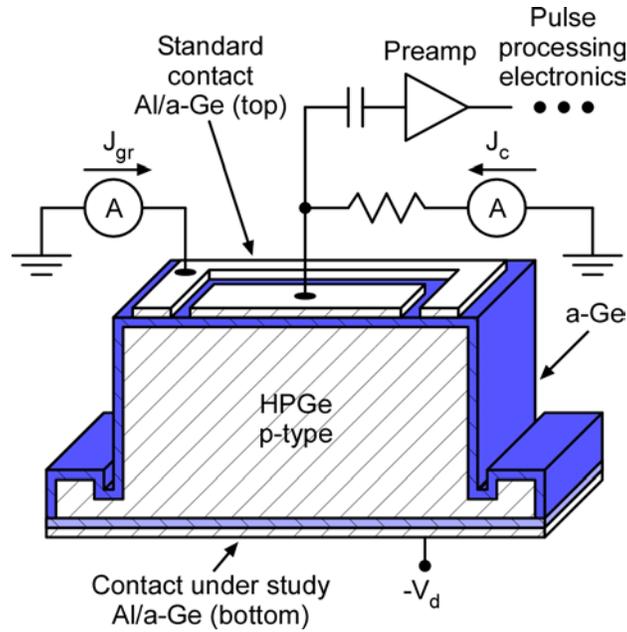

**Figure 3.3** Schematic cross-sectional drawing of the a-Ge contact HPGe detector configuration and electrical circuit used for detector capacitance and bulk injected leakage current measurements. The top contact (anode) had its Al layer patterned into separate center and guard ring electrodes. The Al layer on the bottom (cathode) covered the entire contact surface.

$$J = J_o T^2 \exp[-(\phi_e - \Delta\phi_e)/kT] \, , \qquad (3.7)$$

where

$$\Delta\phi_e = (\varepsilon/N_f)^{1/2} E \, . \qquad (3.8)$$

The above equations explicitly express the current density in terms of the electron barrier lowering term $\Delta\phi_e$. Note also that the above equations also apply to the case of a fully depleted or over depleted detector as long as the appropriate equation for $E$ is used. This is a direct consequence of the field continuity boundary condition at the contact that is applied during the solution of Poisson's equation when deriving $f(V_d)$ [Brodsky 1975b].

One important implication of the contact theory just described is that the contact can potentially block both electron and hole injection. This bipolar blocking behavior is illustrated in the diagrams of Figure 3.1b and Figure 3.2. In the Figure 3.1b diagram, the a-Ge contact is at a relatively negative voltage, the HPGe depletion begins and extends out from that contact, and electron injection is inhibited by $\phi_e$. In the Figure 3.2 diagram, the voltage polarity is reversed, and the voltage magnitude is assumed to be that needed to fully deplete the HPGe. In this case, the depletion begins at the electrical contact not shown in the diagram and extends out through the HPGe until it reaches the contact shown in the diagram. At full depletion and beyond, the a-Ge contact shown in the diagram experiences an electric field that drives hole injection. This hole injection is inhibited by the hole energy barrier $\phi_h$ associated with the contact. From the diagrams, it is clear that this simple model of the contact predicts that the sum of the energy barrier heights for both of the charge injection processes, $\phi_h + \phi_e$, should be equal to the bandgap energy of crystalline Ge (neglecting barrier height lowering). This predicted behavior that the improvement of one energy barrier would be at the expense of the other one for a particular contact was observed in a previous study of a-Ge and a-Si contacts on HPGe [Amman 2007]. From the point of view of assessing the blocking ability of a specific contact, this means that it should be sufficient to characterize the blocking ability of the contact under one polarity. For example, a contact that has been determined to have a large barrier to electron injection and produces a small electron injection current when negatively biased will have a small hole barrier and greater hole injection when positively biased. Furthermore, an energy barrier of approximately half the band gap energy will block equally well for electrons and holes and would be appropriate when it is desirable to produce a detector using only a single a-Ge recipe for all electrical contacts on the detector.

Motivated by the above contact physics, the study of this paper focused only on characterizing the electron blocking properties of the contacts since the hole blocking behavior can be inferred from the electron blocking results. Specifically, detectors of the type shown schematically in Figure 3.3 were produced and characterized. The detectors were of a rectangular geometry and made use of a top (anode) electrical contact separated into a center electrode surrounded by a guard ring. The purpose of the guard ring is to take up the leakage current flowing along the detector surface between the top and bottom contacts [Goulding 1961]. Consequently, the leakage current from the center electrode is free of this surface current contribution and can be used for the characterization of bulk injected leakage current. The electrical behavior of the Figure 3.3 detector can be understood through the electron energy diagrams of Figure 3.4. The HPGe material used for the detectors was p-type, and a negative detector voltage -$V_d$ was applied to the bottom contact. With this voltage polarity, the depletion of the hole carriers in the HPGe begins at the top contact, thereby creating an electric field at the top contact that will drive hole injection into the HPGe. This is shown schematically in Figure 3.4a. At this low detector voltage (below the full depletion voltage), the leakage current density measured from the center contact $J_c$ predominately consists of that from hole injection at the top contact. When the detector voltage is increased to the level that fully depletes the HPGe of hole carriers, the electric field extends completely through the HPGe giving rise to a field at the bottom contact that drives electron injection. Because of this, a step in the center contact leakage current density $\Delta J_c$ will occur at full depletion, and this step can be used to characterize the electron injection behavior of the bottom contact. This full depletion state is shown in the energy diagrams of Figure 3.4b. The detector results shown in this paper are primarily based on the measurement of $\Delta J_c$ as a function of the bottom contact a-Ge fabrication recipe.

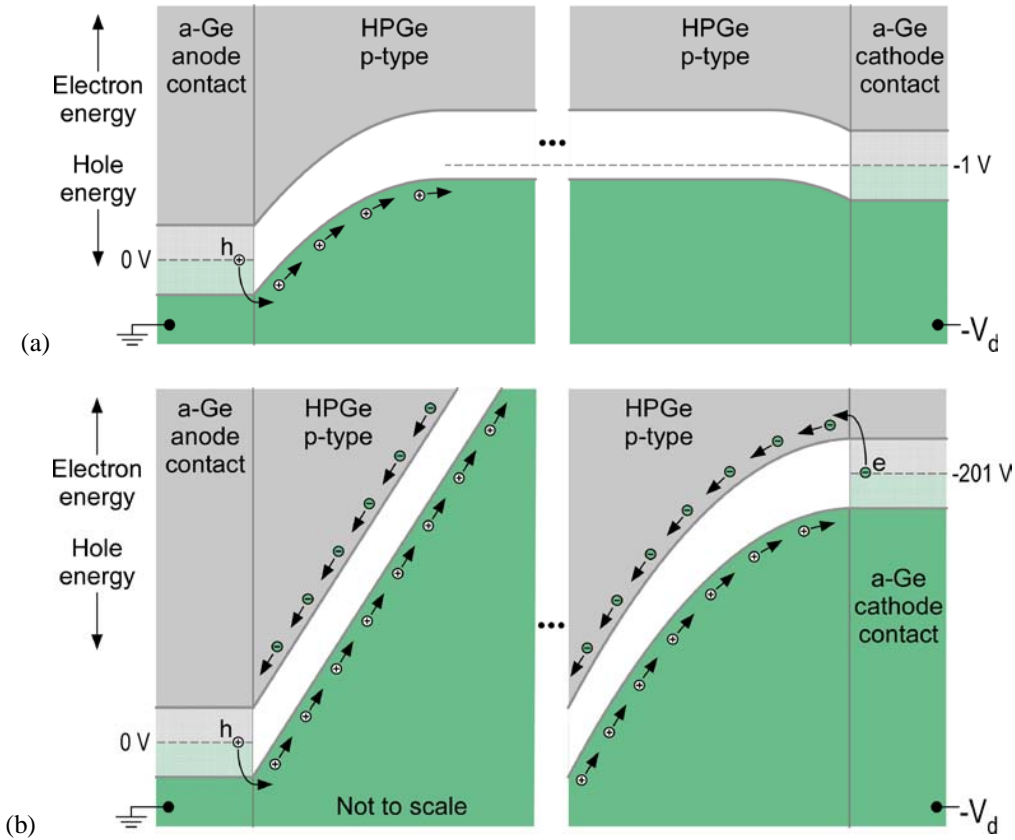

**Figure 3.4** Schematic electron energy diagrams of the a-Ge contact detector shown in Figure 3.3. **(a)** Energy diagram when a small negative detector voltage is applied to the bottom contact. The center electrode leakage current predominantly results from hole injection at the anode contact. **(b)** Energy diagram when the negative detector voltage applied to the bottom contact is large enough to fully deplete the HPGe. At this voltage, a step in the leakage current occurs that results from electron injection at the cathode contact.

 Example measurements from a detector of the type shown in Figure 3.3 are shown in Figure 3.5. The center contact current density measured as a function of the detector voltage is shown for two different detector temperatures. The measured detector capacitance is also plotted along with the leakage current in order to provide an indication of full depletion. As the detector voltage is increased, the voltage at which the capacitance first becomes constant is the full depletion voltage (approximately 200 V for the Figure 3.5 detector). At this voltage, there is a clear step in the leakage current that is solely a result of electron injection from the bottom contact. The barrier lowering predicted by the ACS contact model is clearly evident in the $J_c$-$V_d$ characteristics as the continual increase in current with increasing detector voltage above full depletion. In contrast to this, the ideal MS model predicts a flat current or one that slowly increases due to image charge lowering of the barrier.

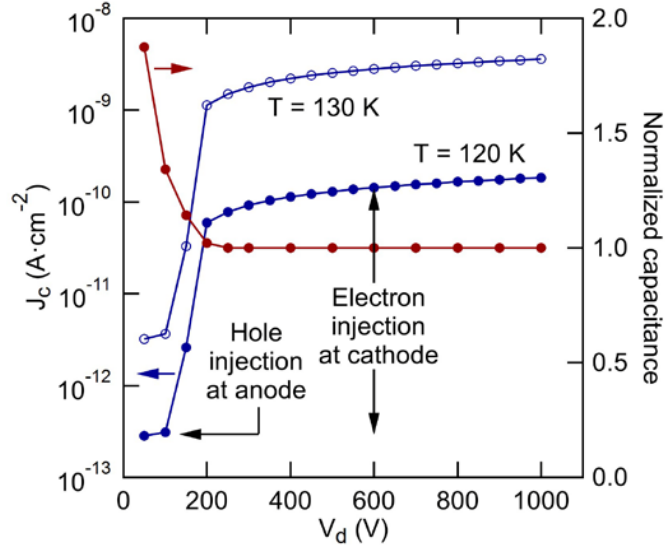

**Figure 3.5** Detector leakage current density and normalized detector capacitance measured as a function of detector voltage for a detector of the type shown in Figure 3.3. The leakage current density is that measured from the center contact of the guard ring detector, and this current is shown for two different detector temperatures ($T$ = 120 K and 130 K). The a-Ge of the top contact and detector sides was sputtered in Ar with 7% $H_2$ at a pressure of 23 mTorr. The a-Ge of the bottom contact was sputtered in pure Ar at a pressure of 23 mTorr.

To gain additional physical insight into the electron injection behavior of the contacts, the ACS model presented above can be used. To accomplish this, Equations (3.7) and (3.8) are rewritten for $\Delta J_c$ assuming that $V_d > V_{fd}$, where $V_{fd}$ is the full depletion voltage:

$$\Delta J_c = J_o T^2 \exp[-(\phi_e - \Delta\phi_e)/kT],\tag{3.9}$$

where

$$\Delta\phi_e = (\varepsilon/N_f)^{1/2}(V_d - V_{fd})/t,\tag{3.10}$$

and $t$ is the detector thickness.

The goal of the data analysis was to determine $\phi_e$, $J_o$, and $N_f$ for the bottom contact. To accomplish this, the following procedure was used. First, $J_c$-$V_d$ characteristics were measured at two different temperatures, $T_1$ and $T_2$. Following this, the hole injection contribution to the current density was removed from each $J_c$-$V_d$ characteristic so that only the electron injection current $\Delta J_c$ remained. Note that this subtraction was typically not needed since the hole injection was normally only a small fraction of the total current for detector voltages above full depletion. From Equations (3.9) and (3.10), it is evident that taking the natural log of the $\Delta J_c$ data will produce a data set that is linear with $V_d$. This is explicitly shown below:

$$\ln(\Delta J_c) = C_1(T) + C_2(T) V_d,\tag{3.11}$$

where

$$C_1(T) = \ln(J_o) + 2\ln(T) - V_{fd}(\varepsilon/N_f)^{1/2}/tkT - \phi_e/kT,\tag{3.12}$$

$$C_2(T) = (\varepsilon/N_f)^{1/2}/tkT.\tag{3.13}$$

The next step of the data analysis was to perform linear fits to the $\ln(\Delta J_c)$ data sets in order to extract $C_1(T_1)$, $C_2(T_1)$, $C_1(T_2)$, and $C_2(T_2)$. These fits were done over a detector voltage range that typically started at a voltage well above full depletion. For example, the data of Figure 3.5, which is from a detector with a full depletion voltage near 200 V, was fit over the range of 600 to 1000 V. The purpose of analyzing data from only a highly over depleted detector was to hopefully avoid any effects associated with non-uniformity in the electric field possibly caused by surface channels. Surface channel effects would be more pronounced at lower detector voltages. Evidence in the $J_c$-$V_d$ characteristic for such undesirable behavior would be a broad rather than sharp step at full depletion. With the linear fitting parameters determined, the ACS contact model parameters were calculated using the following equations:

$$\phi_e = kT_1T_2\{C_1(T_1) - C_1(T_2) + 2[\ln(T_2) - \ln(T_1)] + [C_2(T_1) - C_2(T_2)]V_{fd}\}/(T_1 - T_2) , \quad (3.14)$$

$$J_o = \exp[C_1(T_1) + C_2(T_1)V_{fd} - 2\ln(T_1) + \phi_e/kT_1] , \quad (3.15)$$

$$N_f = \varepsilon/[tkT_1C_2(T_1)]^2 . \quad (3.16)$$

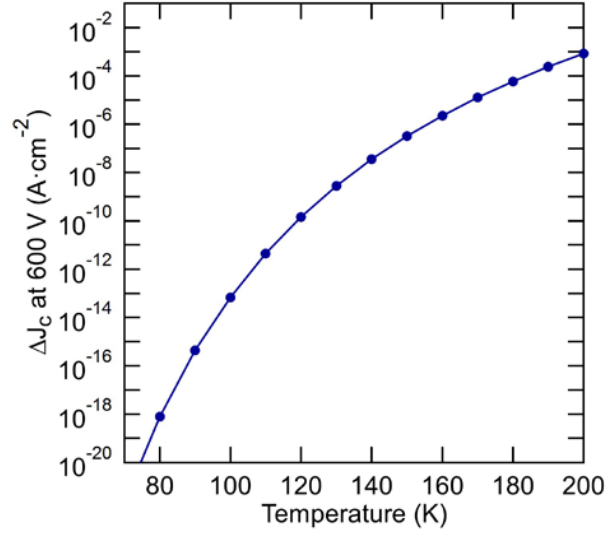

**Figure 3.6** Calculated center contact leakage current density resultant from electron injection at the bottom contact of a detector of the type shown in Figure 3.3. The leakage current density was estimated assuming a detector voltage of 600 V and the ACS contact model parameters extracted based on fits to the measured data shown in Figure 3.5.

Applying the above described analysis to the data of Figure 3.5, the following were obtained for the ACS parameters: $\phi_e = 0.379$ eV, $J_o = 62.9$ A·cm$^{-2}$·K$^{-2}$, and $N_f = 1.76 \times 10^{17}$ eV$^{-1}$·cm$^{-3}$. With these parameters, the electron injection leakage current associated with the bottom contact a-Ge recipe can be estimated for a wide range of temperatures using Equations (3.9) and (3.10). The result of such a calculation is shown in Figure 3.6.

## 4. Detector Physics and Performance Optimization

In the previous section, the basic physics of the amorphous semiconductor electrical contact was covered. In this section, the connection between this basic contact physics and detector performance is made. Detector performance is quantified through properties such as energy resolution, energy threshold, entrance window thickness, spatial resolution, efficiency, and response uniformity. These performance characteristics are impacted by a number of physical properties and processes related to the electrical contacts, surface coatings, and geometry of the detector. Some of the properties and processes capable of degrading detector performance are schematically illustrated in Figure 4.1. These include (a) charge carrier injected leakage current and the addition of finite resistances between the charge sensing electrical contact and other contacts, and (b) through (d) charge collection to surfaces other than the electrical contacts. The total injected leakage current ($I_t = I_e + I_h$) and parallel resistances $R_p$ degrade the detector's energy resolution, which is often the most important performance specification for HPGe detector applications. In order to determine the suitability of a-Ge electrical contacts and surface coatings for a particular application, it is therefore necessary to quantify the impact that the leakage current and parallel resistances have on energy resolution. This is done in the following paragraphs.

The energy resolution of a detector results from signal fluctuations that originate from a number of different sources. Assuming statistical independence, the effects of the various broadening sources add in quadrature so that the total energy resolution of a detector $\Delta E_{total}$ is given by [Knoll 1989]

$$\Delta E_{total}^2 = \Delta E_s^2 + \Delta E_c^2 + \Delta E_n^2 . \quad (4.1)$$

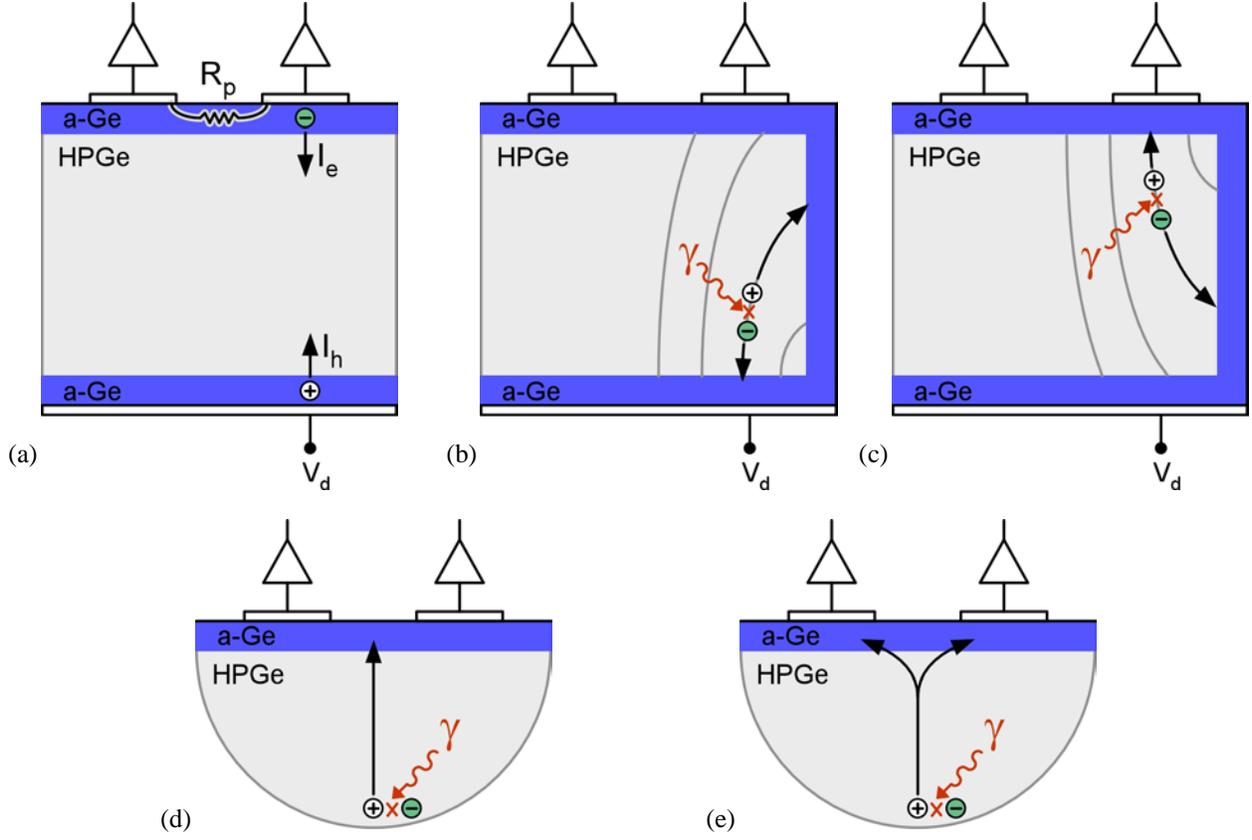

**Figure 4.1** Schematic detector diagrams illustrating physical properties and processes related to the electrical contacts, surface coatings, and geometry of the detector that can affect the detector performance. **(a)** Leakage current in the form of electron injection at cathode contacts ($I_e$) and hole injection at anode contacts ($I_h$), and the addition of finite resistances ($R_p$) between the signal readout electrical contact and other contacts all introduce current noise that can degrade the energy resolution of a detector. **(b)** and **(c)** Charge collection to the side surface of a planar geometry detector produces detected signals with pulse height deficits. In addition to degrading energy resolution, such events may contribute to a background continuum rather than a photopeak and will therefore degrade the detector's efficiency. **(d)** Charge collection to the surface between two signal readout electrical contacts leads to a signal deficit as compared to the signals generated when the charge is completely collected to the contacts, which is shown in **(e)**. The incomplete charge collection can degrade energy resolution.

The component $\Delta E_s$ represents the fluctuations due to charge carrier generation statistics, and its FWHM value is given by

$$\Delta E_s^2 = (2.35)^2 F E_{eh} E_\gamma , \qquad (4.2)$$

where $F$ is the Fano factor and approximately equals 0.1 at 77 K, $E_{eh}$ is the average energy needed to create an electron-hole pair and equals 2.96 eV at 77 K, and $E_\gamma$ is the gamma-ray energy. The energy resolution component $\Delta E_c$ results from incomplete charge collection in the detector and includes broadening caused by imperfections in the HPGe material such as those that lead to charge carrier trapping and non-ideal device operation such as charge collection to non-contact detector surfaces. Finally, the term $\Delta E_n$ represents the fluctuations (also referred to as noise) introduced by the detector and signal readout electronics. The FWHM electronic noise for a typical detector signal readout electronics chain consisting of an FET input charge sensitive preamplifier followed by a pulse shaping amplifier is given by [Spieler 2005] [Radeka 1972] [Goulding 1972] [Goulding 1982]

$$\Delta E_n^2 = \left(2.35\, E_{eh}/q\right)^2 \left\{ F_v\left(4kT_{fet}R_{fet} + 4kT_s R_s\right) C_t^2/\tau + F_f A_f C_t^2 + F_i\left(2qI_t + \frac{4kT_p}{R_p}\right)\tau \right\}, \qquad (4.3)$$

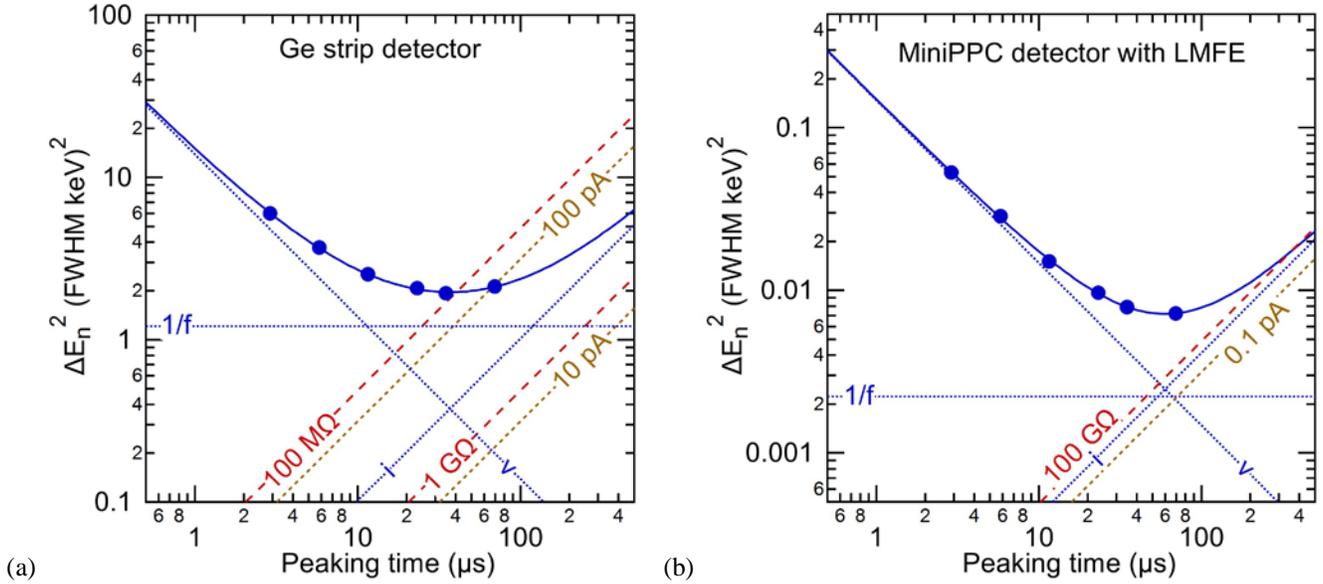

**Figure 4.2** Example measured noise characteristics obtained with HPGe detectors. Measured data are shown with blue circles, fitted noise components with dotted blue lines (labeled $v$, $1/f$, and $i$ for the voltage, $1/f$, and current noise components, respectively), and total fitted noise with a solid blue line. Calculated current noise contributions for various leakage currents and parallel resistances are shown with dashed orange and red lines, respectively. **(a)** Noise characteristic from a GRIPS strip detector of the type shown in Figure 2.1a and described in Section 2. For these measurements, the GRIPS detector was operated in a test cryostat that had only a partial IR shield covering the detector and achieved a detector cold stage temperature of about 81 K. The electronics used to read out a single strip of this detector consisted of a compact, low power preamplifier of the type described by Fabris et al. [Fabris 1999] followed by a commercial semi-Gaussian analog shaping amplifier. The preamplifier was located outside of the detector cryostat and was operated at room temperature. **(b)** Noise characteristic from the work of Barton et al. [Barton 2011] on the low mass front end (LMFE) for the signal readout of p-type point contact detectors. The detector was a small p-type point contact (mini-PPC) detector of size 2 cm diameter and 1 cm thick with an approximately 1.5 mm diameter point contact. The electronics used to read out the point contact consisted of the cooled LMFE (containing the preamplifier FET and feedback components) located inside the cryostat near the detector with the remainder of the preamplifier located outside the cryostat at room temperature followed by a commercial semi-Gaussian analog shaping amplifier.

where $q$ is the magnitude of the electron charge, $k$ is the Boltzmann constant, $T_{fet}$ is the temperature of the front end FET, $R_{fet}$ is the equivalent series resistance associated with the front end FET, $T_s$ is the temperature of $R_s$, $R_s$ is any series resistance between the detector and readout electronics, $C_t$ is the total capacitance at the input of the readout electronics (which includes the detector capacitance $C_d$), $\tau$ is the characteristic time of the shaping amplifier, $A_f$ is a device specific noise coefficient for $1/f$ noise, $I_t$ is the total detector bulk injected leakage current, $T_p$ is the temperature of $R_p$, and $R_p$ is any resistance in parallel with the input of the readout electronics. The parameters $F_v$, $F_f$, and $F_i$ are shape factors that are dictated by the pulse shape of the shaping amplifier. From this expression, it is evident, based upon the different shaping time dependencies, that the electronic noise can be subdivided into three components that are added in quadrature to form the total electronic noise. The component whose square is inversely related to the shaping time is referred to as voltage or series noise and results from the amplifier's voltage noise as well as the thermally induced voltage noise generated by any series resistance between the detector and the readout electronics. The one that is independent of the shaping time is $1/f$ noise and results from dielectrics near the preamplifier input and a component of the noise associated with the preamplifier input transistor [Bertuccio 1996]. Finally, the third component whose square is proportional to the shaping time is referred to as current or parallel noise and consists of the shot noise associated with the detector leakage current and the thermally induced current noise generated by any resistances in parallel with the input of the readout electronics.

The contributions to the electronic noise that are associated with the a-Ge are shot noise from the leakage current injected through the a-Ge contacts and thermally induced current noise from the finite resistance of the a-Ge surface coating as depicted in Figure 4.1a. The a-Ge could also potentially contribute to the $1/f$ component through dielectric noise or through some non-ideal noise generation mechanism, but such noise contributions do not appear to be significant. Furthermore, RF sputtered a-Ge resistors have been used to produce extremely low noise front end preamplifier stages, thereby attesting to the low $1/f$ noise of the a-Ge films [Barton 2011]. The current noise contributions to the total electronic noise resultant from $I_t$ and $R_p$ can be quantified using Equation (4.3). To do this, the $F_i$ shape factor is needed. For a semi-Gaussian shaping amplifier consisting of a single differentiator and $n$ number of integrators all with the same characteristic time constant (CR-RC$^n$ shaping amplifier, where $\tau = nRC$ is the peaking time), $F_i$ is the following [Seller 1999]:

$$F_i = \frac{1}{4\pi n}\left(\frac{n!\exp(n)}{n^n}\right)^2 B\left(\frac{1}{2}, n+\frac{1}{2}\right), \quad (4.4)$$

where $B()$ is the beta function. For the current noise estimates in this paper, a three integrator amplifier ($n = 3$) was assumed.

Using Equations (4.3) and (4.4) in combination with the noise specification dictated by the detector application, it is possible to set upper limits on $I_t$ and $R_p$ and thereby guide the optimization and selection of the a-Ge fabrication processes. Alternatively, in the absence of a firm noise specification, one can look at the measured noise performance of a specific detector configuration and determine upper limits on $I_t$ and $R_p$ based on the condition that they not add significantly to the total electronic noise. As examples of this process to set these limits, consider the two detector noise characteristics shown in Figure 4.2. A noise characteristic is a plot of the square of the electronic noise as a function of the shaping amplifier characteristic time (in this case, peaking time of the semi-Gaussian shaping amplifier). The plot is typically shown on a log-log axes graph with equal ranges for each axis. On such a plot, it is relatively easy to visually identify the presence of the three noise components and to determine the relative significance of each component at a specific amplifier peaking time. A noise characteristic is shown for a planar strip detector in Figure 4.2a and for a small p-type point contact detector (mini-PPC) in Figure 4.2b. In each plot, the measured noise data is shown with blue circles. A fit was made to this data in order to extract out the separate voltage, 1/f, and current noise components. These extracted components are shown as blue dotted lines. The sum of these three fitted components is shown as the solid blue line, which matches well with the measured data for both detector examples.

The electronic noise data of Figure 4.2a was obtained with a single strip on a GRIPS orthogonal strip detector (Section 2) instrumented with a compact, low power, room temperature preamplifier [Fabris 1999] followed by a commercial semi-Gaussian analog shaping amplifier. Three characteristics associated with this detector system markedly impact its noise. First, the total input capacitance $C_t$ to the readout electronics is relatively large at about 20 pF primarily because of the large inter-strip capacitance of the detector. Second, the signal wiring between the detector and preamplifier is routed along a significant length of rigid circuit board and flexible circuit as well as passes through multiple fine pitch connectors and a vacuum feedthrough, all of which contribute dielectric noise. Finally, the third factor is that, due to the difficulties of implementing a cooled front-end preamplifier stage for a high channel count detector, the preamplifier is operated at room temperature. Because of these three characteristics, at the typically used peaking time of 8 µs, the noise is dominated by a combination of the voltage component associated with the preamplifier (square of the noise scales with $T_{fet}C_t^2$, see Equation (4.3)) and the 1/f dielectric component (scales with $C_t^2$). The current noise component of this detector at 8 µs peaking time is not significant. For comparison purposes, the square of the current noise associated with leakage currents 10 pA and 100 pA, and parallel resistances of 100 MΩ and 1 GΩ have also been plotted in Figure 4.2a. Based on this, it is clear that an upper limit on $I_t$ can be as high as about 100 pA and a lower limit on $R_p$ be as low as about 100 MΩ at a peaking time of 8 µs. Consequently, a-Ge processes that at a minimum meet these limits will be sufficient in that their current noise will not significantly add to the overall electronic noise of the detector system.

A more challenging detector system is that of a low capacitance detector combined with a preamplifier that has a cooled front end signal readout stage. An example of this is a mini-PPC detector combined with a low mass front end (LMFE) readout [Barton 2011]. The noise characteristic for this example is plotted in Figure 4.2b. In comparison to the GRIPS detector system, substantially lower noise is achieved with the mini-PPC instrumented with the LMFE. This is a result of the small detector capacitance of less than 1 pF, the minimization of wiring and noisy dielectrics between the detector and the preamplifier, and the cooling of the preamplifier's first stage FET and feedback components. For this detector system, depending on the peaking time used, the upper limit on the leakage current may be as low as 0.1 pA, and the lower limit on the parallel resistances may be as high as 100 GΩ.

In addition to affecting the current noise of a detector, the a-Ge can influence the charge collection within the detector. Examples of this are schematically illustrated in Figures 4.1b through 4.1e. Non-ideal charge collection in which the gamma-ray generated charge is not fully collected to the electrical contacts on the detector can degrade energy resolution as a contribution to the $\Delta E_c$ term of Equation (4.1). In addition, depending on the extent of the collection deviation, detected events may be moved out of the photopeak of the gamma spectrum and into a background continuum. This then can degrade detector efficiency and sensitivity. The examples shown in Figures 4.1b and 4.1c are of charge collection for events taking place near the edge of a planar geometry detector. For these events, either the holes or electrons are deflected to the side detector surface as a result of the charge state of that surface. Such behavior has long been observed and studied in semiconductor based radiation detectors [Kingston 1956] [Llacer 1964] [Dinger 1975] [Malm 1976] [Hansen 1980] [Hull 1995]. Hansen et al. [Hansen 1980], in their work that first demonstrated the value of sputtered deposited a-Ge as a passivating surface coating on HPGe detectors, explored the ability of the coating to control the electrical state of planar detector side surfaces. They showed that the hydrogen concentration of the sputter gas during a-Ge deposition could be used to adjust and optimize the electrical state on the side of the detector and thereby obtain a condition in which charge collection to the side is substantially eliminated. A concentration of 7% $H_2$ in Ar produced good results with p-type HPGe detectors. For this reason, the work presented later in this paper is based on a-Ge films sputtered with this $H_2$ concentration or with pure Ar.

Another charge collection issue that can occur in segmented contact detectors is shown in Figures 4.1d and 4.1e. The figures illustrate the situation in which a gamma-ray interaction occurs between two contact segments. Since the contacts are at approximately the same electrical potential, it is possible for the gamma generated charge to initially collect directly to the surface between the two contacts (Figure 4.1d) rather than to the contacts themselves (Figure 4.1e). If the charge remains between the two contacts during the signal measurement time, the induced charge signal from the event will be shared between the two contact segments, other nearby contact segments, as well as with nearby conductors that are not used for signal readout. Because of this, a signal deficit occurs even if all contact signals from the event are summed, and the energy resolution can be impacted. Such non-ideal behavior has been observed in a-Ge contact strip detectors [Luke 2000] [Amman 2000a] [Amman 2000b]. There are a number of strategies aimed at addressing this charge collection issue. One is to use signal processing to correct for the signal loss. This has the disadvantage of increasing the complexity of the readout electronics. Another is to reduce the spacing between the contact segments. This both reduces the fraction of events that collect to the inter-contact surface and the amount of signal lost by these events. The downside of this strategy is that the inter-contact capacitance is increased, which may then increase the electronic noise and degrade energy resolution. The use of field shaping contacts has been demonstrated as a means to improve the charge collection in segmented contact detectors [Amman 2000a] [Amman 2000b] [Cooper 2015] [Cooper 2018]. However, the addition and operation of the field contacts have the disadvantage of increasing the complexity of the detector system. Another strategy that has been successfully employed is to etch away the a-Ge between the contact segments [Protic 2003] [Protic 2004]. This method increases the complexity of the fabrication process and potentially results in detectors that will be more susceptible to performance changes resulting from environmental exposure unless the inter-contact surfaces are passivated. The ideal strategy, which is the one of interest to the work of this paper, is to optimize the a-Ge process for minimum inter-contact charge collection. It has been previously shown that a-Ge RF sputtered in Ar with $H_2$ under appropriate sputter conditions will produce a very high resistivity a-Ge film and, when used for the segmented contacts, substantially eliminates the inter-contact charge collection [Amman 2000b] [Looker 2015b]. However, the selection of an optimum a-Ge recipe will not likely be solely dictated by the inter-contact charge collection behavior since other properties such as charge injection are often critical for the detector application. For this reason, the film resistivity and electrical contact properties associated with the a-Ge recipes previously shown to eliminate inter-contact charge collection were investigated and are presented and discussed later in this paper.

Other important characteristics of the a-Ge film that must be considered when optimizing the process recipes include the maximum sustainable electric field before breakdown, passivation capability, and performance stability with temperature cycling, storage at room temperature, and annealing. The maximum electric field that the a-Ge contact can withstand and the passivation effectiveness of the a-Ge coating both would be worthwhile topics of study but have not been explored as part of the work of this paper. However, for a well-made detector, the a-Ge contact has been consistently shown to be capable of operating at typical detector fields without suffering electrical breakdown. The contact stability with temperature cycling appears to not be an issue, but the stability with room temperature storage can be and has been investigated previously [Looker 2015a] and is also a subject of this paper. The a-Ge contact and passivation coating performance stabilities with annealing are relevant for applications where significant radiation damage of the detector is anticipated and the damage must be annealed away. Some anneal testing has been done in the past [Hansen 1980], but a thorough study has not appeared in the literature. However, the data on detector stability with extended room temperature storage provided in this paper should be of relevance to the stability with typical radiation damage annealing.

## 5. Experimental Methods

For this study, both thin film a-Ge resistors and small planar HPGe detectors with a-Ge contacts were fabricated and characterized. Illustrations of these devices along with the electrical circuits used to carry out the characterization measurements are shown in Figure 5.1. The resistors were used to extract the a-Ge electrical resistance, and the detectors were used to measure the charge injection properties of the a-Ge electrical contacts. In this section, the device structures, fabrication processes, and characterization methods associated with these devices are described.

Each thin film a-Ge resistor, as depicted in Figure 5.1a, consisted of a-Ge deposited onto a glass substrate along with two Al interdigitated grid electrodes deposited on top of the a-Ge layer. These devices were used to extract the a-Ge electrical resistance, and the chosen Al electrode pattern allowed for a sensitive measurement of this property. Each resistor was produced using the following procedure. A standard microscope glass slide was first ultrasonically cleaned in detergent, rinsed in deionized water, and blown dry with dry nitrogen. This cleaned substrate was then placed into an RF sputter deposition system. The sputter deposition system was typically pumped overnight prior to the deposition of an approximately 250 nm thick a-Ge film onto the glass substrate. Following the sputter deposition, the coated substrate was loaded into a thermal evaporation vacuum system, and an evaporation shadow mask was placed over the a-Ge film. After pumping the system to a pressure at the low $10^{-7}$ Torr level, an Al layer of roughly 100 nm in thickness was deposited through the shadow mask onto the a-Ge film. Two separate Al evaporations, each with a different shadow mask, allowed the electrode pattern of Figure 5.1a to be created. After the completion of the Al grid electrode depositions, the resistor was ready for testing and was loaded into a variable temperature cryostat. This cryostat is of the same design as that used for the

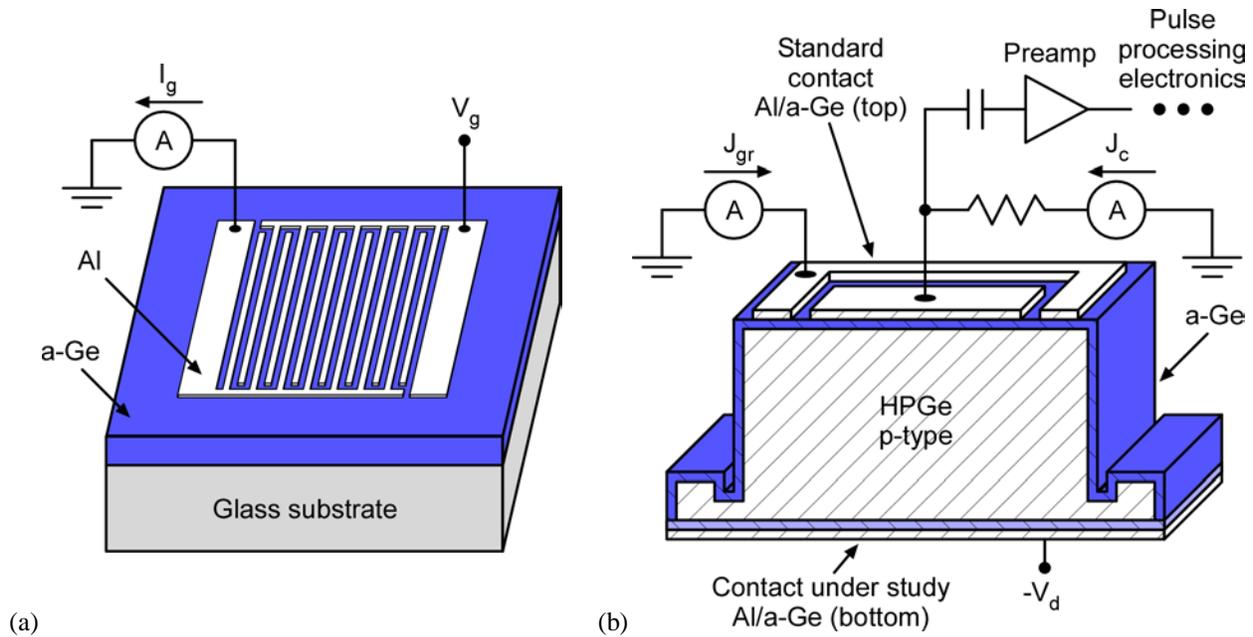

**Figure 5.1 (a)** Schematic drawing showing the thin film a-Ge resistor configuration and electrical circuit used for the determination of the a-Ge electrical resistance. The a-Ge thickness was typically 250 nm. The Al electrodes deposited on top of the a-Ge consisted of two interdigitated sets of strips. The strip length, width, and separation to neighboring strip were 5.4 mm, 0.2 mm, and 0.2 mm, respectively. **(b)** Schematic cross-sectional drawing of the HPGe detector configuration and electrical circuit used for detector capacitance and bulk injected leakage current measurements. The top electrical contact on the detector was produced using the same process for all detectors, whereas the bottom contact varied from detector to detector and was the contact being studied. The top contact had its Al layer patterned into separate center and guard ring electrodes. The Al layer on the bottom covered the entire contact surface.

HPGe detector testing. The variable temperature sample stage of this cryostat is enclosed by an infrared shield held at a temperature near 77 K. The temperature of the sample stage can be controlled from 79 K to a maximum that is greater than 200 K. The measurement electronics used to characterize the a-Ge samples were Keithley picoammeters including a model with a high resolution voltage source. The measurement procedure consisted of setting the sample temperature, waiting for the sample temperature to stabilize, and acquiring a current versus voltage ($I_g$-$V_g$) characteristic between the two Al grid electrodes. For the $I_g$-$V_g$ characteristic, the voltage range was from 0 to 50 V, and the voltage step size was 10 V. This procedure was then repeated for a range of sample temperatures. Since the $I_g$-$V_g$ characteristics from all of the samples tested were linear, the resistance between the two grid electrodes of each sample was extracted through a simple linear fit to the $I_g$-$V_g$ data. Finally, the a-Ge sheet resistance and resistivity were determined from the inter-grid resistance by multiplying the resistance with the appropriate geometric factors.

    In order to measure the charge injection properties of a-Ge electrical contacts on HPGe, small planar HPGe detectors were produced and characterized. The detector geometry along with the measurement setup used to test the detectors is shown in Figure 5.1b. Several HPGe crystals were used for this study, but all were cut from a single HPGe slice purchased from Ortec [Ortec 2018]. Because of this, all of the crystals were of the same net impurity type and approximately the same net impurity concentration. To convert each HPGe crystal into a detector, the following procedure was used. The crystal was first cut to the shape shown in Figure 5.1b using linear cuts with an outer diameter disk diamond saw. The main body of the cut crystal consisted of a square contact area measuring 18 mm on a side and a thickness perpendicular to the contact faces of about 10 mm. In addition to this active volume of the detector crystal were two thin extensions (handles) protruding from the bottom side of the crystal. The geometry of the crystal and contacts is such that, during operation as a detector, the depletion region within the crystal never extends significantly into the handles. Since the handles remain undepleted, surface damage to the handles will not introduce leakage current. Consequently, the handles simplify detector fabrication and mounting during testing by providing an area of the crystal that can be handled without negatively affecting the detector performance. This geometry is similar to the top hat geometry that has been extensively used in the past both to provide the handle (brim of the top hat) for processing and to introduce geometric control of any surface channel along the side of the detector [Llacer 1966]. However, for the surface channel geometric control to be fully effective, there would need to be a handle around the entire bottom perimeter of the detector rather than on just two sides. A full perimeter handle would also ease fabrication difficulties since its presence makes it easier to eliminate undercoating during the first a-Ge deposition. Nonetheless, the two handle

geometry of Figure 5.1b was chosen instead for this study because it mimics the geometry used for the large area strip detectors of the type shown in Figure 2.1. For these large area detectors, this geometry helps achieve the design goals of minimizing both dead material and the spacing between detectors.

Following the crystal cutting, each of the exposed flat surfaces of the cut crystal was then lapped in order to remove any blade marks left by the cutting operation. The surface damage introduced by these mechanical processes was then removed by etching the crystal in a concentrated 4:1 nitric to hydrofluoric acid mixture. An etching time of several minutes was used so that the surfaces would become smooth and shiny. The crystal was then rinsed and dried, and then inspected for the presence of any cracks or deep scratches. Any identified defect was then removed through additional lapping and etching. Following this surface polish etch, the crystal was again etched briefly in fresh 4:1 etchant, quenched in deionized water, rinsed in methanol, and blown dry with nitrogen in order to prepare the surfaces for the electrical contact depositions. During this etch and all subsequent processing steps, only the crystal handles were used to hold and manipulate the crystal. The crystal was then immediately loaded into an RF sputter deposition system. The first of two sputter depositions consisted of coating the top contact face and the sides of the crystal with a-Ge. The deposition was done with the crystal offset from the sputter target center and with rotation to ensure adequate coating of the crystal sides. The crystal was also placed on an Al block sized specifically for the crystal shape so that the crystal sides would be coated, the bottom face would receive little to no coating, and the bottom face would not be in contact with the block. The recipe used for this top contact was approximately the same for all detectors and is described below and referred to as the standard process. After the top and sides were coated, the crystal was flipped over and loaded onto a different Al holder, and the bottom contact face was sputter coated with a-Ge. This bottom contact was the focus of the study, and the process used to deposit the film was varied from detector to detector. With the detector crystal completely coated with a-Ge, Al was thermally evaporated onto both the top and bottom surfaces of the crystal. A metal ring shadow mask was used to define a guard ring structure on the top face, and a full area coating was deposited onto the bottom face as shown in Figure 5.1b. The Al thickness used for both faces was approximately 100 nm. Since the Al evaporations potentially could partially coat the detector sides, the top and bottom contacts were covered with acid resistant tape, and the detector was soaked in a 100:1 deionized water to hydrofluoric acid mixture in order to remove any unwanted Al. Following this, the detector was rinsed, untaped, rinsed again, and dried. This then completed the detector fabrication.

Each crystal was fabricated into a detector many times and electrically tested after each fabrication run. Detector reprocessing consisted of removing the Al and a-Ge layers with appropriate etchants and then following the fabrication sequence described above starting at the brief 4:1 etching step. Each fabrication iteration typically utilized a slightly different recipe for the bottom a-Ge electrical contact, and the measurements from different runs were compared in order to gain a better understanding of the contact behavior and its impact on the detector performance.

After each detector was fabricated, it was loaded onto the variable temperature sample stage of a test cryostat and then cooled so that capacitance and leakage currents could be measured. The design of the cryostat is the same as that used for the testing of the a-Ge resistors. The measurement electronics used to characterize the detectors included Keithley picoammeters for current measurements on both the guard ring and center contacts, and signal processing electronics for the readout of the signals from the center contact (see Figure 5.1b). The signal readout electronics consisted of an AC coupled charge sensitive preamplifier followed by a commercial analog pulse shaping amplifier. The signal readout allowed for the spectral characterization of the detector and the measurement of detector capacitance as a function of the applied detector voltage ($C$-$V_d$ characteristic).

For each detector crystal, a $C$-$V_d$ characteristic was measured in order to determine the full depletion voltage and the impurity concentration of the crystal. This characteristic was obtained by applying a constant voltage $V_d$ plus a small voltage step to the detector. In our measurements, we applied a negative voltage to the bottom contact so that depletion began at the top contact (assuming p-type HPGe). The voltage step riding on top of $V_d$ leads to a measured charge pulse that is proportional to the product of the voltage step amplitude and the capacitance between the center contact of the detector and the undepleted HPGe (or the bottom contact if the detector is fully depleted). Since the voltage step amplitude is known, the capacitance is determined by measuring the charge pulse amplitude. This measurement is repeated for a range of detector voltages to generate a $C$-$V_d$ characteristic. As $V_d$ is increased from zero, the depletion region grows, and the capacitance decreases. Once the depletion region reaches the bottom contact and the detector becomes fully depleted, the capacitance becomes constant with further increases in $V_d$. Assuming a constant impurity concentration throughout the crystal, the impurity concentration is then estimated from the extracted full depletion voltage $V_{fd}$ using the equation $N = 2\varepsilon V_{fd}/qt^2$, where $N$ is the net ionized impurity concentration, $\varepsilon$ is the dielectric constant of Ge, $q$ is the magnitude of the electron charge, and $t$ is the detector thickness [Bertolini 1968] [Knoll 1989]. The crystals used in this study were all p-type, 10 mm thick, and fully depleted at roughly 200 V. Based on this, the net impurity concentration for these crystals was determined to be about $3 \times 10^9$ impurities/cm$^3$.

The leakage current in a fully depleted HPGe detector arises from several mechanisms including hole injection at the positive electrical contact, electron injection at the negative contact, charge flow along the surface of the detector, and thermal generation within the HPGe. The objective of the contact study was to focus on the charge carrier injection behavior of one of the contacts on the detectors (typically electron injection at the bottom). The use of the guard ring structure of Figure 5.1b along with the measurement of the center contact leakage current as a function of detector voltage allowed this to be done. With the guard ring configuration and using only the leakage current measured from the center contact, the surface leakage is eliminated since this component is taken up by the guard ring [Goulding 1961]. Furthermore, measuring the step increase in the center contact current at full depletion quantifies the

# M. Amman, 2018, "Optimization of Amorphous Germanium Electrical Contacts and Surface Coatings on High Purity Germanium Radiation Detectors"

electron injection at the bottom contact. An example illustrating this is the measured $J_c$-$V_d$ characteristics plotted along with the $C$-$V_d$ characteristic shown previously in Figure 3.5 and discussed in Section 3. At detector voltages below full depletion, there is no electric field capable of injecting electrons at the bottom contact and hence no leakage current from injection at that contact. The leakage current, in this case, is predominantly from hole injection at the top center contact. However, once full depletion is reached (as indicated by the $C$-$V_d$ characteristic) and an electric field appears at the bottom contact, a step in the current occurs. This step increase is solely the result of electron injection at the bottom contact. By measuring this electron injection current step as a function of the fabrication process variables and detector storage time at room temperature, it is possible to identify and optimize parameters that affect this source of leakage current. Furthermore, measurement of this electron injection as a function of temperature along with analysis using the ACS theory of Section 3, allows one to determine the electron barrier height of the contact $\varphi_e$ as well as $J_o$ and $N_f$. This analysis was done for all of the detectors produced. Additionally, the hole barrier height of the contact can be estimated using $\varphi_h = E_g - \varphi_e$, where $E_g$ is the energy bandgap of Ge. This means that all of the ACS parameters for the contact are known, and, as a result, the charge carrier injection of the contact under either voltage bias polarity and at any temperature can be estimated. Consequently, the suitability of the contact for a particular application from a leakage current perspective can easily be assessed.

As mentioned previously, the a-Ge thin films of the resistor samples and on the detectors were deposited using RF sputtering. The properties of the a-Ge substantially depend on the sputter conditions possibly including sputter chamber geometry, sputter target composition, sputter target configuration (for example, diode or magnetron), sputter gas composition, sputter gas pressure, sputter gas flow rate, RF power, sample temperature, and deposition rate. For this reason, additional details are provided here concerning the sputter deposition systems and processes. Two different sputter deposition systems were used for the study. This was done so that the results from the two systems could be compared and common behaviors identified. The first of the two sputter deposition systems, referred to as Sputterer 1 in this paper, was customized and then used at LBNL over the last several decades. The vacuum system of this sputterer consists of a 61 cm diameter turbo-pumped chamber capable of achieving a base pressure in the mid to low $10^{-6}$ Torr range. Inside the chamber are three 20 cm diameter sputter targets configured for diode sputtering. The Ge target used in this study is composed of 99.999% purity Ge obtained from American Elements [American Elements 2018]. Also inside the vacuum chamber is a water cooled sample stage that could be positioned to be directly beneath the active target. However, this cooling stage was not used when sputtering a-Ge on the top and sides of the detectors. Instead, the detector was placed on a rotation stage offset from the center of the target. This was done to improve the thickness uniformity of the a-Ge coating on the side of the detector. The resistor samples and the deposition on the bottom face of the detectors made use of the water cooled stage. Despite this, it was observed that the detector temperature increased significantly during the a-Ge deposition in Sputterer 1 as evidenced by the elevated temperature of the detector holder when it and the detector were removed from the sputter deposition system shortly after the deposition was completed. In an attempt to reduce this heating, a multistep deposition was used for all depositions made in the Sputterer 1 system. This consisted of sputtering for 2 minutes, waiting for a cool down period of at least 5 minutes, and then repeating the sputtering and cool down until the desired film thickness was achieved. The process parameters used for depositing a-Ge in the Sputterer 1 system consisted of a power of 300 W, sputter gas of pure Ar or Ar with 7% $H_2$, and sputter gas pressure ranging from about 5 to 23 mTorr. The sputter gas pressure was measured using a Kurt J. Lesker Company 275 convection enhanced Pirani vacuum gauge. This gauge measures pressure indirectly and was factory calibrated for $N_2$. Consequently, a correction was made based on a conversion table provided by the gauge manufacturer in order to obtain the true sputter gas pressure [Lesker 2018]. In recent previous publications [Looker 2015a] [Looker 2015b], this correction was not made to the sputter pressures provided but should have been.

The second of the two sputter deposition systems, referred to as Sputterer 2 in this paper, was a model ATC-1800 UHV unit manufactured by AJA International [AJA 2018]. The vacuum system of the AJA sputterer consists of a 46 cm diameter cryo-pumped chamber capable of achieving a base pressure in the low $10^{-8}$ Torr range. Inside the chamber are three 7.6 cm diameter magnetron sputter source targets in a confocal configuration. The Ge target used in this study is composed of 99.999% purity Ge provided by AJA. The sample stage at the focal point of the sputter sources is water cooled and rotates during deposition. In this system, there appeared to be little to no detector heating during the a-Ge sputter deposition. This is expected since the magnetron sputter source traps free electrons in a magnetic field near the target thereby reducing the electron induced sample heating and damage that would normally occur in the diode target configuration. The process parameters used to deposit a-Ge in the magnetron sputter system consisted of a power of 200 W, sputter gas of pure Ar or Ar with 7% $H_2$, and sputter gas pressure ranging from about 1 to 15 mTorr. The sputter gas pressure was measured using a capacitance manometer gauge, which provides a direct measure of the pressure and therefore requires no correction based on gas composition.

As described above, the first of the two sputter depositions on a detector typically consisted of coating the top contact face and the sides of the crystal with a-Ge using a standard process that was the same for all detectors. This deposition was done in the Sputterer 1 system using the offset rotation stage. The a-Ge sputter process parameters consisted of a power of 300 W, sputter gas of Ar with 7% $H_2$, sputter gas pressure of 23 mTorr, and total deposition time of 16 min. This produced a film thickness on the top face of about 300 nm. The bottom face of the detector received an a-Ge coating deposited using parameters that varied from detector to detector. The thickness of the a-Ge on the bottom was approximately 250 to 300 nm.

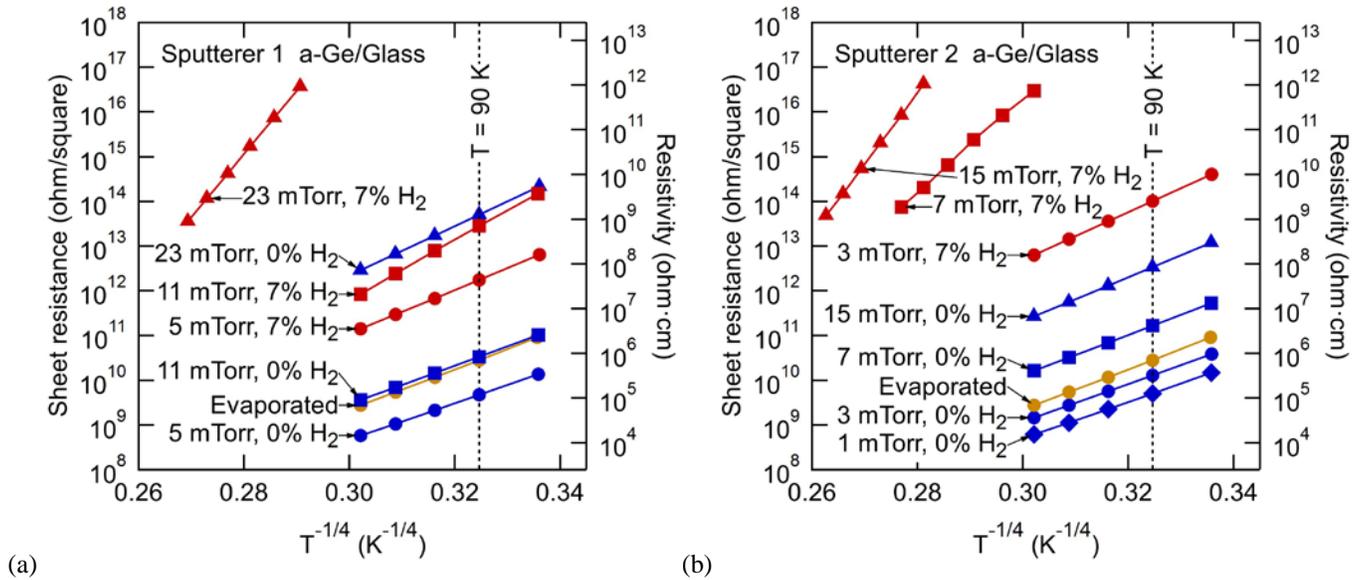

**Figure 6.1** Measured sheet resistances of a-Ge thin films deposited onto glass plotted as a function of temperature to the power of -1/4. The thin films were deposited using RF sputtering with various sputter gas pressures and either pure Ar gas or a mixture of Ar with 7% $H_2$ as indicated in the figure. For comparison, measurements obtained from a thermally evaporated a-Ge film are also plotted in the graphs. The film resistivity is equal to the product of the film sheet resistance and the film thickness. The resistivity axis range for the graphs was set assuming a film thickness of 250 nm, which is approximately correct for all of the sputtered films. The evaporated film however was approximately 190 nm thick. A typical temperature for HPGe detector operation is 90 K. This temperature is indicated with a black dashed line in the graphs. **(a)** Data from a-Ge films deposited in the Sputterer 1 system. **(b)** Data from a-Ge films deposited in the Sputterer 2 system.

## 6. Amorphous Germanium Electrical Resistance Measurements

In Section 3, the electronic energy band structure of a-Ge and the physics of the charge transport through the material were described. Two key facts from that section are relevant to the electrical resistance of the a-Ge typically used to produce HPGe detectors. First, there is a significant density of localized defect states within the bandgap of this material. Second, at the low temperatures used for HPGe detector operation, the charge transport through the a-Ge is typically dominated by conduction through these localized states and specifically through those located near the Fermi energy [Mott 1979 and references therein]. Beyond this homogeneous view of the a-Ge film are the realizations that sputtered films are typically heterogeneous with a microstructure containing voids and that this microstructure plays a role in determining the material properties including that of the charge transport. Consequently, fabrication process modifications known to affect the defect states and the film microstructure can be used to modify the charge conduction and hence the a-Ge film resistance. In Section 4, for the a-Ge thin film layer used to coat the inter-contact detector surfaces, it was shown that the charge transport in this film is important to the performance of the detector. Specifically, it was argued that the resistance of the a-Ge coating can both impact the current noise of the detector and the extent to which undesirable charge collection to the inter-contact surfaces occurs. Due to the significance of this film property on the detector performance, in this section, a summary of electrical resistance measurements made on thin films of RF sputtered a-Ge of the type used to produce HPGe detectors is presented.

Extensive studies of the charge transport within a-Ge exist in the literature [Mott 1979 and references therein] [Lewis 1976] [Connell 1976] [Moustakas 1977]. These studies have contributed to an understanding of the electrical conduction in the films and the effect of process parameters on this material property. As described in Section 5, two different sputter deposition systems were used to create the thin film a-Ge resistor samples studied as part of the work of this paper. Comparing the results from the two systems allows common process dependencies to be identified. This combined with the existing literature on this subject should assist in determining the critical process parameters that can be used to adjust and optimize the properties of the films produced in other deposition systems. A summary of the resistance measurements made as a function of temperature on a-Ge films sputter deposited onto glass is contained in Figure 6.1. The measurements associated with the two different deposition systems (Sputterer 1 and Sputterer 2 described previously in Section 5) are plotted in separate graphs. In each graph, the sheet resistance measured as a function of temperature is plotted for a number of samples. To obtain the film sheet resistance, an $I_g$-$V_g$ characteristic was measured between the Al interdigitated grid electrodes deposited onto the a-Ge (Figure 5.1a). All samples produced linear $I_g$-$V_g$ characteristics from which the resistance was

M. Amman, 2018, "Optimization of Amorphous Germanium Electrical Contacts and Surface Coatings on High Purity Germanium Radiation Detectors"

extracted. This resistance was then converted to a sheet resistance using a multiplicative factor dictated by the grid electrode geometry. This extracted sheet resistance is plotted on a log scale as a function of the measurement temperature to the power of -1/4. The motivation for this type of plot is a charge transport model based on variable range hopping that predicts such a dependence on the temperature [Mott 1979 and references therein].

Each sample whose data is plotted in one of the graphs of Figure 6.1 was produced using different sputter process parameters. There are many parameters that could potentially impact the resistance of the a-Ge including sputter chamber geometry, sputter target composition, sputter target configuration (diode or magnetron), sputter gas composition, sputter gas pressure, sputter gas flow rate, RF power, sample temperature, and deposition rate. The work of this paper focused on two process parameters: sputter gas $H_2$ content and sputter gas pressure. The $H_2$ content of the sputter gas was investigated because measurements previously done at LBNL as well as those in the literature [Lewis 1976] [Connell 1976] [Moustakas 1977] demonstrated that adding $H_2$ to the Ar sputter gas can increase the resistance of the resultant a-Ge film by several orders of magnitude. An explanation for this change is that localized electron energy band gap states in a-Ge can arise from Ge atoms that have incomplete electron bonding with neighboring Ge atoms. With the addition of $H_2$ to the sputter gas, H is incorporated into the film and can bond with the Ge atoms that would otherwise have incomplete bonding. That is, the incorporated H can compensate individual dangling Ge bonds, thereby modifying or removing the associated energy gap defect states. The gap states are reduced, and, as a result, the conduction that relies on the existence of these states is reduced. A further reason for studying the impact of the sputter gas $H_2$ concentration is the work of Hansen et al. [Hansen 1980] that demonstrated the adjustment of the electrical state of planar detector side surfaces coated with a-Ge by varying the sputter gas $H_2$ content. In their work, they obtained desirable results on p-type HPGe detectors (minimum charge collection to the detector side surfaces) with a concentration of 7% $H_2$ in the Ar sputter gas.

The considerable impact of $H_2$ on the film resistance of a-Ge is clear from the measurements plotted in the graphs of Figure 6.1. Results are shown for samples produced with pure Ar (0% $H_2$) and Ar with 7% $H_2$. Comparing samples sputtered in the same sputter deposition system and at the same gas pressure, the addition of $H_2$ increased the electrical resistance of the resultant film by about three orders of magnitude or more. This was true regardless of the sputter deposition system or the sputter gas pressure used to produce the samples. A further observation of note that is not evident from the figure was that residual $H_2O$ in the sputter chamber could also contribute H to the film and increase its resistance. Substantial sample to sample variation can result for samples sputtered in pure Ar as a result of the variable nature of the residual $H_2O$. The $H_2O$ remaining in the chamber at the time of sputtering will depend in part on the humidity of the air at the time the chamber was last open, the length of time the chamber was left open to air, the pump down time, the pumping method and process, the base pressure of the deposition system, and the duration of the presputter step. In comparing the two sputter deposition systems used for this study, Sputterer 1 was likely much more susceptible to suffer from residual $H_2O$ induced sample variation. This is because of the much higher base pressure of Sputterer 1 (mid to low $10^{-6}$ Torr compared to the low $10^{-8}$ Torr of Sputterer 2) and the poorer $H_2O$ pumping ability of the Sputterer 1 pumping system (turbopump as compared to the cryopump of Sputterer 2). Significant variation in sample resistance was observed early on in the work at LBNL from resistors produced in Sputterer 1 with pure Ar when a short pump down time of several hours was used. Reproducibility was later improved with the adoption of an overnight chamber pump out along with minimizing the time the chamber was open to air. All samples produced for this study were sputtered using this more careful pump out procedure. Nonetheless, some variable amount of residual $H_2O$ will remain that potentially can measurably increase the resistance of the films sputtered in pure Ar. This should be less of a concern when sputtering in Ar with 7% $H_2$ since the uncontrolled H addition from $H_2O$ will only be a fraction of the total amount. The influence of the O contributed by residual $H_2O$ is also of note. Oxygen can reduce the surface mobility of the deposited Ge atoms, as has been shown for other materials [Thornton 1986], thereby leading to a more porous film microstructure, and the O will also react with the Ge to form an oxide. Both processes will act to increase the resistivity of the resultant film over that of a film deposited in an O free system.

Previous studies of HPGe detectors with a-Ge contacts revealed that the sputter gas pressure of the a-Ge deposition affects and could be used to optimize the room temperature storage stability of the electrical contacts [Amman 2007] [Looker 2015a]. For this reason, the study of this paper evaluated the impact of sputter gas pressure on a-Ge film resistance. The measurements of Figure 6.1 demonstrate that, when all settable process parameters other than the gas pressure are fixed, the a-Ge resistance consistently increases with increasing gas pressure regardless of the gas $H_2$ content or which of the two deposition systems was used to produce the film. The sputter gas pressure can directly affect the properties of the deposited film or indirectly by changing other process parameters such as deposition rate and substrate temperature that then, in turn, influence the film properties [Messier 1976] [Fahnline 1989] [Windischmann 1992]. A physical mechanism for the role that the pressure plays is based on the pressure dependence of the mean free path of the atoms and ions in the sputter chamber and the impingement of these particles on the growing film [Windischmann 1992 and references therein]. The deposited film is bombarded by secondary electrons, sputtered Ge atoms, and back reflected neutralized sputter gas atoms. The secondary electrons primarily act to heat the film and substrate. This heating is of greater importance for diode sputter systems such as Sputterer 1 as compared to magnetron configured target systems such as Sputterer 2 that confine the electrons near the target. Despite this difference in substrate heating between Sputterer 1 and Sputterer 2, the sputter gas pressure dependence of the two systems is similar. In contrast to the effect of electron bombardment is that of the Ge and Ar atoms impinging on the a-Ge film. The Ge atoms sputtered off of the target can have an energy of order 10 eV, whereas the back reflected Ar atoms near the target can have an energy that is greater by an order of magnitude. Consequently, this flux of atoms, particularly the back reflected Ar, can induce structural changes in the a-Ge film. The energy of these energetic atoms will be reduced and their direction modified by collisions with the sputter gas during their transit from the target to the substrate. The extent to which this takes place depends on the mean free path of the atoms, which is inversely proportional to the gas pressure in the sputter chamber. For example, the mean free path of Ar atoms in a background of Ar at a pressure of 5 mTorr is about 1 cm, whereas that of 23 mTorr is about 0.2 cm [Weissler 1979]. At low sputter gas pressures, scattering is less, thereby leading to back reflected Ar and sputtered Ge atoms with a high normal flux component and high energy striking the a-Ge film. Through a process referred to as "shot peening" this acts to modify the growing a-Ge film creating a denser material. At higher sputter gas pressures, gas scattering is increased with the effect that the atoms reaching the a-Ge film have a decreased energy and flux, and an increased oblique component of the flux. This promotes a column like, porous film structure, thereby resulting in a lower density material. At high enough sputter gas pressures, the energetic atoms will all thermalize before reaching the a-Ge film. It should also be noted that for the case of a planar diode sputter source such as that of Sputterer 1, the substrate is in contact with the plasma. The a-Ge film will thus also see an energetic ion flux that will promote the generation of a denser film [Thornton 1986]. The energy of these ions is inversely related to the sputter gas pressure, thereby reinforcing the generation of a denser film at lower gas pressures.

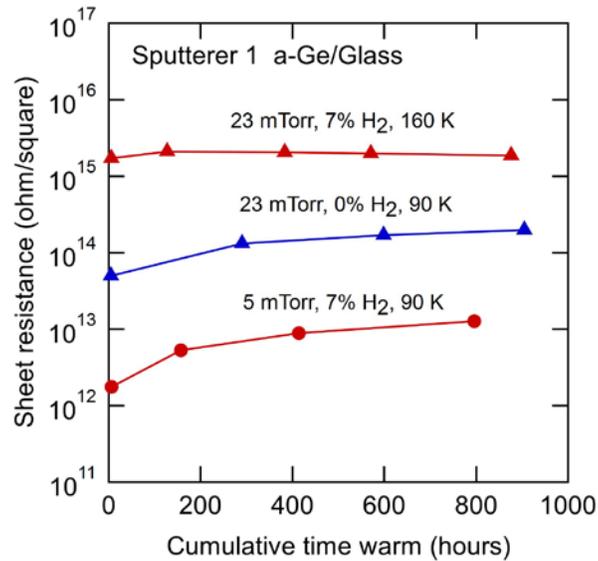

**Figure 6.2** Measured sheet resistances of a-Ge thin films deposited onto glass plotted as a function of cumulative time stored at room temperature. The thin films were deposited using RF sputtering with different sputter gas pressures and either pure Ar gas or a mixture of Ar with 7% $H_2$ as indicated in the figure.

If the a-Ge film were homogeneous, a possible mechanism for the resistance dependence on sputter gas pressure could be that the more energetic bombardment of particles onto the film at lower sputter pressures introduces more film disorder. This greater disorder results in more electron defect states within the energy band gap of the material and, consequently, a lower resistance. This explanation, however, neglects the known film microstructure as described above. The consequences of the film microstructure dependence on the sputter gas pressure are that at lower sputter gas pressures the a-Ge film will be denser, have a lower void fraction, have more incorporated Ar, and tend to be under compressive stress, whereas at higher pressures the film will be less dense, have a higher void fraction, have less incorporated Ar, and be more likely to be under tensile stress [Fahnline 1989] [Windischmann 1992]. This connection between sputter gas pressure, film microstructure, and film properties has been shown to be common to a wide variety of sputter deposited materials and sputter deposition configurations [Thornton 1986]. It has also been shown that the resistivity of these other materials typically increases with increasing sputter gas pressure. This is what is seen in the a-Ge resistor data of Figure 6.1. An explanation put forth for this behavior when observed with other materials is that the more open structure produced at higher sputter gas pressures is more significantly oxidized during growth and after exposure to the atmosphere which then increases the film resistivity [Thornton 1986].

Several fundamental factors play a role in the physical process of film growth just described. These include the flux, energy, oblique component, and composition of the particles impinging on the growing film, and the substrate temperature. These factors are not only affected by the sputter pressure as described above, but also by sputter RF power, substrate bias, intentional ion bombardment of the substrate, active substrate heating or cooling, substrate orientation, substrate to target distance, and cathode shape [Thornton 1986] [Windischmann 1992]. Consequently, these other factors could also be used to adjust the film microstructure and optimize the resultant properties of the film.

The stability of the film resistance with storage at room temperature was also evaluated since HPGe detectors must often be stored for extended periods at room temperature between uses or during assembly of complex multi-detector systems. Several a-Ge resistor samples were characterized through a repetitive sequence of cooling, measuring $I_g$-$V_g$ characteristics at low temperatures, warming, and storing for a period at room temperature. The extracted sheet resistance data from this measurement sequence performed on three samples are plotted in Figure 6.2. Each of the three samples was produced with a different sputter recipe using Sputterer 1, yet all exhibited a resistance that at least initially increased with room temperature storage time. The sample sputtered in 23 mTorr of Ar with 7% $H_2$ had a resistance that initially increased with storage time followed by a slow decrease. The resistances of the other two samples continuously increased with time, and the rate of this increase appeared to lessen over time. Possible processes that could produce a resistance change would include those involving the movement and bonding changes of the Ge, O, and H in the film. It was demonstrated early on in the a-Ge literature that annealing decreases the conductivity (increases the resistivity) of sputter deposited a-Ge films [Mott 1979 and references therein]. This was attributed to a reduction of Ge bond angle distortion and improvement in coordination number that then would result in a reduction of the localized electron energy states in the band gap of a-Ge. Structural changes in sputter deposited a-Ge have also been observed by Okugawa et al. [Okugawa 2016] to occur during aging at room temperature. Their measurements relied on electron diffraction to characterize the film structure and identified changes occurring over periods of several months in duration with these continued changes diminishing after about seven months. Regarding O, its possible role in affecting the resistance of a-Ge, as mentioned previously, is based on the oxidation of the exposed a-Ge surface, which would be more significant for the high pressure sputter gas process recipes that lead to a more open microstructure [Ghazala 1991]. An increased resistance with storage would be expected as atmospheric $O_2$ and $H_2O$ further oxidize the a-Ge film. However, the measurements associated with Figure 6.2 were all made with the samples left in vacuum for the duration of the measurement sequence. Finally, the ability of H to increase the resistance of a-Ge has clearly been established. The diffusion of H in a-Ge is a possible mechanism capable of creating a time dependent film resistance. Hydrogen diffusion and desorption in a-Ge is a research area that has been investigated by others in the past, though much of this work has been for samples at temperatures above that of room temperature [Wu 1991] [Graeff 1993] [Beyer 1996]. Regardless of the mechanism causing the resistance increase, for most applications, a very high resistance is needed, so an increasing value should not be a problem unless the physical changes in the film also lead to additional noise such as that of a $1/f$ type seen in noisy dielectrics.

In addition to the a-Ge thin film resistor samples, sputtered a-Ge films were studied when deposited onto the cathode face of HPGe detectors. The detector configuration was similar to that of Figure 5.1b except with the cathode full area Al metallization replaced with the interdigitated grid electrode pattern of Figure 5.1a plus a surrounding guard ring. As with the resistor samples, the measurements consisted of acquiring $I_g$-$V_g$ characteristics as a function of temperature. These measurements were made with the detector fully depleted. All of the detector samples produced linear $I_g$-$V_g$ characteristics from which resistances were extracted and then converted to sheet resistances. The results from four different samples are shown in Figure 6.3. Each of the samples was characterized as a function of room temperature storage time, and the measurements made shortly after the detector fabrication, as well as those acquired after more than a month of room temperature storage, are plotted in the figure. Overall, the resistance behavior of these a-Ge films on the HPGe detectors qualitatively matches that of the films deposited on glass. Common dependencies include increasing resistance with the addition of $H_2$ to the sputter gas and with increasing sputter gas pressure, and a resistance that typically increases with storage time. Several possible explanations for the lack of a close quantitative match exist. First, with depleted HPGe as the substrate, there is the possibility of parallel conduction pathways including that along the interface between the a-Ge and HPGe, and that injected into and transported through the HPGe. Second, the substrate material and surface preparation may affect the a-Ge film nucleation, growth, and microstructure, thereby impacting the film resistance. Third, the substrate temperature during a-Ge deposition will likely be different between glass and HPGe since different fixtures and connection to a cooling plate were used for the two types of substrates. Substrate temperature directly affects the film microstructure. Finally, the sample to sample variation was not assessed and could be significant enough to account for the differences.

Some additional observations of interest can be made regarding the resistance plots obtained from both the resistors and the detectors. First, the slopes of the plots are roughly all the same for the samples that have relatively low resistance values regardless of the sputter deposition system used to produce the a-Ge film, the $H_2$ content of the sputter gas, or the substrate type. This constancy of the slope has also been noted in the literature [Lewis 1976] [Theye 1980]. A further observation regarding the slopes of the plots is that the higher resistance samples depart from this constant slope behavior, and the slopes become steeper. Possibly the most curious behavior is that of the a-Ge sputtered in 11 mTorr of Ar with 7% $H_2$ onto HPGe whose data is plotted with square symbols in Figure 6.3. For the measurements made shortly after fabrication (filled squares), the resistance plot is not linear and has a slope that varies between that of the low resistance samples and that of the high resistance ones. After the sample was stored for about a month at room temperature (open squares), the resistance plot has transitioned to one that is substantially linear with a slope similar to that of the high resistance samples. Furthermore, this sample, unlike nearly all of the others, exhibited resistance values that did not consistently increase with storage time. The resistance change depended on the measurement temperature and could be seen to either decrease or increase over the one month period that it was studied.

The measurements presented in this section have demonstrated the ability to dramatically change (by many orders of magnitude) the sheet resistance of sputtered a-Ge by altering the sputter gas $H_2$ content and pressure. From these measurements, a-Ge sputter deposition recipes can be chosen based on the need to meet specific detector specifications. For example, consider the GRIPS strip

**M. Amman, 2018, "Optimization of Amorphous Germanium Electrical Contacts and Surface Coatings on High Purity Germanium Radiation Detectors"**

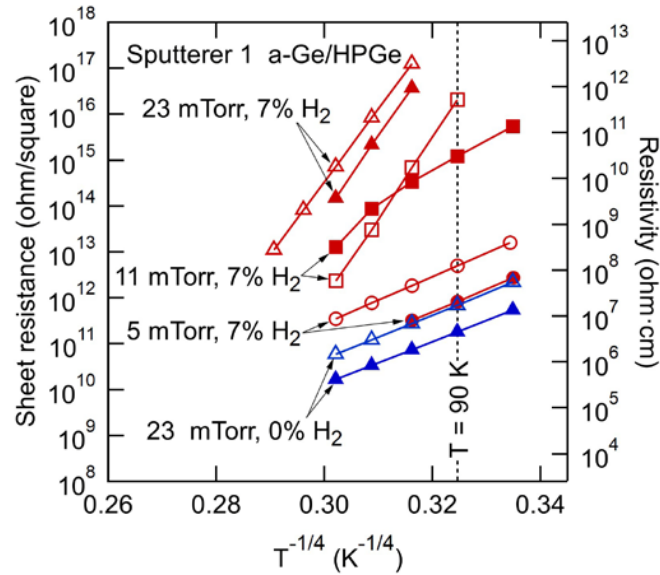

**Figure 6.3** Measured sheet resistances of a-Ge thin films deposited onto HPGe detectors plotted as a function of temperature to the power of -1/4. The thin films were deposited using RF sputtering with various sputter gas pressures and either pure Ar gas or a mixture of Ar with 7% $H_2$ as indicated in the figure. The detector configuration consisted of the a-Ge film and interdigitated grid electrode pattern of Figure 5.1a plus a surrounding guard ring deposited onto the cathode face of a detector similar to that of Figure 5.1b. The measurements were made with the detector fully depleted at $V_d$ = 600 V. The full depletion voltage of the detector was approximately 200 V. The measurements were made both shortly after the detector fabrication was completed (filled symbols) and after the detector was stored at room temperature for a cumulative time between 750 and 1000 hours (open symbols). The film resistivity is equal to the product of the film sheet resistance and the film thickness. The resistivity axis range for the graph was set assuming a film thickness of 250 nm, which is approximately correct for all of the sputtered films. A typical temperature for HPGe detector operation is 90 K. This temperature is indicated with a black dashed line.

detectors presented in Sections 2 and 4. A reasonable requirement would be for the inter-strip resistance as dictated by the a-Ge to have a negligible contribution to the current noise of the detector. Noise measurements from one of these detectors were presented and discussed in Section 4. Based on the noise analysis, it is clear that setting a lower limit of 1 GΩ on the resistance between a strip and all surrounding electrodes would conservatively ensure that this requirement is met. From the strip geometry, this minimum resistance of 1 GΩ translates to a minimum sheet resistance of about 2.5 x $10^{12}$ Ω/square. As can be determined from Figures 6.1 and 6.3, to meet this requirement at a typical detector temperature of 90 K, the a-Ge should be sputtered with $H_2$ added to the Ar sputter gas, and a moderate to high sputter gas pressure should be used. A more challenging requirement for the strip detectors that is possibly dictated by the a-Ge resistance is to eliminate the inter-strip charge collection as discussed in Section 4. Previous studies of the inter-strip collection have shown that it can substantially be eliminated through the use of a-Ge sputtered at high pressures of Ar with 7% $H_2$ [Amman 2000b] [Looker 2015b]. This is the sputter parameter regime producing the highest film resistances over the set of samples characterized in this study.

In contrast to the strip detector and its a-Ge property requirements is the proximity electrode signal readout detector, of which an example is schematically shown in Figure 1.4b [Luke 2009] [Amman 2013]. With the proximity readout technology, the signal readout electrodes are not electrically connected to the a-Ge layer but instead are separated from the a-Ge surface by a small gap. This configuration guarantees that multiple electrodes will register a finite integrated signal for each detected event, and, as a result, enables interpolation to improve spatial resolution. Since the readout electrodes are not in contact with the a-Ge, they cannot serve to maintain the electric field in the detector and remove the collected charge. This function must instead be taken care of by the a-Ge layer and necessitates a film surface resistance of about $10^9$ Ω/square. Consequently, the a-Ge surface coating should be sputtered in pure Ar at a low sputter gas pressure.

A final example is that of the PPC detector schematically shown in Figure 1.3 in which the a-Ge is used to produce a thin window area on the front face of the detector. Since, in this case, the a-Ge is solely used as a hole blocking contact and not for surface passivation, its resistance is not a critical property and therefore will likely not be used as a criterion for selecting the deposition recipe. Instead, the a-Ge recipe should be chosen based on hole injection blocking ability and performance stability.

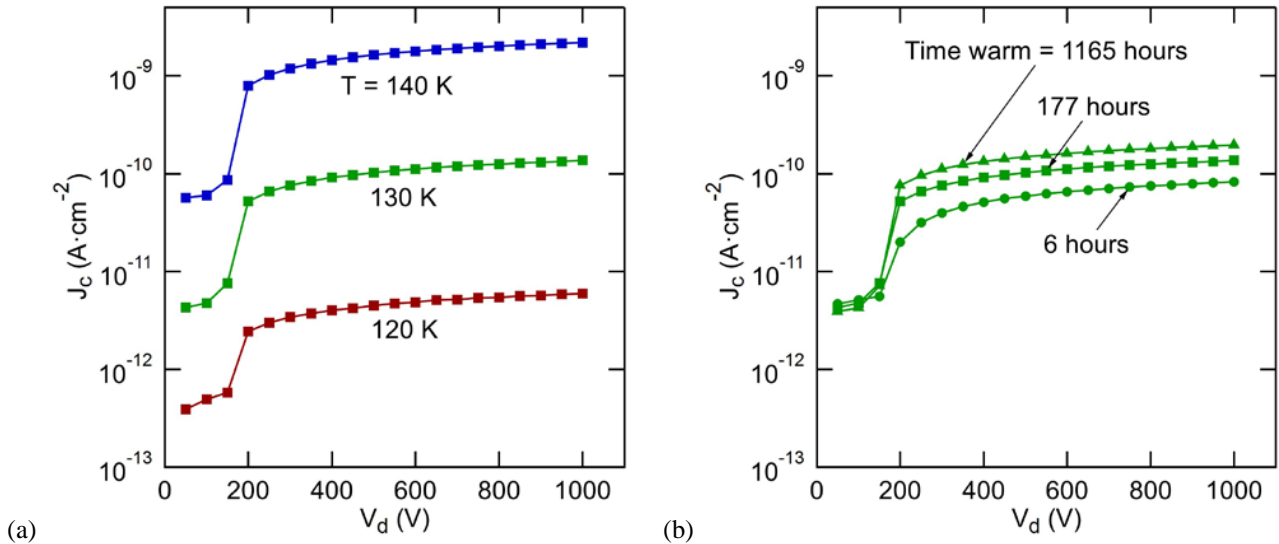

**Figure 7.1** Detector leakage current density measured as a function of detector voltage for a HPGe detector of the type shown in Figure 5.1b. The leakage current density is that measured from the center contact of the guard ring detector. The a-Ge of the top contact (anode) and detector sides was sputtered using Sputterer 1 with a sputter gas of Ar containing 7% $H_2$ at a pressure of 23 mTorr. The a-Ge of the bottom contact (cathode) was sputtered using Sputterer 1 with a sputter gas of pure Ar at a pressure of 5 mTorr. **(a)** Leakage current density as a function of detector voltage characteristic for three different detector temperatures of 120 K (red), 130 K (green), and 140 K (blue). The characteristics were all measured shortly after the detector was fabricated. **(b)** Leakage current density as a function of detector voltage characteristic measured after three different cumulative room temperature storage times. The storage times were 6 hours (circles), 177 hours (squares), and 1165 hours (triangles). The detector temperature was 130 K for all three characteristics.

## 7. Amorphous Germanium Charge Blocking Measurements

A theoretical model of the a-Ge electrical contact based on MS contact theory as first proposed by Dohler and Brodsky [Dohler 1974] [Brodsky 1975a] [Brodsky 1975b] was presented in Section 3. For the work of this paper, this theory (ACS model) was used as a means to understand the basic physics governing the injection of charge at the contact and to identify possible methods to control this component of the detector leakage current. As presented in Section 4, bulk injected leakage current contributes to the electronic noise of a detector and, if large enough, can significantly degrade the energy resolution of the detector. Consequently, charge injection blocking is a crucially important property associated with the contact. There are a couple important points from Section 3 that are worth restating here due to their relevance to the measurements of this section. First, in the model of the leakage current, it is useful to focus on two parameters. One is the potential energy barrier that acts to inhibit charge injection. For charge to be injected, it must gain enough thermal energy to overcome this barrier, thereby leading to a current that exponentially decreases with the energy barrier divided by the thermal energy. In the simple model, this energy barrier is determined by the position of the Fermi energy within the a-Ge energy band gap. The other model parameter is a proportionality prefactor that in principle depends on transition probabilities between energy states on each side of the contact. These parameters can be affected by the a-Ge contact fabrication process recipe. A second important point from Section 3 is that when a contact is negatively voltage biased, its electron injection is inhibited by its electron energy barrier, whereas when this same contact is positively biased, the hole injection is inhibited by the contact's hole barrier. The sum of the electron barrier and hole barrier is equal to the band gap energy of Ge. As a result, studying the charge injection behavior of only one of the carrier types is, in principle, sufficient since the blocking of the other carrier type can be inferred based on the contact model. Because of this, the measurements of this section primarily focused on electron injection.

As detailed in Section 5, planar HPGe detectors were produced with a guard ring electrical contact configuration so that measurements of only the bulk injected leakage current (excluding surface leakage) could be made. Through the use of p-type HPGe and placing the a-Ge contact to be studied on the cathode side of the detector, the electron injection of this contact could be quantified based on the current step that occurs at the point of full depletion. Such a method was previously employed in the study of amorphous semiconductor contacts [Amman 2007] [Looker 2015a]. For this paper, numerous detectors of the type shown in Figure 5.1b were fabricated and their current-voltage characteristics measured as a function of temperature and room temperature storage time. The fabrication process used to produce the cathode a-Ge contact was varied in order to determine the dependence of the electron injection and stability on sputter gas $H_2$ concentration and pressure. Two different sputter deposition systems were used to create the detectors, thereby allowing common process dependencies to be identified and improving the relevancy of the results to other sputter deposition

**M. Amman, 2018, "Optimization of Amorphous Germanium Electrical Contacts and Surface Coatings on High Purity Germanium Radiation Detectors"**

systems. Using the contact theory described in Section 3, the measurements were also analyzed, and the electron barrier heights and prefactors were extracted. In this section, the results of this leakage current study are summarized and discussed.

In Figure 7.1, example current-voltage characteristics are shown that were obtained from a detector with an a-Ge cathode (bottom contact, Figure 5.1b) deposited in Sputterer 1 using pure Ar at a pressure of 5 mTorr. Since the detector material is p-type, the depletion of the hole charge carriers in the material begins at the anode (top contact) as the detector voltage is increased from zero. At low voltages, the leakage current measured from the center contact of the anode is due to hole injection at the anode. This detector first becomes fully depleted at a voltage of about 200 V, and this voltage is the location of the current step associated with electron injection from the cathode that is clearly visible in all of the characteristics. Current-voltage characteristics acquired at three different temperatures (120 K, 130 K, and 140 K) are plotted in Figure 7.1a. Note that these temperatures are significantly higher than the operating temperature of a typical cryogenically cooled HPGe detector. At a more typical detector temperature of 90 K, the current densities were smaller than the sensitivity limit of the measurement ($\sim 0.1$ pA·cm$^{-2}$) and thereby demonstrate the excellent charge blocking ability of both detector contacts. Using the temperature dependence of the electron injection current step and the analysis procedure described in Section 3, the ACS model electron barrier and electron injection prefactor values for the cathode contact were determined to be: $\phi_e = 0.400$ eV and $J_o = 18.4$ A·cm$^{-2}$·K$^{-2}$. Assuming the validity of this simple ACS model and a band gap energy for Ge at these temperatures of 0.72 eV [Sze 1981], it is anticipated that the hole barrier for this contact would be $\phi_h = E_g - \phi_e = 0.32$ eV. Consequently, the Sputterer 1 recipe of 5 mTorr of pure Ar produces an a-Ge contact that is better for electron blocking than it is for hole blocking, that is, it will produce lower leakage current as a cathode as compared to when positively biased as an anode. Though this study focused primarily on electron injection, it is also possible to analyze the Figure 7.1a data in terms of hole injection at the top (anode) contact. This is accomplished by applying the ACS analysis to the low voltage (below full depletion) leakage current. After carrying out such an analysis, the following were obtained: $\phi_h = 0.383$ eV and $J_o = 0.143$ A·cm$^{-2}$·K$^{-2}$. Since this implies that the electron barrier for this contact is 0.34 eV, it can be concluded that the top contact recipe (Sputterer 1, 23 mTorr of Ar with 7% H$_2$) is better for hole blocking than it is for blocking electrons. With the ACS model parameters determined, the leakage current can be estimated for other detector operating temperatures. For example, at the temperature of 90 K, both the hole injection from the anode and the electron injection from the cathode at a detector voltage of 600 V are estimated to be extremely low at just under $10^{-17}$ A·cm$^{-2}$.

In addition to the variable temperature measurements just discussed, the detector current-voltage characteristics were also studied as a function of cumulative time stored at room temperature. The characteristics shown in Figure 7.1b are an example of this. A typical measurement sequence consisted of cooling and testing the detector shortly after its fabrication was completed. After variable temperature current-voltage characteristics were acquired, the detector was warmed and stored at room temperature for a period of time before the cooling and measurement procedure was repeated. This measurement and storage sequence was repeated many times over a typical period of about one month. From these measurements, the room temperature storage stability of the electrical contacts could be determined. A number of conclusions can be made from Figure 7.1b. By looking at the behavior of the leakage current of the partially depleted detector ($V_d < 200$ V), the stability of the hole injection at the top contact can be assessed. Since the leakage current at $\sim 100$ V experienced relatively little change with storage, it is apparent that the Sputterer 1 a-Ge recipe of 23 mTorr of Ar with 7% H$_2$ produces electrical contacts that have storage stable hole injection. Similarly, the stability of the electron blocking of the bottom contact can be determined based on the behavior of the electron injection current step at full depletion. As is clear from the current-voltage characteristics of Figure 7.1b, this current step increased with each measurement cycle and thereby indicates that the Sputterer 1 a-Ge recipe of 5 mTorr of Ar with 0% H$_2$ lacks good storage stability. However, it can be further noted that the rate at which the electron injection increased diminished the longer that the detector was stored at room temperature. Eventually, the electron injected leakage current from this contact would become reasonably stable but at a leakage current value much higher than that measured just after fabrication.

A large set of HPGe detectors was characterized through measurements of the type shown in Figure 7.1. Each of the detectors in this set was produced with a bottom cathode electrical contact whose a-Ge layer was sputtered with a different recipe. Since the top electrical contact recipe on all of the detectors was fabricated using a standard, low hole injection a-Ge process, the leakage current above full depletion was always dominated by electron injection at the bottom contact. Furthermore, the same HPGe crystal or crystals from the same slice of HPGe were used as the starting point for all of the detectors. Because of this, the electron injection associated with the various a-Ge recipes could simply be assessed and compared using the leakage current density measured at a specific detector voltage whose value is above full depletion. A detector voltage of 600 V was chosen for this comparison since it was well above the full depletion voltage of 200 V and, as such, would lessen the impact caused by any surface channel induced field distortions. The effect of sputter gas pressure and H$_2$ content on the electron injection and the injection stability with room temperature storage were studied. A summary of this study is contained in Figures 7.2 and 7.3. In these figures the center contact leakage current density at the detector voltage of 600 V is plotted as a function of the cumulative time the detector was stored at room temperature. The detectors whose data is plotted in Figure 7.2 had bottom contacts sputtered in the sputter deposition system Sputterer 1, whereas those of Figure 7.3 were all sputtered in Sputterer 2. Part (a) of each figure contains data from detectors with bottom contacts sputtered at three different pressures of pure Ar, and, likewise, part (b) has the data from those sputtered in Ar containing 7% H$_2$. Note that all of the measurements were made at a detector temperature of 120 K, which is well above that typically used for the operation of a HPGe detector.

M. Amman, 2018, "Optimization of Amorphous Germanium Electrical Contacts and Surface Coatings on High Purity Germanium Radiation Detectors"

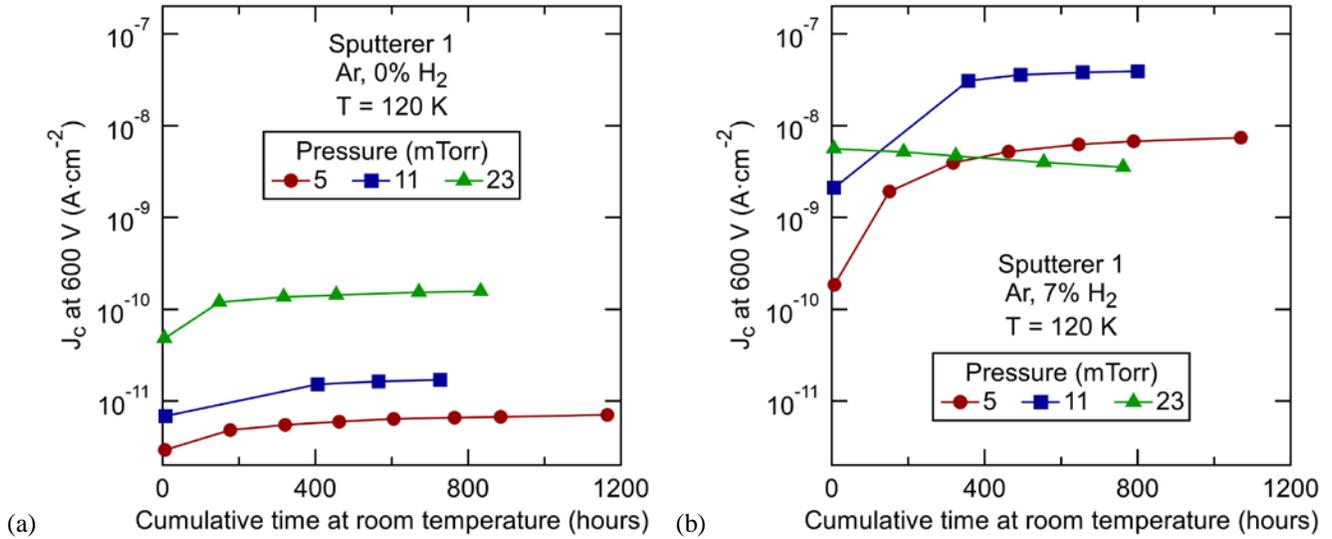

**Figure 7.2** Detector leakage current density measured as a function of storage time at room temperature for HPGe detectors of the type shown in Figure 5.1b. The leakage current density is that measured from the center contact of the guard ring detector when the detector voltage was set to 600 V and the detector temperature to 120 K. This leakage current was dominated by electron injection from the bottom electrical contact. Measurements from multiple detectors are shown, and each detector was produced using a different recipe for the a-Ge of the bottom contact. **(a)** Detectors with bottom contacts sputtered in Sputterer 1 using different pressures of pure Ar. **(b)** Detectors with bottom contacts sputtered in Sputterer 1 using different pressures of Ar containing 7% $H_2$.

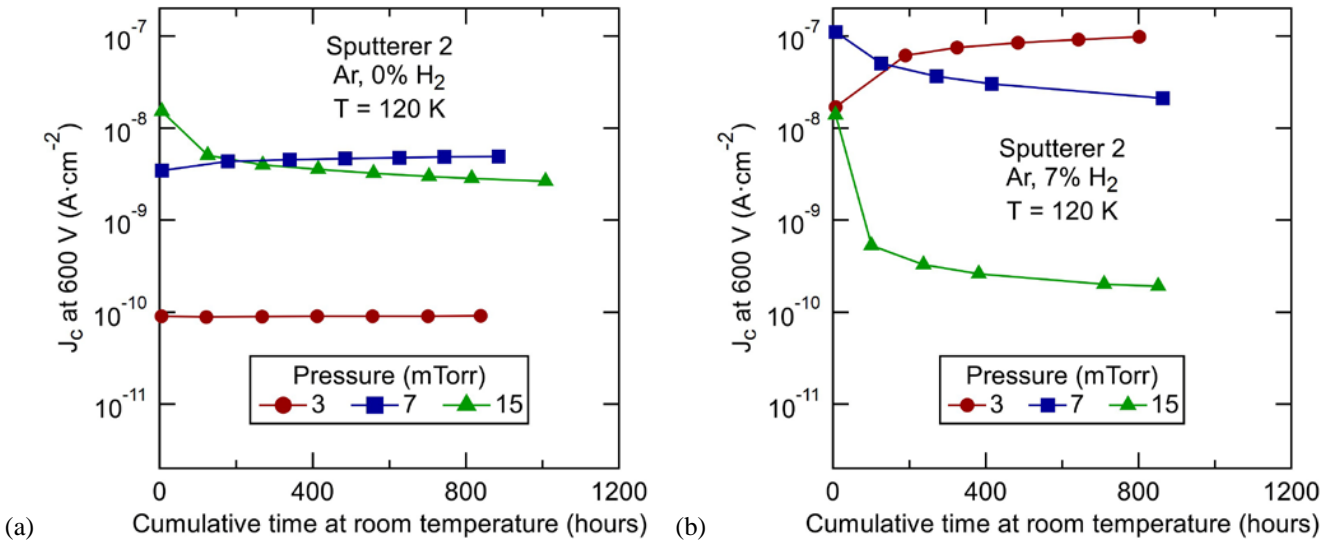

**Figure 7.3** Detector leakage current density measured as a function of storage time at room temperature for HPGe detectors of the type shown in Figure 5.1b. The leakage current density is that measured from the center contact of the guard ring detector when the detector voltage was set to 600 V and the detector temperature to 120 K. This leakage current was dominated by electron injection from the bottom electrical contact. Measurements from multiple detectors are shown, and each detector was produced using a different recipe for the a-Ge of the bottom contact. **(a)** Detectors with bottom contacts sputtered in Sputterer 2 using different pressures of pure Ar. **(b)** Detectors with bottom contacts sputtered in Sputterer 2 using different pressures of Ar containing 7% $H_2$.

M. Amman, 2018, "Optimization of Amorphous Germanium Electrical Contacts and Surface Coatings on High Purity Germanium Radiation Detectors"

Several notable characteristics can be identified in the data of Figures 7.2 and 7.3. Consider first the electron injected leakage current measured shortly after the completion of the detector fabrication and its variation with the various process parameters. From the Sputterer 1 data of Figure 7.2, it is evident that the leakage current increased with increasing sputter gas pressure. This trend existed both for a sputter gas of pure Ar as well as for Ar with 7% $H_2$. The contacts produced with Sputterer 2 whose measurements are in Figure 7.3 exhibited a similar dependency on sputter gas pressure with the exception of the 15 mTorr Ar with 7% $H_2$ detector. However, the leakage current of this detector decreased rapidly after fabrication, and, therefore, its initial value at the time of fabrication was not accurately determined due to the typical 7 hour delay between the a-Ge contact deposition and the detector cooling for the measurements. The leakage current just after fabrication for this detector could have been much higher than that shown in Figure 7.3b and could account for its outlier status. Another finding that can be ascertained from Figure 7.2 is that the addition of $H_2$ to the sputter gas increased the electron injected leakage current measured just after detector fabrication. The magnitude of this change depended on the sputter gas pressure but could be as much as two orders of magnitude or more. A similar though less pronounced effect of adding $H_2$ to the sputter gas was also seen in the data from Sputterer 2 (Figure 7.3) with the exception of the rapidly changing 15 mTorr Ar with 7% $H_2$ detector.

A third observation that can be made regarding the initial leakage current is that overall Sputterer 1 produced electrical contacts with lower electron injection than those produced using Sputterer 2. Numerous differences exist between the two systems as detailed in the previous two sections. Many of these differences, such as target configuration (diode versus magnetron), RF power, DC self bias voltage, deposition rate, substrate temperature, etc., will impact the a-Ge film properties. This highlights a research area in need of further work. Additional exploration of the sputter parameter space is necessary in order to identify parameters that can be used to more extensively adjust the charge injection properties. For example, it might be desirable to identify a set of parameters for Sputterer 2 that would produce lower electron injection in a contact sputtered with Ar containing $H_2$ that also has good room temperature storage stability.

Turning now to the stability of the electron injected leakage current with room temperature storage time, additional relationships between the leakage current and a-Ge sputter process parameters can be identified. To begin, electrical contacts sputtered in Ar with 7% $H_2$ were found to have a stability that depended on the sputter gas pressure. This behavior is apparent from Figures 7.2b and 7.3b. In going from a low sputter gas pressure to a high one, the leakage current transitioned from one that increased with storage time to one that decreased with storage time. Room temperature storage induced changes in the leakage current of more than an order of magnitude were possible depending on the sputter gas pressure. A further observation is that the rate of this change with storage time diminished the longer the detector was stored, and typically the leakage current approached a stable value after long term storage. These observations were also reported previously and were based on an earlier study of HPGe detectors produced using Sputterer 1 at LBNL [Looker 2015a]. The work of this paper has gone further in this aspect of the contact behavior in that the pressure dependent stability has been demonstrated in contacts deposited in two very different sputter deposition systems and, as a result, indicates that such behavior may be characteristic of the sputter deposition process and the resultant film microstructure. A benefit of this pressure dependence is that it provides the means to optimize the contact recipe in order to achieve good storage time stability. For example, based on the measurements shown in Figures 7.2b and 7.3b, relatively stable contacts should be obtained with Sputterer 1 at a pressure of about 23 mTorr and with Sputterer 2 at about 5 mTorr. Note that to compare the Sputterer 1 pressure values of the Looker et al. paper [Looker 2015a] to those of this paper, the pressures of the earlier paper should be multiplied by 1.55 [Lesker 2018]. This is because the Looker et al. paper gave the direct readings from an $N_2$ calibrated convection enhanced Pirani vacuum gauge rather than those with an Ar correction factor applied. This correction has already been applied as needed to the pressures listed in this paper.

Another observation of significance from Figures 7.2 and 7.3 is that as a whole the contacts sputtered in pure Ar produced leakage currents that were more stable with storage at room temperature than those sputtered in the Ar-$H_2$ gas mixture. As discussed previously in Section 6, bonding changes associated with the Ge and H in the film are likely the basis for much of the electrical changes in the a-Ge films and would also reasonably affect the electrical contact charge injection stability. Structural changes in sputter deposited a-Ge have been observed to occur during aging at room temperature [Okugawa 2016] and could play a role in the stability of contacts deposited both with and without $H_2$ in the sputter gas. However, the addition of H to the film provides a further mechanism through which the stability could be affected. Hydrogen diffusion and desorption from the H containing a-Ge film will occur and have been studied by others [Wu 1991] [Graeff 1993] [Beyer 1996], though these studies were at elevated temperatures. It is expected that these processes and their effect on the contact stability would be dependent on the a-Ge sputter deposition conditions. Specifically, the sputter gas pressure affects the amount of sputter gas (both Ar and $H_2$) incorporated into the film as well as the film microstructure as discussed in Section 6. Both of these properties would impact the storage time dependent changes in the film caused by H and could account for the sputter pressure dependent stability of the H containing contacts whose data is plotted in Figures 7.2b and 7.3b.

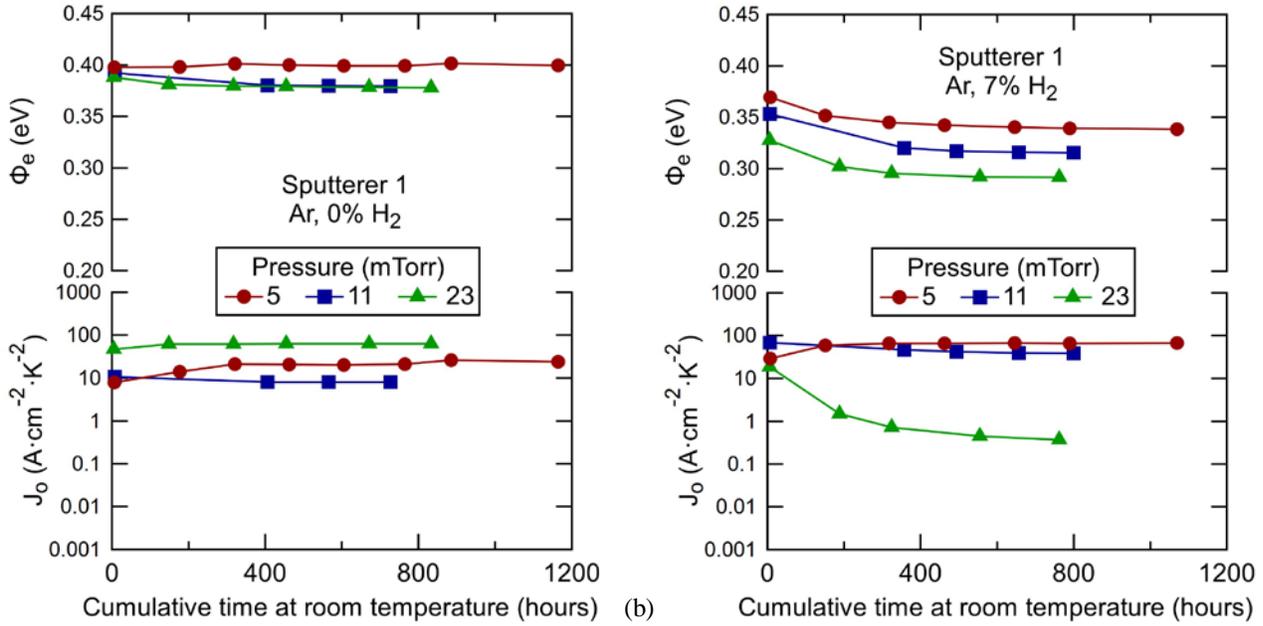

**Figure 7.4** ACS contact model parameters of electron barrier height and electron injection prefactor extracted from leakage current measurements made on HPGe detectors of the type shown in Figure 5.1b. These parameters are those associated with the bottom electrical contact of the detectors. The leakage current measurements were made at the temperatures of 120 K and 130 K. Extracted parameters from multiple detectors are shown, and each detector was produced using a different recipe for the a-Ge of the bottom contact. **(a)** Detectors with bottom contacts sputtered in Sputterer 1 using different pressures of pure Ar. **(b)** Detectors with bottom contacts sputtered in Sputterer 1 using different pressures of Ar containing 7% $H_2$.

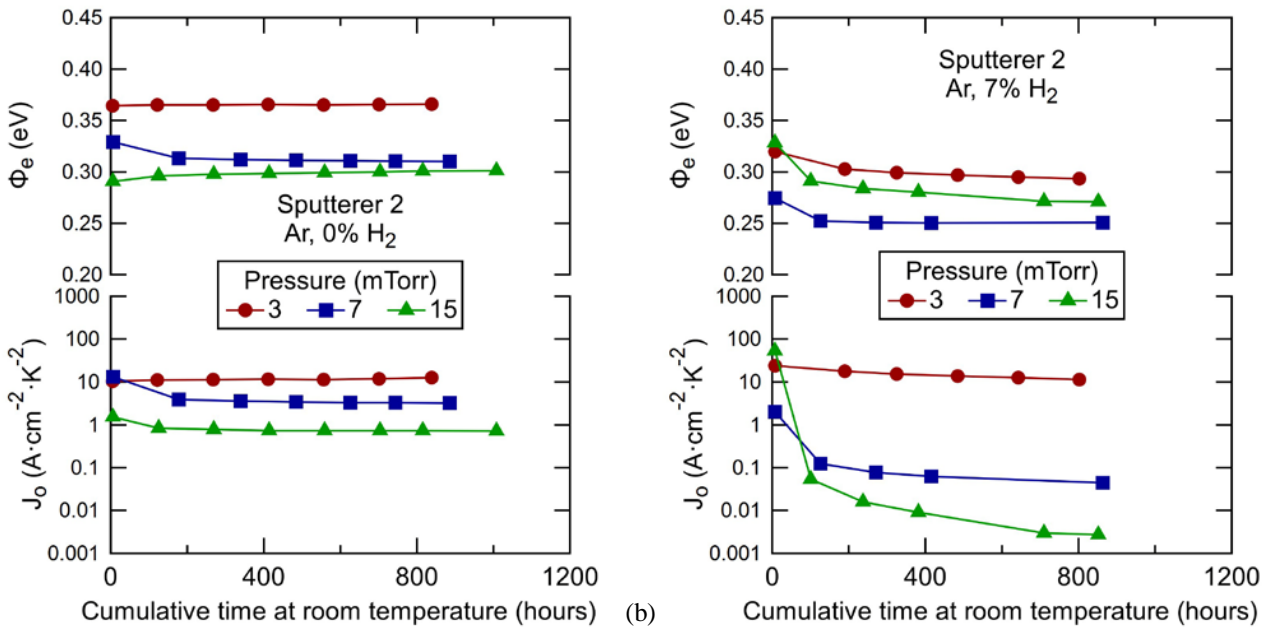

**Figure 7.5** ACS contact model parameters of electron barrier height and electron injection prefactor extracted from leakage current measurements made on HPGe detectors of the type shown in Figure 5.1b. These parameters are those associated with the bottom electrical contact of the detectors. The leakage current measurements were made at the temperatures of 110 K and 120 K. Extracted parameters from multiple detectors are shown, and each detector was produced using a different recipe for the a-Ge of the bottom contact. **(a)** Detectors with bottom contacts sputtered in Sputterer 2 using different pressures of pure Ar. **(b)** Detectors with bottom contacts sputtered in Sputterer 2 using different pressures of Ar containing 7% $H_2$.

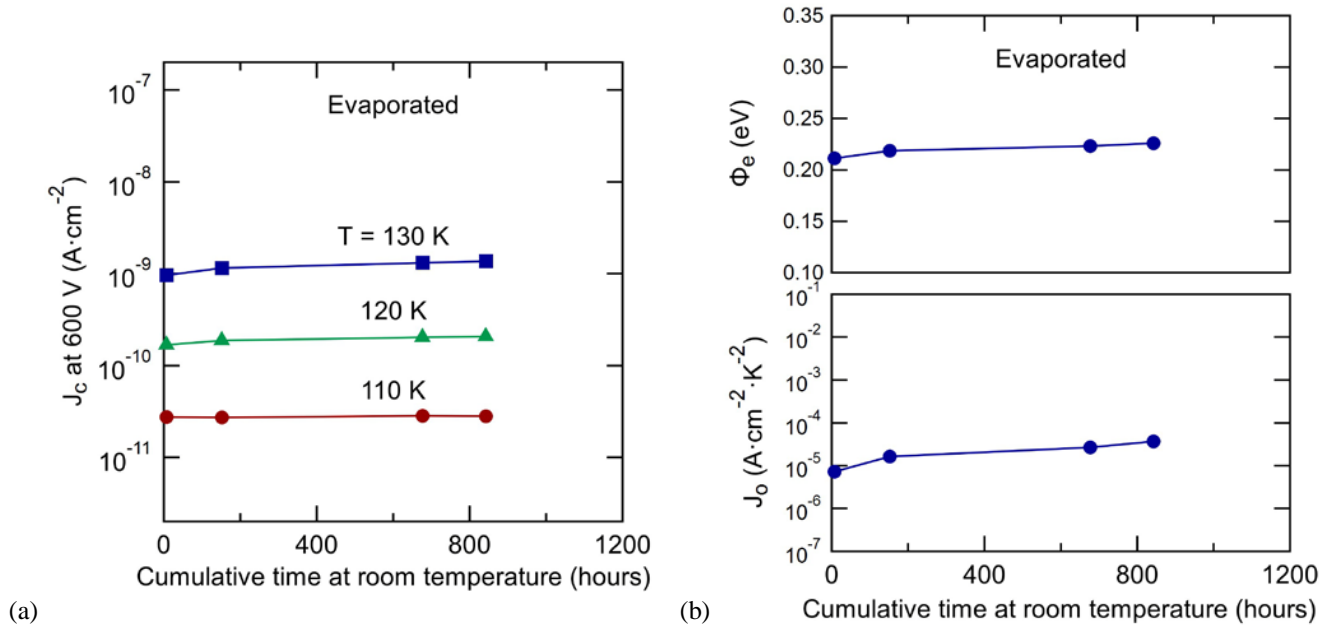

**Figure 7.6 (a)** Measured detector leakage current density plotted as a function of storage time at room temperature for a HPGe detector of the type shown in Figure 5.1b. The leakage current density is that measured from the center contact of the guard ring detector when the detector voltage was set to 600 V. This leakage current was dominated by electron injection from the bottom a-Ge electrical contact that was produced through thermal evaporation of the a-Ge. Measurements are shown for three different detector temperatures: 110 K, 120 K, and 130 K. **(b)** ACS contact model parameters of electron barrier height and electron injection prefactor extracted from leakage current measurements made on the HPGe detector whose leakage current data are shown in part (a) of this figure. These parameters are those associated with the bottom thermally evaporated a-Ge electrical contact.

In order to gain additional physical insight into the electrical contact charge injection and its dependence on the sputter process parameters, the leakage current data from the detectors were analyzed based on the ACS model as described in Section 3. Through this analysis, the contact parameters of electron barrier height, electron injection prefactor, and density of localized energy states near the Fermi level ($\phi_e$, $J_o$, and $N_f$, respectively) were extracted for each a-Ge contact recipe. Plots of the barrier height and prefactor as a function of cumulative storage time at room temperature for the different recipes are given in Figures 7.4 and 7.5. As might be expected based upon the leakage current plots of Figures 7.2 and 7.3, the electron barrier heights of the contacts sputtered in Sputterer 1 were generally larger than those of the contacts sputtered in Sputterer 2, and the barrier heights of contacts sputtered in pure Ar were generally larger than those of contacts sputtered in Ar containing $H_2$. A barrier height reduction with the addition of $H_2$ was also identified in a previous study [Looker 2015a]. A less than expected result, though, is that all contacts sputtered with $H_2$ in the sputter gas exhibited a decreasing barrier height with storage time. This indicates that none of these contacts were truly stable. This is surprising since a relatively stable leakage current could be obtained with this type of contact through optimization of the sputter gas pressure. The plots in Figures 7.4 and 7.5 demonstrate that this leakage current stability at a particular detector temperature is achieved by a fortuitous reduction of the prefactor that compensates for the lowering of the barrier height with storage time.

The final set of leakage current measurements and associated ACS model analysis results to be discussed are given in Figure 7.6. These are from a single HPGe detector with the bottom electrical contact consisting of thermally evaporated Ge with an overlayer of thermally evaporated Al. Thermal evaporation of Ge is easily done, the required deposition system can be simple, and early work in the area of a-Ge contacts was based on such contacts [England 1971] [Hansen 1977]. The downsides of thermal evaporation as compared to sputtering include its directional nature making it difficult to conformally coat surfaces, fewer available process parameters for the adjustment of film properties including microstructure, and the lack of ability to incorporate H into the film as needed to obtain the high resistivities required for many applications. The electron injected leakage current densities of Figure 7.6a indicate that the thermally evaporated contact functions fairly well at blocking electron injection, though not quite as well as the contacts sputtered at low pressures in pure Ar. At a typical detector operating temperature of 90 K, the electron injected current density of the evaporated contact was below the sensitivity limit of the measurement (~ 0.1 pA·cm$^{-2}$). The contact also showed good stability with room temperature storage. Despite the decent electron blocking, the ACS analysis came out with a relatively small electron barrier height for the contact of only about 0.22 eV. The low leakage current resulted from a low electron injection prefactor ($J_o \sim 2.2 \times 10^{-5}$ A·cm$^{-2}$·K$^{-2}$) rather than from a large electron barrier.

M. Amman, 2018, "Optimization of Amorphous Germanium Electrical Contacts and Surface Coatings on High Purity Germanium Radiation Detectors"

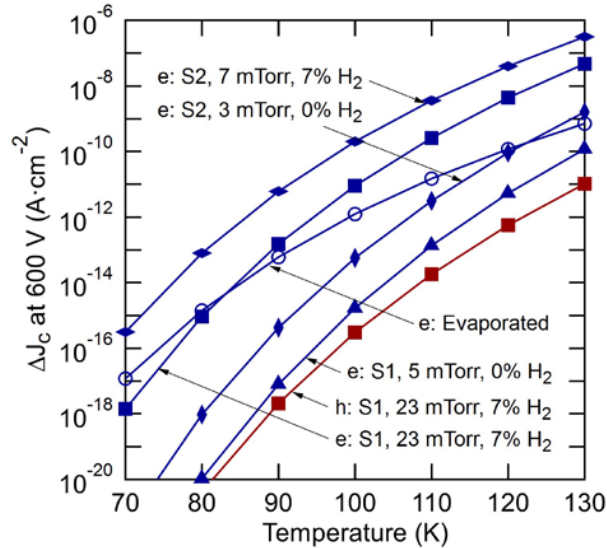

**Figure 7.7** Calculated center contact leakage current densities resultant from either electron injection at the bottom contact (plots labeled with e and shown in blue) or hole injection at the top contact (plot labeled with h and shown in red) of a detector of the type shown in Figure 5.1b. The leakage current densities were estimated assuming a detector voltage of 600 V, a full depletion voltage of 200 V, and the ACS contact model parameters extracted based on fits to the leakage current data measured from detectors with electrical contacts formed using various a-Ge deposition recipes. The a-Ge recipes are designated in the figure with S1 or S2 to specify either Sputterer 1 or Sputterer 2, a pressure specifying the sputter gas pressure, and the percent $H_2$ content of the sputter gas.

    To end this section, the selection of a-Ge contact process recipes for two different detector configurations will be discussed based on the knowledge gained from this and the previous sections. To aid this discussion, estimated leakage currents for various recipes are plotted as a function of temperature in Figure 7.7. These estimates were calculated using Equations 3.9 and 3.10, and the ACS model parameters of Figures 7.4, 7.5, and 7.6b, as well other model parameters extracted from the leakage current data.

    As the first example, consider the large area orthogonal strip detectors developed for the GRIPS and COSI instruments that were presented in Section 2. The most straightforward processing sequence for the production of these detectors would be to completely sputter coat all detector surfaces with a-Ge and then deposit and pattern the Al electrodes. For this simple process, an a-Ge coating of a single recipe (or at most two) provides the charge injection blocking of the electrical contacts, the passivation coating on the sides of the detector, and the passivation coating between the strips. As concluded in Section 6, the a-Ge should be sputtered in a high pressure of Ar containing $H_2$ so that the electronic noise from the parallel resistance associated with the film will be negligible and to eliminate charge collection to the inter-strip surface. A sputter gas mixture with 7% $H_2$ should also produce a desirable electrical state on the sides of the detector that will substantially eliminate charge collection to these surfaces [Hansen 1980]. As detailed in Section 4, the noise associated with the bulk injected leakage current must also be considered when selecting an electrical contact technology. In that section, from an electronic noise perspective, it was concluded that strip leakage currents as high as 100 pA would be acceptable for the GRIPS and COSI detectors because the noise contribution from this current would not significantly contribute to the total electronic noise. These detectors have strip areas of order 1 $cm^2$ and are operated at temperatures ranging from about 70 to 100 K. Assuming the highest operating temperature of 100 K and referring to the leakage current density plots of Figure 7.7, it can be concluded that a recipe of Sputterer 1, 23 mTorr Ar, and 7% $H_2$ will satisfy the leakage current requirement. To make this conclusion, both the hole injection at the anode strips and the electron injection at the cathode strips must be considered since it is the total leakage current at a strip that is of concern. The hole injection can be determined from the curve shown with red squares in Figure 7.7 while that of the electrons from the curve shown with blue squares. It is clear from the two curves that the hole injection is several orders of magnitude lower than that of the electrons for this a-Ge recipe and is therefore not of concern. At a temperature of 100 K, the injected hole current for a 1 $cm^2$ area strip would be about $3 \times 10^{-4}$ pA while that from electron injection would be about 9 pA. This easily meets the electronic noise related leakage current requirement. For all of the reasons just discussed along with the demonstrated leakage current stability associated with the a-Ge process of Sputterer 1, 23 mTorr Ar, and 7% $H_2$, this process is a good choice for the strip detectors and was used to produce the COSI detectors. The GRIPS detectors, however, presented an additional challenge. The readout electronics could not accommodate a leakage current above about 10 pA on any channel, and it was uncertain as to the operating temperature of the detectors since the cooling system was not completed until after the detectors were produced. This meant that electron injection could become a problem. For this reason, the cathode face of the GRIPS detectors used a-Si sputtered in Ar with 7%

$H_2$ rather than a-Ge. It was previously shown that a-Si deposited with Sputterer 1 using Ar containing $H_2$ has lower electron injection than a-Ge sputtered in the Ar-$H_2$ gas mixture [Amman 2007] [Looker 2015a]. All surfaces other than the cathode face of the GRIPS detectors were coated with a-Ge deposited using the COSI detector process of Sputterer 1, 23 mTorr Ar, and 7% $H_2$.

The second example is that of a cylindrical geometry detector with a thin entrance window contact. Specific configurations include the p-type point contact (PPC) detector schematically shown in Figure 1.3 or a coaxial detector constructed from p-type HPGe. For the thin window electrical contact, which will be at a positive voltage bias, the hole injection should be low at a level that the associated current noise does not significantly contribute to the total electronic noise of the detector. Meeting this requirement is more challenging with the point contact detector than it is with the large area strip detectors. As discussed in Section 4, the low capacitance of the point contact readout electrode along with the typically cooled front end electronics used for signal readout leads to a much lower voltage noise component of the electronic noise for the PPC detector relative to that typically achievable with the strip detectors. This means that the current noise and, therefore, the injected leakage current of the PPC detector should also be much less than what is acceptable for the strip detectors. Furthermore, the window contact area of the PPC can be very large, close to 100 $cm^2$. For a large area PPC detector, reasonable limits on the leakage current and operating temperature would be about 1 pA and 100 K. For a 100 $cm^2$ area of an a-Ge contact of recipe Sputterer 1, 23 mTorr Ar, and 7% $H_2$ operated at 100 K, the hole injected current would be about 0.03 pA based on the red plot of Figure 7.7. Consequently, the low hole injection requirement should be easily met with this a-Ge recipe. Potentially, other recipes should also be able to meet this requirement and may have lower hole injection. Though the study of this paper focused on electron injection of the a-Ge contacts, as discussed previously, hole injection blocking ability can be inferred based on the extracted electron ACS model parameters. A poor electron blocking contact with a low electron potential energy barrier should have a large hole barrier and be a good hole blocking contact. Note that, unlike the strip detector example, the a-Ge recipe for the thin window contact on the PPC detector can in principle be chosen without regard to the resistivity of the a-Ge film. This means that low resistivity film recipes can be considered. Based on an inspection of the electron barrier heights given in Figures 7.4, 7.5, and 7.6, the recipes of a-Ge sputtered using Sputterer 2 with Ar containing $H_2$ and evaporated a-Ge look to be interesting candidates for further exploration of their hole blocking characteristics.

## 8. Summary and Possible Future Work

Driven primarily by the need for HPGe based radiation detectors capable of good position resolution in addition to excellent energy resolution, the use of amorphous semiconductor electrical contacts and surface passivation coatings to produce such detectors have expanded substantially over the last decade. The amorphous semiconductor technologies have been successfully used to implement a variety of detector configurations, and the resultant detectors have been utilized in a wide range of application areas including space science, material science, nuclear and particle physics, medical imaging, nuclear nonproliferation and homeland security, and environmental remediation. The demonstrated advantages of the amorphous semiconductor contacts are that the contacts are simple to fabricate, are easily segmented as needed for radiation interaction position readout, produce thin entrance windows, block both electron and hole injection, naturally lead to a fully passivated detector, and enable simple implementation of unique detector configurations. These contacts have been shown to be particularly of value as a replacement for the thick and difficult to segment Li diffused hole blocking contact that is typically used on conventional HPGe detectors.

Deposited films of both a-Ge and a-Si are successfully employed as electrical contacts and passivation surface coatings on HPGe detectors. An extensive literature dating back many decades exists regarding these films. Of particular interest are the measured thin film material properties along with their relationships to deposition conditions, the developed understanding of the electrical charge transport through the films, and the application of metal to semiconductor contact theory to the amorphous semiconductor to crystalline semiconductor contact. The knowledge gained from this literature combined with fabrication process studies directed at improving HPGe detector performance have advanced and continue to advance the detector technology. The study of this paper focused on RF sputtered a-Ge and its application as both an electrical contact and surface coating on HPGe. The a-Ge can impact the performance of the detector through a number of processes including charge distribution fluctuations within the a-Ge surface coating (current noise associated with the parallel resistance of the film and $1/f$ noise), charge injection at the a-Ge contact (current noise associated with the detector leakage current), and undesirable charge collection to the a-Ge surface coating rather than to the a-Ge electrical contact. The a-Ge properties relevant to these processes include the a-Ge film resistance, electron and hole injected leakage current of the a-Ge to HPGe contact, and the stability of the resistance and current with room temperature storage. These properties were measured as a function of the sputter deposition conditions of sputter gas pressure and $H_2$ content. Two different sputter deposition systems were used to produce a-Ge thin film resistors and HPGe detectors so that common dependencies of the a-Ge properties on the deposition conditions could be identified.

The measurements from the a-Ge resistor samples of this study in part supported the findings of other previous studies. With respect to the operation of an HPGe detector, the charge transport within the a-Ge at low temperatures near about 90 K is relevant. Based on the literature, the charge transport at this temperature is dominated by that taking place through localized electron energy states near the Fermi energy level within the energy band gap of a-Ge. The density of these defect states can be reduced and the film resistance increased through the incorporation of H into the film since H can compensate dangling Ge bonds and thereby modify or remove the associated energy gap defect states. Comparing samples sputtered in the same sputter deposition system and at the same

M. Amman, 2018, "Optimization of Amorphous Germanium Electrical Contacts and Surface Coatings on High Purity Germanium Radiation Detectors"

gas pressure, the addition of 7% $H_2$ to Ar increased the low temperature electrical resistance of the resultant film by about three orders of magnitude or more compared to that sputtered in pure Ar. This was observed to be true regardless of the sputter deposition system or the sputter gas pressure used to produce the samples. The influence of sputter gas pressure on film resistance was also studied since pressure is one of many factors capable of affecting the film microstructure, and it was demonstrated in a previous LBNL study to affect the room temperature storage stability of the a-Ge electrical contact leakage current. Others have previously shown that at lower sputter gas pressures the a-Ge film will be denser, have a lower void fraction, have more incorporated Ar, and tend to be under compressive stress, whereas at higher pressures the film will be less dense, have a higher void fraction, have less incorporated Ar, and be more likely to be under tensile stress. The measurements presented in this paper demonstrated that, when all settable process parameters other than the gas pressure were fixed, the a-Ge resistance at low temperatures consistently increased with increasing gas pressure regardless of the gas $H_2$ content or which of the two sputter deposition systems was used to produce the film. Possible explanations for this include the more energetic bombardment of particles onto the film at lower sputter pressures introducing more film disorder or the dense microstructure obtained at low pressures being less susceptible to oxidation as compared to that of the more open microstructure of high pressures. Another interesting observation was related to the temperature dependence of the film resistance. Since a variable range hopping charge transport model applicable to a-Ge predicts that the low temperature resistance should have an exponential dependence on temperature to the power of -1/4, the log of the resistance is typically plotted as a function of temperature to the power of -1/4. The slopes of the resultant linear resistance plots were found to be roughly the same for all of the samples that had relatively low to moderate resistance values regardless of the sputter deposition system used to produce the a-Ge film, the $H_2$ content of the sputter gas, or the substrate type (glass or depleted HPGe). This was true for resistors that differed in resistance by many orders of magnitude. Such a constancy of the slope has also been noted in the literature. In contrast to this, however, were the high resistance samples. These resistors, which were sputtered in a high pressure of Ar containing $H_2$, exhibited steeper slopes (more rapid resistance change with temperature) than that of the low to moderate resistance samples. The resistor stability with extended room temperature storage was also measured with a subset of the resistor samples. For most of these samples, the low temperature resistance was observed to increase with room temperature storage time, and the rate of this increase appeared to lessen over time. However, in one of the relatively high resistance samples, the resistance was seen to both increase and decrease depending on the measurement temperature. This particular sample was at the transition point between lower resistance samples that all had nearly the same slope in their resistance plots and the high resistance samples that had a greater rate of change in resistance with temperature. Room temperature storage appeared to convert the temperature behavior of this resistor from one that was near that of the moderate resistance samples to that characteristic of the high resistance samples.

The study of the a-Ge contact charge injection was carried out with planar p-type HPGe detectors incorporating a guard ring so that surface leakage current could be eliminated from the measurements. The measurements focused primarily on electron injection. This property was quantified using the step increase in the leakage current that occurs when the detector depletion first reaches through the HPGe to the cathode a-Ge electrical contact. By varying the a-Ge recipe of the cathode, the electron injection was examined as a function of the a-Ge sputter deposition process. In most cases, at a typical detector temperature of 90 K, the current densities were smaller than the sensitivity limit of the measurement (~ 0.1 pA·cm$^{-2}$) and thereby demonstrated the good electron blocking ability of the contacts. For this reason, leakage current measurements were made at significantly higher temperatures. As was done in the study of the a-Ge film resistance, two different RF sputter deposition systems were used to deposit the a-Ge contacts. Sputterer 1 was an older unit with a diode configured target, and Sputterer 2 was a relatively modern system with a magnetron configured target. In addition to the target geometry and configuration differences, there were also many other significant dissimilarities between the two systems. The base pressure, unintentional sample heating, RF power, DC self bias voltage, and deposition rate were all greater in Sputterer 1. Despite the numerous differences between the two deposition systems, several common behaviors between the two were identified in the a-Ge contacts that they produced. One was that the electron injected leakage current measured soon after detector fabrication generally increased with increasing sputter gas pressure. This trend existed both for a sputter gas of pure Ar as well as for Ar with 7% $H_2$. Another common behavior was that the addition of $H_2$ to the sputter gas generally increased the electron injected leakage current measured soon after detector fabrication. The magnitude of this increase depended on the sputter deposition system and sputter gas pressure but could be as much as two orders of magnitude or more.

The stability of the electron injection with room temperature storage was also assessed. One behavior in common between the two sputter deposition systems that was related to this stability was that the set of HPGe detectors as a whole exhibited better stability of the electron injected leakage current when the contacts were sputtered in pure Ar as compared to sputtering in Ar containing 7% $H_2$. Such a behavior is not unreasonable to expect since diffusion, desorption, or any bonding changes associated with H in the Ar-$H_2$ sputtered contact provide additional mechanisms by which the contact can change that would not exist in a contact produced with pure Ar. Another stability related behavior explored in the study was the dependence on the sputter gas pressure for contacts sputtered in Ar containing 7% $H_2$. Regardless of the sputter deposition system used to deposit the a-Ge, when going from a low sputter gas pressure to a high one, the electron injected leakage current transitioned from one that increased with storage time to one that decreased with storage time. Room temperature storage induced changes in the leakage current of more than an order of magnitude were possible depending on the sputter gas pressure. Furthermore, it was observed that the rate of this change with storage time diminished the longer the detector was stored. A benefit of this pressure dependence is that it provides a means to optimize the contact recipe in order to achieve good storage time stability. These observations concerning the stability of the a-Ge contacts sputtered in Ar containing $H_2$ confirmed those noted in work previously published by LBNL.

To this point, this summary of the electrical contact study has covered the relationships between the measurements and sputter process parameters that were observed regardless of which of the two sputter deposition systems was used to deposit the a-Ge. However, differences between the contacts produced by the two systems were also noted. One of significance was that, overall, Sputterer 1 produced electrical contacts with lower electron injection when measured soon after fabrication as compared to those produced using Sputterer 2.

For comparison purposes, a single thermally evaporated a-Ge contact was also fabricated and tested. This contact functioned fairly well at blocking electron injection. At a typical detector operating temperature of 90 K, the electron injected current density was below the sensitivity limit of the measurement. Furthermore, the electron injection from this contact was found to be generally lower than that of the a-Ge contacts sputtered in the Ar-$H_2$ gas mixture but not quite as low as that obtained with the contacts sputtered at low pressures in pure Ar. The evaporated contact also exhibited good stability with room temperature storage.

In order to gain additional insight into the physics of the contacts, the leakage current measurements were analyzed based on the ACS contact model. The end product of this analysis was a set of extracted model parameters including electron potential energy barrier height and electron injection current proportionality prefactor. One interesting result from this analysis was that all contacts sputtered with $H_2$ in the sputter gas exhibited a decreasing barrier height with room temperature storage time. This meant that none of the contacts were truly stable. This was surprising since a relatively stable leakage current was obtained with this type of contact through optimization of the sputter gas pressure. It turned out that the leakage current stability at a particular detector temperature was achieved by a fortuitous reduction of the prefactor with storage time that compensated for the lowering of the barrier height. Another result from the analysis was that the electron barrier heights of the contacts sputtered in Sputterer 1 were generally larger than those of the contacts sputtered in Sputterer 2. Regarding the evaporated a-Ge contact, despite its decent electron blocking, the analysis revealed a relatively small electron barrier height for the contact of only about 0.22 eV. The low leakage current resulted from a low electron injection prefactor rather than from a large electron barrier.

Under the assumption that the ACS contact model accurately accounts for the a-Ge contact physics, the hole blocking ability of select contacts was assessed based on the electron injection current measurements and the ACS analysis. This was possible since, in the ACS model, the sum of the hole barrier and electron barrier should be approximately equal to the energy bandgap of Ge. As an example, the a-Ge contact produced with Sputterer 1 using pure Ar at a low pressure of 5 mTorr had an electron barrier of about 0.40 eV. This is greater than half the bandgap of Ge and indicates that the contact would be better at blocking electrons when negatively biased than it would be at blocking holes when positively biased. In contrast to this is the contact with the recipe of Sputterer 1, 23 mTorr Ar, and 7% $H_2$. This contact had an electron barrier of about 0.30 eV, which is significantly less than half the Ge bandgap and implies that it should operate with lower leakage current as a hole blocking, positive contact. This superiority of the hole blocking was directly demonstrated by its use as the anode contact recipe for most all of the detectors used in this study, which all had low hole injection. Furthermore, a hole barrier height of 0.38 eV was extracted from hole injection measurements made with one of these detectors. Overall, the contact analysis revealed that the sputtered contacts tended to have barrier heights near half the bandgap of Ge, thereby making them reasonably good at blocking the injection of both electrons and holes. Additionally, it was shown that by adjusting the sputter process parameters, such as sputter gas pressure and $H_2$ content, the balance between the barrier heights could be shifted, and, as a result, the blocking of one carrier type could be improved at the expense of the other. In contrast to the sputtered contacts was the evaporated Ge contact that had a small electron barrier height of only 0.22 eV. This result implies that the contact should have a much larger hole barrier and be an excellent hole blocking contact. The hole blocking capability of the contact, however, was not evaluated.

To end this paper, some thoughts on possible future work will be given. The study reported in this paper was not wide-ranging, and there remain many possible areas for further exploration. To begin, the applicability of this paper's findings (as well as those of other recent LBNL papers in this area) to other a-Ge deposition systems and processes should be explored. Additionally, the LBNL work to date has examined only a small number of the possible processing variables. Further exploration of process parameters could be aimed at specific goals such as establishing a greater ability to adjust and shift the blocking behavior between that favoring holes and that favoring electrons, or achieving specific combinations of film resistivity, charge carrier blocking, and room temperature stability. The process parameters of interest encompass the variables directly associated with the sputter deposition of the film such as sputter system geometry, target composition, target configuration, gas composition, gas pressure, gas flow rate, RF power, substrate temperature, deposition rate, ion bombardment, etc., as well as those related to the preparation of the HPGe surface prior to the sputter deposition. The surface preparation would include processes such as mechanical polishing, chemical polishing, and chemical treatment. As an example of the type of work that could be done, consider the contact and surface coating requirements of the COSI and GRIPS strip detectors along with the properties of the a-Ge films and contacts produced with the two LBNL sputter deposition systems. As discussed previously, this application requires a very high resistivity a-Ge surface coating, good electron blocking of the cathode strips, good hole blocking of the anode strips, and stability of these properties with room temperature storage. The recipe of Sputterer 1, 23 mTorr Ar, and 7% $H_2$ was capable of meeting all of these requirements and was used to produce the COSI detectors. However, accomplishing this with Sputterer 2 appears to be more of a challenge primarily because of greater electron injection. Though a recipe such as Sputterer 2, 15 mTorr Ar, and 7% $H_2$ has low electron injection after several days of room temperature

M. Amman, 2018, "Optimization of Amorphous Germanium Electrical Contacts and Surface Coatings on High Purity Germanium Radiation Detectors"storage, the substantial change in the property with time is of concern. Furthermore, the current-voltage characteristics associated with this recipe typically exhibited an abnormally shaped step at full depletion, possibly indicating a non-uniform depletion caused by an undesirable surface channel on the side of the detector. Consequently, implementation of an orthogonal strip detector process with Sputterer 2 that is equivalent to the successful one demonstrated with Sputterer 1 would require further study of the sputter parameter space.

This paper concentrated on a subset of the properties important to the performance of an HPGe detector. Other properties, such as the surface channel charge state, inter-contact charge collection, passivation capability, and maximum sustainable electric field before the breakdown of the contact, should also be evaluated as a function of the a-Ge process parameters. As described earlier in this paper, some work has been done in the past related to these a-Ge properties, but more extensive studies would be of value. In addition to examining these properties as a function of the process recipe, the possible influence of the HPGe net impurity type and concentration should also be explored. For example, the first work showing the use of a-Ge as an HPGe surface coating demonstrated that the adjustment of the surface channel state was dependent on the HPGe impurity type and concentration.

Another avenue of study is the stability of the a-Ge film and contact properties with detector annealing. Depending on the application, an HPGe detector may require periodic annealing in order to repair the damage caused by absorbed radiation. A typical anneal can consist of raising the detector temperature to a value of 100 to 120°C for a duration ranging from a day to a week. Even if significant damage is not anticipated during the use of the detector, some detector configurations necessitate elevated temperature fabrication processes. For example, the patterning of the fine strips of the GRIPS detectors was done with a photolithography process that included an oven bake of the detector at a temperature of 90°C for 30 minutes. As mentioned earlier in this paper, a body of literature exists regarding the significant impact of annealing on a-Ge film resistivity. However, the effect of above room temperature anneals on the other a-Ge properties of importance to HPGe detectors has not been well studied. As has been clearly shown in this and previous papers, room temperature provides sufficient thermal energy to cause changes in the charge injection properties of the a-Ge contacts, and the significance of these changes are recipe dependent. It is therefore also reasonable to anticipate that the impact of elevated temperature anneals will be recipe dependent and that recipes could be identified that minimize any undesirable changes. It is also possible that for some a-Ge recipes a post detector fabrication anneal could be used to adjust specific properties for the betterment of the detector performance.

The results from the HPGe detector with a thermally evaporated a-Ge contact that was tested as part of this study point to another area worth further exploration. The small electron barrier that was extracted for this contact implies a large hole barrier. This combined with the small charge injection prefactor means that the contact should be excellent for blocking holes. Possible additional work with this contact recipe includes testing the reproducibility and consistency of the electron blocking behavior and verifying the expected hole blocking. The low resistivity of the film would, however, present a problem for applications that require the film to also be used as a surface coating between the electrical contacts.

Beyond the above described work intended to directly impact detector performance would be research aimed at elucidating the basic science of the a-Ge technology. Many possible avenues for research in this area exist, and only a couple will be mentioned here. From a big picture standpoint, one objective would be to strengthen the understanding of the interconnection between the a-Ge fabrication process parameters, the microscopic film characteristics such as the film microstructure and H incorporation, and the properties of interest for detectors including resistivity, charge carrier injection, and surface passivation. As a specific example, there is the apparent dependence of the a-Ge film resistivity on the film microstructure. The high resistance films obtained with high sputter gas pressures are likely to have a significant density of voids and have a column like structure. Confirming that the microstructure is one of the key attributes controlling the film resistivity and clarifying the mechanism by which the microstructure dictates the resistivity would be important contributions. If the very high resistivity films are columnar and more open in structure, it would follow that the film resistivity could exhibit anisotropy. A higher resistivity might be expected along the direction perpendicular to the film growth as compared to that parallel to this direction. The more open film structure could also negatively affect the passivation ability of the film when used as a surface coating. Testing for this anisotropy in various a-Ge films including those with very high resistivities as well as evaluating the passivation effectiveness of the films would be interesting and useful research investigations.

Another area of work relates to the a-Ge contacts sputtered with $H_2$ in the sputter gas. These contacts exhibited changes with room temperature storage that included electron barrier height reduction and sputter pressure dependent leakage current stability. A further study of this behavior with a goal of identifying the physical mechanisms behind these changes would be of interest.

A final basic research topic to be mentioned relates to the physics behind the charge transport within the a-Ge film. A significantly large literature exists that relates to the experimental and theoretical study of the charge transport within a-Ge thin films. At the temperatures of typical interest for the operation of HPGe detectors, this charge transport has been shown to take place through spatially localized energy states near the Fermi energy located within the energy bandgap of the a-Ge. A theoretical model for this transport is variable range hopping. When these same a-Ge films are used as an electrical contact on HPGe, a theoretical model based on metal to semiconductor contact theory has been used to explain the behavior of the charge transport between the a-Ge and the HPGe. In this paper, this theory was referred to as the ACS theory. These two charge transport models have associated model parameters that can be extracted from measurements made on resistor and HPGe detector samples. Since the parameters of the two models should in some cases be related to each other, the study of the two different device structures provides a unique method to explore the physics of the charge transport and test the expected relationship and consistency between the two models.

M. Amman, 2018, "Optimization of Amorphous Germanium Electrical Contacts and Surface Coatings on High Purity Germanium Radiation Detectors"

## Acknowledgments

The author thanks Paul Luke for useful discussions throughout this project and Julie Lee for help in producing this paper. The author also thanks the numerous collaborators who were involved in projects that contributed to the knowledge needed to write this paper. They include Steve Boggs and the NCT/COSI team, Albert Shih, Pascal Saint-Hilaire and the GRIPS team, Alan Poon and the Majorana team, Dongming Mei and the University of South Dakota team, Steve Asztalos, Craig Tindall, Quinn Looker, Anders Priest, Ren Cooper, Paul Barton, and Kai Vetter.

This work was supported by the U.S. Department of Energy, Office of Science, under contract number DE-AC02-05CH11231.

M. Amman, 2018, "Optimization of Amorphous Germanium Electrical Contacts and Surface Coatings on High Purity Germanium Radiation Detectors"

**M. Amman, 2018, "Optimization of Amorphous Germanium Electrical Contacts and Surface Coatings on High Purity Germanium Radiation Detectors"**